\documentclass{article}
\usepackage{graphics,amsmath,amsfonts,revsymb,latexsym, enumerate}
\newtheorem{theorem}{Theorem}[section]
\newtheorem{corollary}[theorem]{Corollary}
\newtheorem{lemma}[theorem]{Lemma}

\newtheorem{proposition}[theorem]{Proposition}

\newtheorem{defi}[theorem]{Definition}
\newtheorem{eg}[theorem]{Example}

\newcommand{\qed}{{\hfill$\Box$}}
\newenvironment{proof}{\noindent \textbf{{Proof~} }}{\qed}

\def\bi{\begin{itemize}}
\def\ei{\end{itemize}}
\def\be{\begin{equation}}
\def\ee{\end{equation}}
\def\bea{\begin{eqnarray}}
\def\eea{\end{eqnarray}}
\def\ben{\begin{eqnarray*}}
\def\een{\end{eqnarray*}}
\def\non{\nonumber}
\def\>{\rangle}
\def\<{\langle}
\def\l{\left}
\def\r{\right}

\def\ra{\rightarrow}

\def\ot{\otimes}

\def\eps{\epsilon}

\def\bR{{\bf R}}
\def\bS{{\bf S}}

\def\bbL{{\mathbb{L}}}
\def\bbM{{\mathbb{M}}}
\def\bbN{{\mathbb{N}}}
\def\bbP{{\mathbb{P}}}

\def\bbT{\mathbb{T}}

\newcommand{\1} I %{{\openone}}

\newcommand{\fig}[1]{Fig.~\ref{fig:#1}}
\newcommand{\eq}[1]{(\ref{eq:#1})}
\newcommand{\peq}[1]{(\ref{eq:#1})}
\newcommand{\eqs}[2]{(\ref{eq:#1}) and (\ref{eq:#2})}
\newcommand{\sect}[1]{Section~\ref{sec:#1}}

\newcommand{\thm}[1]{Theorem~\ref{thm:#1}}

\newcommand{\lem}[1]{Lemma~\ref{lemma:#1}}
\newcommand{\cor}[1]{Corollary~\ref{cor:#1}}
\newcommand{\mscite}[1]{Ref.~\cite{#1}}

\newcommand{\bra}[1]{\langle #1 |}
\newcommand{\ket}[1]{| #1 \rangle}

\newcommand{\proj}[1]{| #1 \>\!\< #1 |}

\newcommand{\smfrac}[2]{\mbox{$\frac{#1}{#2}$}}

\newcommand{\sourceRI}{{\stackrel{{\rm s}}{\geq}}}

\def\half{\smfrac{1}{2}}

\DeclareMathOperator{\id}{id}

\DeclareMathOperator{\tr}{Tr}

\def\*{\star}
\def\ext{\supseteq}
\def\reduction{\stackrel{{}_{\scriptstyle *}}{\geq}}
\def\tilde{\widetilde}
\def\bar{\overline}
\newcommand{\abs}{{\rm abs}}
\newcommand{\rel}{{\rm rel}}
\newcommand{\aux}{{\rm aux}}
\newcommand\app[1]{{\cal A}^{#1}}

\def\cA{{\cal A}}		 \def\cB{{\cal B}}		 
\def\cC{{\cal C}}
\def\cD{{\cal D}}		 \def\cE{{\cal E}}		 
\def\cF{{\cal F}}

		 \def\cK{{\cal K}}		 
\def\cL{{\cal L}}

\def\cM{{\cal M}}		 \def\cN{{\cal N}}		 

\def\cP{{\cal P}}		 		 
\def\cR{{\cal R}}
\def\cS{{\cal S}}		 \def\cT{{\cal T}}		 
\def\cU{{\cal U}}
		 		 
\def\cX{{\cal X}}
\def\cY{{\cal Y}}		 

\def\qq{[q\, q]}
\def\qtq{[q\ra q]}

\def\ctc{[c\ra c]}

\def\cc{[c\,c]}

\def\coftau{[q \ra qq : \tau]}
\def\ctctau{[c \ra c : \tau]}
\def\qtqtau{[q \ra q : \tau]}

\begin{document}

%\title{A High Level Framework for Quantum Shannon Theory
%\protect\vspace{1cm}}
\title{A Resource Framework for Quantum Shannon Theory
\protect\vspace{1cm}}

\author{I. Devetak\footnote{\tt devetak@usc.edu}\protect\\
\it{Department of Electrical Engineering--Systems,} 
\it{University of Southern California,}\protect\\
\it{Los Angeles, CA 90089, USA}\protect\\
\protect\\
A. W. Harrow\footnote{\tt a.harrow@bris.ac.uk}\protect\\
\it{MIT Physics Dept., 77 Massachusetts Ave, Cambridge, MA 02139, USA}\protect\\ 
\it{Department of Computer Science, University of Bristol, BS8 1UB, UK}\protect\\ 
\protect\\
A. Winter\footnote{\tt a.j.winter@bris.ac.uk}\protect\\
\it{Department of Mathematics, University of Bristol, Bristol BS8 1TW, UK}}

%\date{\small (Dated 26 November 2005)} 
\date{\today}

\maketitle

\thispagestyle{empty}

\begin{abstract}
Quantum Shannon theory is loosely defined as
a collection of coding theorems,
such as classical and quantum source compression, 
noisy channel coding theorems, entanglement distillation, etc.,
which characterize asymptotic properties of quantum 
and classical channels and states. In this paper
we advocate a unified approach to an important class of 
problems in quantum Shannon theory, consisting of those that
are bipartite, unidirectional and memoryless.

We formalize two principles that have long been tacitly understood.
First, we describe how the Church of the larger Hilbert space allows
us to move flexibly between states, channels, ensembles and their
purifications.  Second, we introduce finite and asymptotic (quantum)
information processing \emph{resources} as the basic objects
of quantum Shannon theory and recast the protocols used in direct
coding theorems as {\em inequalities} between resources.
We develop the rules of a \emph{resource calculus} which allows us 
to manipulate and combine resource inequalities. This framework
simplifies many coding theorem proofs and    
provides structural insights into the logical dependencies among coding theorems.

We review  the above-mentioned basic coding
results and show how a subset of them can be
unified into a family of related resource inequalities. Finally, we
use this family to find optimal trade-off curves for
all protocols involving one noisy quantum resource and two noiseless ones.
\end{abstract}

\vfill
\pagebreak

\section{Introduction}
\label{sec:intro}

Hitherto quantum and classical 
information theory have been developed using a %n inefficient 
``{first principles}'' approach. Each new coding theorem requires importing or
re-deriving the basic tools from a previous communication scenario
and then applying them in a new, usually more sophisticated, way. 
This may be compared to computer programming directly 
in assembly language as opposed to using a high-level programming language like C$++$.
In this work we advocate an alternative to the first principles approach,
stemming from the view that all quantum and classical coding theorems
are \emph{quantitative statements regarding inter-conversions 
between non-local information processing resources} \cite{DW03a}. 
As an example, 
consider the scenario in which the sender Alice and receiver Bob have
the predefined goal of perfect transmission of a classical
message, but have at their disposal only ``imperfect''
resources such as a noisy channel. This is Shannon's channel coding
problem~\cite{Shannon48}: allowing the parties arbitrary
local operations, they can perform encoding and decoding of the
message to effectively reduce the noise level of the given channel.
Their performance is measured by two parameters: the error
probability and the number of bits in the message, and 
they want to minimize the former while maximizing the latter.
In Shannon theory, we are particularly interested in the \emph{memoryless}
case in which the message is long and the channel is a number of
independent realizations of the same noisy channel $\cN^{A \rightarrow B}$.
The efficiency of the code is then measured by the \emph{rate} $R$,
the ratio of the
number of bits in a message to the number of channel uses.
We are specifically concerned with the \emph{asymptotic regime} of
arbitrarily long messages and vanishing error probability. 
%[In subsection 3.5 we will move away from this asymptotic regime]. 
Note that not only the given channel, but also the goal of the
parties, noiseless communication, is a resource: the channel
which transmits one bit perfectly. The latter resource we 
call  a \emph{cbit} (``classical bit'') and 
denote by the symbol $[c\to c]$.
Thus coding can be described more generally as
the conversion of one resource into another, i.e.,
simulation of the target resource by using the given resource
together with local processing. We express such an 
asymptotically faithful conversion as a 
\emph{resource inequality} (RI)
$$\< \cN^{A \rightarrow B} \> \geq R \, [c\to c].$$
%which we would like to think of as a sort of chemical reaction,
The left hand side we call the \emph{input resource} (or consumed resource)
and the right hand side the \emph{output resource} (or created resource). 
In the asymptotic
setting, $R$ can be any real number, and the supremum of $R$ is the
\emph{capacity} of the channel.

Obviously, there exist other useful or desirable resources, such as
perfect correlation in the form of a uniformly random bit
(abbreviated \emph{rbit}) known
to both parties, denoted by $[c \, c]$, or more generally some noisy correlation.
In quantum information theory, we have further resources:
noisy quantum channels and quantum correlations (a.k.a. \emph{entanglement}) 
between the parties. Again of patricular interest are the noiseless unit
resources; $[q\to q]$ is an ideal quantum bit
channel (\emph{qubit} for short), and
$[q \, q]$ is a unit of maximal entanglement, a two-qubit
singlet state (\emph{ebit}). 

To illustrate our goals, it is instructive to look
at the conversions permitted by the unit resources $[c\to c]$,
$[q\to q]$ and $[q \, q]$, where resource inequalities are finite
and exact. The following inequalities always refer to a specific
integral number of available resources of a given type, and the
protocol introduces no error. % We mark such inequalities by a $*$ above the $\geq$ sign.
For example, it is always possible
to use a qubit to send one classical bit,
$[q\to q]\geq [c\to c]$, and to distribute
one ebit using a qubit channel, $[q\to q] \geq [q\, q]$.
The latter is referred to as entanglement distribution.
More inequalities are obtained by combining resources.
Super-dense coding (SD) \cite{BW92} is a coding protocol
to send two classical bits using one qubit and one ebit:
\begin{equation}
  [q\to q]  + [q\, q]\geq 2[c\to c]. 
\label{sd}
\end{equation}
Teleportation  (TP) \cite{BBCJPW98} is expressed as
\begin{equation}
  2[c\to c] + [q\, q]\geq  [q\to q].
\end{equation}

In~\cite{BBCJPW98} the following argument was used
that the ratio of $1:2$ between
$[q\ra q]$ and $[c\ra c]$ in these protocols is optimal. %, even with
%unlimited entanglement, and even asymptotically: 
Assume, with $R>1$,
$[q\ra q] + \infty \, [q\, q] \geq 2 R \, [c\ra c]$; then chaining this
with (TP) gives $[q\ra q] + \infty \,[qq] \geq R \, [q\ra q]$. Hence by
iteration $[q\ra q] + \infty \, [q\, q] \geq R^k \, [q\ra q] \geq R^k \, [c\ra c]$
for arbitrary $k$, which can make $R^k$ arbitrarily large,
and this is easily disproved. Analogously,
$2[c\ra c] + \infty \, [q\, q] \geq R[q\ra q]$, with $R>1$, gives, when
chained with SD, $2[c\ra c] + \infty \, [q\, q] \geq 2R \, [c\ra c]$,
which also easily leads to a contradiction. In a similar way, the optimality
of the one ebit involved in both SD and TP can be seen.

\medskip

While the above demonstration looks as if we did nothing but introduce
a fancy notation for things understood perfectly otherwise,
in this paper we want to make the case for a systematic
theory of resource inequalities. We will present a framework
general enough to include all unidirectional two-player setups,
specifically designed for the asymptotic memoryless regime.
There are three main issues there: first, a suitably flexible
definition of a \emph{protocol}, i.e., a way of combining resources
(and with it a mathematically precise notion of a resource inequality);
second, a justification of the composition (chaining) of resource
inequalities; and third, general tools to produce new protocols
(and hence resource inequalities) from existing ones.

The benefit of such a theory should be clear. While it
does not mean that we get coding theorems ``for free'', we \emph{do}
get many protocols by canonical modifications from others,
which saves effort and provides structural insights into
the logical dependencies among coding theorems. As the above
example shows, we also can relate (and sometimes actually
prove) the converses, i.e. the statements of optimality,
using the resource calculus.

\bigskip\noindent
From here, the paper is structured as follows.
\begin{description}
  \item[Section \ref{sec:prelim}] (p.~\pageref{sec:prelim})
    covers the preliminaries and
    describes several complementary formalisms
    for quantum mechanics, which serve diverse purposes in 
    the study of quantum information processing. Here also
    some basic facts are collected.

  \item[Section \ref{sec:resources}] (p.~\pageref{sec:resources})
    sets up the basic communication scenario
    we will be interested in. It contains definitions and
    basic properties of so-called finite resources, and
    how they can be used in protocols.
    Building upon these we define asymptotic resources and
    inequalities between them, in such a way as to ensure 
    natural composability properties.

  \item[Section \ref{sec:general-inequalities}] (p.~\pageref{sec:general-inequalities})
    contains a number of general and useful resource inequalities.

  \item[Section \ref{sec:known}] (p.~\pageref{sec:known})
    compiles most of the hitherto discovered 
    coding theorems, rewritten as resource inequalities.
    %The role of ``coherent classical communication''~\cite{Har03}
    %becomes apparent.

  \item[Section \ref{sec:family}] (p.~\pageref{sec:family}):
    Armed with these we give rigorous proofs of a family
    of resource inequalities from~\cite{DHW03}, as 
    well as of two general rules for ``making protocols coherent''. 

  \item[Section \ref{sec:trade-off}] (p.~\pageref{sec:trade-off}):
    Here we discover the sense in which this 
    family of resource inequalities is optimal by exhibiting
    an entropic characterization of five new resource trade-offs.

  \item[Section~\ref{sec:conclusion}] (p.~\pageref{sec:conclusion})
    concludes the paper with some remarks on open problems
    and possible future work.

\end{description}

\vfill\pagebreak

\section{Preliminaries}
\label{sec:prelim}

This section is intended to introduce notation and ways of
speaking about quantum mechanical information scenarios.
We also state several key lemmas needed for the technical proofs.
Most of the facts and the spirit of this section can be
found in~\cite{Holevo01b}; a presentation slightly
more on the algebraic side is~\cite{WinterPhD}, appendix A.

\subsection{Variations on the formalism of quantum mechanics}
\label{subsec:formalism}

We start by reviewing several equivalent
formulations of quantum mechanics and
discussing their relevance for the study of quantum information
processing. As we shall be using several of them in different
contexts, it is useful to present them in a systematic way.
The main two observations are,
first, that a classical random variable can be identified with a
quantum system equipped with a preferred basis, and
second, that a quantum Hilbert space can always
be extended to render all states pure (via a reference system)
and all operations unitary  
(via an environment system) on the larger Hilbert space.

Both have been part of the quantum information processing 
folklore for at least a decade (the second of course
goes back much farther: the GNS construction, Naimark's
and Stinespring's theorems, see \cite{Holevo01b}),
and roughly correspond to
the ``Church of the Larger Hilbert Space'' viewpoint. 

Based on this hierarchy of embeddings  
${\rm C(lassical) \Rightarrow Q(uantum) \Rightarrow P(ure)}$,
in the above sense,
we shall see how the basic ``CQ'' formalism of
quantum mechanics gets modified to (embedded into)
CP, QQ, QP, PQ and PP formalisms. (The second letter refers
to the way quantum information is presented; the first,
how knowledge about this information is presented.)
We stress that from an
operational perspective they are all equivalent --- they are
just of variable expressive convenience in different situations.

\medskip
Throughout the paper we shall use labels such as $A$ 
(similarly, $B$, $C$, etc.) 
to denote not only a particular quantum system but also the 
corresponding Hilbert space (and to some degree even
the set of bounded linear operators on that Hilbert space).
When talking about tensor products of spaces, we will
habitually omit the tensor sign, so $A\ot B = AB$, etc.
Labels such as $X$, $Y$, etc. will be used for
classical random variables. For simplicity, all spaces and
ranges of variables will be assumed to be finite.

\paragraph{The CQ formalism.} This formalism is
the most commonly used one in the literature,
as it captures most of the operational features of a
Copenhagen style  quantum mechanics in the Schr\"odinger picture .
The postulates of quantum mechanics can
be classified into static and dynamic ones.
The static postulates define the static entities of the
theory, while the dynamic postulates describe the physically allowed 
evolution of the static entities.
 
The most general static entity is an \emph{ensemble} of
quantum states $(p_x, \rho_x)_{x \in \cX}$.
The probability distribution $(p_x)_{x \in \cX}$ is defined on some set $\cX$
and is associated with the random variable $X$.
The $\rho_x$ are density operators (positive Hermitian 
operators of unit trace) on
the Hilbert space of a quantum system $A$.  
The state of the quantum system $A$ is 
thus correlated with the classical index random variable $X$.
We refer to $XA$ as a hybrid classical-quantum system, 
and the ensemble $(p_x, \rho_x)_{x \in \cX}$ is the ``state'' 
of $XA$. 
We will occasionally refer to a classical-quantum 
system  as a ``$\{ c \, q \}$ {entity}''. 
Special cases of
$\{ c \, q \}$ entities are $\{ c \}$ entities 
(``classical systems'', i.e. random variables) 
and $\{ q \}$ entities (quantum systems).

The most general dynamic entity would be a map between two $\{ c \,q \}$
entities. Let us highlight only a few special cases:

A map between two $\{ c \}$ entities is a stochastic map, or a 
 $\{ c \rightarrow c \}$ entity. It is defined by a conditional
probability distribution $Q(y|x)$,  where ${x \in \cX}$ and  ${y \in \cY}$.

The most general map from a $\{ c \}$ entity to a $\{ q \}$ 
entity is a \emph{state preparation map} or a ``$\{ c \rightarrow q \}$
entity''. It is defined by a \emph{quantum alphabet} $(\rho_x)_{x \in \cX}$
and maps the classical index $x$ to the quantum state $\rho_x$.

Next we have a $\{ q \rightarrow c \}$ entity, a 
\emph{quantum measurement}, defined by a 
positive operator-valued measure (POVM) $(\Lambda_x)_{x \in \cX}$, where 
$\Lambda_x$ are positive operators satisfying $\sum_x \Lambda_x = \1$,
with the identity operator $\1$ on the underlying Hilbert space.
The action of the POVM $(\Lambda_x)_{x \in \cX}$ on  some
quantum system $\rho$ results in the random variable defined
by the probability distribution $(\tr \rho \Lambda_x)_{x \in \cX}$ on $\cX$.
POVMs will throughout the paper be denoted by greek capitals.

A $\{q \rightarrow q \}$ entity is a \emph{quantum operation}, a
completely positive and trace
preserving (CPTP) map $\cN: A \rightarrow B$, described (non-uniquely)
by its \emph{Kraus representation}:  a set of operators
$\{ N_x \}_{x \in \cX}$, 
$\sum_x N_x^\dagger N_x = \1^B$, 
whose action is given by
$$
\cN(\rho) = \sum_x N_x \rho N_x^\dagger. 
$$ 
(In this paper, $\dagger$ indicates the adjoint,
while $*$ is reserved for the complex conjugate.)
A CP map is defined as above, but with the weaker restriction 
$\sum_x A_x^\dagger A_x \leq \1^B$, and by itself is unphysical
(or rather, it includes a postselection of the system).
Throughout, we will denote CP and CPTP maps by calligraphic letters:
$\cL$, $\cM$, $\cN$, $\cP$, etc. A special CPTP map
is the identity on a system $A$, $\id^A:A\rightarrow A$,
with $\id^A(\rho)=\rho$. More generally, for an isometry
$U:A\rightarrow B$, we denote --- for once deviating from the
notation scheme outlined here --- the corresponding CPTP map by the
same letter: $U(\rho) = U\rho U^\dagger$.

A $\{q \rightarrow c q \}$ entity is an \emph{instrument} $\bbP$,
described by an ordered set of CP maps $(\cP_x)_x$
that add up to a CPTP map.
$\bbP$ maps a quantum state $\rho$ to the
ensemble $(p_{x}, \cP_{x}(\rho)/p_x)_x$, with $p_x = \tr \cP_{x}(\rho)$.
A special case of an instrument is one in which $\cP_x = p_x \cN_x$,
and the $\cN_x$ are CPTP; it is equivalent to 
an ensemble of CPTP maps, $(p_x, \cN_x)_{x \in \cX}$. 
Instruments will be denoted by blackboard style capitals:
$\bbL$, $\bbM$, $\bbN$, $\bbP$, etc.

A $\{cq \rightarrow q \}$ entity is given by an ordered set of CPTP
maps $(\cN_{x})_x$, and maps the ensemble $(p_x, \rho_x)_{x \in \cX}$
to $\sum_x p_x \cN_x(\rho_x)$. % By contrast, a $\{c,q\ra q\}$ map
%saves the classical label, mapping $(p_x, \rho_x)_{x \in \cX}$ to
%$(p_x, \cN_x(\rho_x))_{x\in\cX}$.

In quantum information theory the CQ formalism
is used for proving direct coding theorems 
of a part classical -- part quantum nature, 
such as the HSW theorem \cite{Holevo98,SW97}. In addition, 
it is most suitable for computational purposes.

\medskip
For two states, we write $ \varphi^{ RA} \ext \rho^{A}$ to mean
that the state $\rho^{A}$ is a \emph{restriction} of 
$ \varphi^{RA}$, namely 
$\rho^{A} = \tr_{\!R}  \varphi^{RA}$. The subsystem $R$ is
possibly null (which we write $R = \emptyset$), i.e., a $1$-dimensional
Hilbert space.
Conversely, $ \varphi^{RA}$ is called an
\emph{extension} of $\rho^{A}$. 
Furthermore, if $ \varphi^{RA} = \proj{ \varphi}^{RA}$ is pure it is called a 
\emph{purification} of $\rho^{R}$.
The purification is unique up to a local isometry on $R$:
this is an elementary consequence of the singular
value decomposition, or Schmidt decomposition.
These notions carry over to dynamic entities as well.
For two quantum operations $\cA: A \rightarrow
B E$ and $\cB: A \rightarrow B$ we write $\cA \ext \cB$ 
if $\cB = \tr_{\!E} \circ \cA$. If $\cA$ is an isometry, 
it is called an \emph{isometric extension} or 
\emph{Stinespring dilation} \cite{Sti55} of $\cB$, 
and is unique up to an isometry on $E$.

Observe that we can safely represent noiseless quantum evolution by
isometries between systems (whereas quantum mechanics demands \emph{unitarity}).
This is because our systems are all finite, and we can embed the
isometries into unitaries on larger systems. Thus we lose no
generality but gain flexibility.

\paragraph{The CP formalism.} In order to define the CP formalism, 
it is necessary to review an alternative representation
of the CQ formalism that involves fewer primitives. For instance,

\begin{itemize}
\item $\{q\}$. A quantum state  $\rho^{A}$ is referred to by 
   its purification $\ket{\phi}^{AR}$.
\item $\{c \, q\}$, $\{c \rightarrow q\}$. 
  The ensemble $(p_x, \rho_x^{A})_x$ 
  [resp.~quantum alphabet  $(\rho_x^{A})_x$] is
  similarly seen as the set of restrictions of a pure state ensemble 
  $(p_x, \ket{\phi_x}^{AR})_x$ [resp.~quantum alphabet 
  $(\ket{\phi_x}^{AR})_x$].
\item $\{ q \rightarrow q \}$. A  CPTP map $\cN: A \rightarrow 
  B$ is referred to by its isometric extension $U_\cN:  A \rightarrow B E$.
\item $\{ q \rightarrow c \}$.
  A  POVM $(\Lambda_x)_x$ on the system $A$ is equivalent to 
  some isometry  $U_M: A \rightarrow A {E_X}$,
  followed by a von Neumann measurement of the system $E_X$ in 
  basis $\{ \ket{x}^{E_X} \}$, and discarding $A$.
\item $\{ q \rightarrow c \, q \}$.
  An instrument $\bbP$ is equivalent to some isometry 
  $U_\bbP: A \rightarrow B {E} {E_X}$,
  followed by a von Neumann measurement of the system $E_X$ 
  in basis $\{ \ket{x}^{E_X} \}$, and discarding $E$.
\item $\{ c\, q \rightarrow  q \}$ The collection
  of CPTP maps $(\cN_x)_x$ is identified with the
  collection of isometric extensions $(U_{\cN_x})_x$.
\end{itemize}

In this alternative representation of the CQ formalism 
all the quantum static entities are  thus seen as restrictions 
of pure states; 
all quantum dynamic entities are combinations of
performing isometries, von Neumann measurements,
and discarding auxiliary subsystems. 
The \emph{CP formalism} is characterized by never discarding (tracing out)
the  auxiliary subsystems (reference systems,
environments, ancillas); 
they are kept in the description of our system.
As for the auxiliary subsystems that get (von-Neumann-) measured,
without loss of generality they may be discarded:
the leftover state of such a subsystem may be set
to a standard state $\ket{0}$
(and hence  decoupled from the rest of the system)
by a local unitary conditional upon the measurement outcome. 

The CP formalism is mainly used in quantum information theory
for proving direct coding theorems 
of a quantum nature, such as the quantum channel
coding theorem (see e.g.~\cite{Devetak03}).

\paragraph{The QP formalism.} 
The QP formalism  differs from CP in that 
the classical random variables, i.e. classical systems,
are embedded into quantum systems,
thus enabling a unified treatment of the two.
 
\begin{itemize}
\item $\{ c \}$. The classical random variable $X$ is identified with
  a dummy quantum system $X$ equipped with preferred basis $\{ \ket{x}^X \}$,
  in the state $\sigma^{X} = \sum_x p_x \proj{x}^X$.
  The main difference between random variables and
  quantum systems is that random variables exist
  without reference to a particular physical implementation,
  or a particular system ``containing'' it. 
  In the QP formalism this is reflected in the fact that
  the state $\sigma^{X}$ remains intact under the ``copying''
  operation $\bar{\Delta}: X \rightarrow X X'$, with Kraus representation
  $\{ \ket{x}^{X}  \ket{x}^{X'} \bra{x}^{X}  \}$. In this way, instances of
  the same random variable may be contained in 
  different physical systems. 
\item $\{ c \rightarrow c \}$.  The stochastic map $Q_{y|x}$ becomes
  the  operation  $\bar{\cN}: X' \rightarrow Y$ with Kraus representation 
  $\{ \sqrt{Q_{y|x}} \, \ket{y}^{Y} \bra{x}^{X'}  \}_{x \in \cX, y \in \cY}$.
  Since the operation $\bar{\cN}$ remains intact under copying the input,
  we can define the \emph{classical extension} of $\bar{\cN}$ by the map
  $\cC_{\bar{\cN}}: X' \rightarrow Y X$,
  $$
  \cC_{\bar{\cN}} =  \bar{\cN} \circ  \bar{\Delta}^{X' \rightarrow X'X}.
  $$
  The operation $\cC_{\bar{\cN}}$ thus  implements $\bar{\cN}$ 
  while storing a copy of the input
  in the system $X$.
\item $\{ c \, q \}$. An ensemble $(p_x, \ket{\phi_x}^{AR})_x$
  is represented by a quantum state
  $$
  \sigma^{XAR} = \sum_x p_x \proj{x}^X \otimes \phi_x^{AR}.
  $$
\item $\{c \rightarrow q \}$. A state preparation map 
  $(\ket{\phi_x}^{AR})_x$ is given by the isometry
  $\sum_x \ket{\phi_x}^{AR}\ket{x}^{X} \bra{x}^{X}$, 
  followed by tracing out $X$.
%\item $\{ c \rightarrow c \}$ is a special case of a 
\item $\{c q \rightarrow q \}$. The collection of isometries
  $(U_x)_x$ is represented by the controlled isometry
  $$
  \sum_x \proj{x}^X \otimes U_x.
  $$
\item $\{ q \rightarrow c \}, \{ q \rightarrow c \, q \}$. 
  POVMs and instruments are treated as in the CP picture, 
  except that the final von Neumann measurement is 
  replaced by a completely dephasing operation 
  $\bar{\id}: {E_X} \rightarrow {X}$, defined
  by the 
  Kraus representation $\{ \ket{x}^X \bra{x}^{E_X} \}_x$.
\end{itemize}

The QP formalism is mainly used in quantum information theory
for proving converse theorems.

\paragraph{Other formalisms.}
The QQ formalism is obtained from the QP formalism by 
tracing out the auxiliary systems, and is also
convenient for proving converse theorems. In this 
formalism the primitives are 
general quantum states (static) and quantum operations 
(dynamic).

The PP formalism involves further ``purifying'' the 
classical systems in the QP formalism; it is distinguished
by its remarkably simple structure: all of quantum 
information processing is described in terms of 
isometries on pure states. There is 
also a PQ formalism, for which we don't see much use; 
one may also conceive of hybrid formalisms, such as 
QQ/QP, in which some but not all auxiliary systems are 
traced out. One should remain flexible.
We will indicate which formalism is  used in a given  section.

\subsection{Quantities, norms, inequalities, and miscellaneous notation}

For a state $\rho^{R A}$ and quantum operation 
$\cN: A \rightarrow B$ we often abuse notation, identifying
$$
\cN(\rho) := (\id^{R} \otimes \cN) \rho^{RA}.
$$
With each state $\rho^{B}$, associate a quantum
operation $\app{\rho} : {A} \rightarrow {AB}$ that appends the state to the input:
$$
\app{\rho} (\sigma^{A}) =  \sigma^{A} \otimes \rho^{B}.  
$$
The state $\rho$ and the operation $\app{\rho}$ are
clearly equivalent in an operational sense. 

\medskip
Given some state, say $\rho^{XAB}$, one may 
define the usual entropic quantities with respect to it.
Recall the definition
of the von Neumann entropy 
$H(A) = H(A)_\rho = H(\rho^A) = -\tr (\rho^A \log \rho^A)$, 
where $\rho^A = \tr_{\!XB} \,{\rho}^{XAB}$.
Further define the conditional entropy \cite{CA95}
$$H(A|B) = H(A|B)_\rho = H(AB) - H(B),$$
the quantum mutual information \cite{CA95}
$$I(A;B) = I(A;B)_\rho = H(A) + H(B) - H(AB),$$ 
the coherent information \cite{Sch96,SN96}
$$I(A\,\rangle B) = -H(A|B)=H(B)-H(AB),$$
and the conditional mutual information
\begin{equation*}\begin{split}
  I(A;B|X) &= H(A|X)+H(B|X)-H(AB|X) \\
           &= H(AX)+H(BX)-H(ABX)-H(X).
\end{split}\end{equation*}
Note that the conditional mutual information is always
non-negative, thanks to strong subadditivity~\cite{LR73}.

It should be noted that conditioning on 
classical variables (systems) amounts to averaging.
For instance, for a state of the form
$$
\sigma^{XA} = \sum_x p_x \proj{x}^X \otimes \rho_x^{A},
$$
$$
H(A|X)_\sigma = \sum_x p_x H(A)_{\rho_x}.
$$
We shall freely make use of standard identities for these
entropic quantities, which are formally identical to their
classical predecessors (see~\cite{CT91}, Ch.~2).
One such identity is the so-called chain rule for mutual information,
$$ I(A;BC) = I(A;B|C) + I(A;C),$$
and using it we can derive an identity will later be useful:
\be I(X;AB) = H(A) + I(A\> BX) - I(A;B) + I(X;B).
\label{eq:trip-entropy}\ee

We shall usually work in situations where the underlying state
is unambiguous, but as shown above, we can emphasize the state
by putting it in the subscript.

\medskip
We measure the distance between  
two quantum states $\rho^A$ and $\sigma^A$ by the trace norm,
$$
\| \rho^A - \sigma^A \|_1,
$$
where $\| \omega \|_1 = \tr\sqrt{\omega^\dagger\omega}$.
%;for Hermitian
%operators this is the sum of absolute values of the eigenvalues.
An important property of the trace distance is its
monotonicity under quantum operations $\cN$:
$$
\| \cN(\rho^A) - \cN(\sigma^A) \|_1 \leq \| \rho^A - \sigma^A \|_1.
$$
The trace distance is operationally connected to the
distinguishability of the states. If $\rho$ and $\sigma$ have
uniform prior, by Helstrom's theorem~\cite{Hel76}  the
maximum probability of correct identification of the state by
a POVM is $\frac{1}{2}+\frac{1}{4}\| \rho - \sigma \|_1$.

%We say that two states $\rho^A$ and $\sigma^A$, are
%\emph{$\epsilon$-close} if 
%$$\| \rho^A - \sigma^A \|_1 \leq \epsilon. $$

The following lemma is a trivial application of Fannes' inequality~\cite{Fannes73}.
\begin{lemma}
\label{lemma:fano}
  For the quantity $I(A \, \> B)$ defined on
  a system $AB$ of total dimension $d$, if
  $\|\rho^{AB} - \sigma^{AB} \|_1 \leq \epsilon$ then
  $$
  | I(A \, \> B)_\rho - 
  I(A \, \> B)_\sigma | \leq \eta(\epsilon) + K \epsilon \log d,
  $$ 
  where $\lim_{\epsilon \rightarrow 0} \eta(\epsilon) = 0$
  and $K$ is some constant. The same holds for
  $I(A;B)$ and other entropic quantities.
  \qed
\end{lemma}

Define a distance measure between  
two quantum operations $\cM, \cN: A_1 A_2 \rightarrow B$ 
with respect to some state
$\omega^{A_1}$
by 
\be
\| \cM - \cN \|_{\omega^{A_1}} := 
\max_{{\xi}^{R A_1A_2} \ext \omega^{A_1}}
\bigl\| (\id^{R} \otimes \cM)\xi^{R A_1A_2}
        - (\id^{R} \otimes \cN)\xi^{R A_1A_2} \bigr\|_1.
\label{eq:lung}
\ee
The maximization may, w.l.o.g., be performed over pure
states $\xi^{R A_1A_2}$. This is due
to the monotonicity of trace distance under the
partial trace map.
Important extremes are when $A_1$ or $A_2$ are null.
The first case measures absolute closeness between the two 
operations (and in fact, $\|\cdot\|_\emptyset$ is the
dual of the cb-norm, see~\cite{KW03}),
while the second measures how similar they are relative
to a particular input state.
\eq{lung} is written more succinctly as
$$
\| \cM - \cN \|_\omega := \max_{\xi \ext \omega}
                            \| (\cM - \cN) \xi \|_1.
$$
We say that $\cM$ and $\cN$ are 
$\epsilon$-close with respect to
$\omega$ if
$$
\| \cM - \cN \|_{\omega} \leq \epsilon.
$$
Note that $\|\cdot\|_\omega$ is a norm only if $\omega$ has full
rank; otherwise, different operations can be at distance $0$.
If $\rho$ and $\sigma$ are $\epsilon$-close
then so are $\app{\rho}$ and $\app{\sigma}$ (with respect to
$\emptyset$, hence every state).

Define
the fidelity of two density operators with respect to 
each other as 
$$
F(\rho, \sigma) = \| \sqrt{\rho} \sqrt{\sigma} \|^2_1
= \l(\tr\sqrt{\sqrt{\sigma}\rho\sqrt{\sigma}}\r)^2.
$$
For two pure states $\ket{\phi}$, $\ket{\psi}$ this amounts to
$$
F({\phi}, {\psi}) = |\langle \phi \ket{\psi} |^2.
$$
We shall need the following relation between 
fidelity and the trace distance~\cite{FG97}
\be
1 - \sqrt{F(\rho, \sigma)} \leq  \frac{1}{2} \| \rho - \sigma \|_1
                                    \leq \sqrt{1 - F(\rho, \sigma)},
\label{eq:fid-trace}
\ee
the second inequality becoming an equality for pure states.
Uhlmann's theorem~\cite{Uhlmann76,Jozsa94} states that, for
any fixed purification $\proj{\phi}$ of $\sigma$,
$$
F(\rho, \sigma)  = \max_{\proj{\psi} \ext \rho} 
F({\psi}, {\phi}).
$$
As the fidelity is only defined between two states living on
the same space, we are, of course, implicitly maximizing over 
extensions ${\psi}$ that live on the same space as
${\phi}$.

\begin{lemma}
  \label{pomoc}
  If $\| \rho - \sigma \|_1 \leq \epsilon$
  and $ \sigma' \ext \sigma$, then there exists
  some $ \rho' \ext \rho$ for which
  $\| \rho' - \sigma' \|_1 \leq 2 \sqrt{\epsilon}$.
\end{lemma}
\begin{proof}
  Fix a purification $\proj{\phi}^{ABC} \ext {\sigma'}^{AB} \ext \sigma^A$.
  By Uhlmann's theorem, there exists some 
  $\proj{\psi}^{ABC} \ext \rho^A$ such that
  $$ 
  F({\psi}, {\phi}) = F(\rho, \sigma) \geq 1 - 2 \epsilon,
  $$
  using also \eq{fid-trace}
  Define ${\rho'}^{AB} = \tr_{\!C}\proj{\psi}^{ABC}$.
  By the monotonicity of trace distance under the partial trace map
  and \eq{fid-trace}, we have
  $$
  \| \rho' - \sigma' \|_1 \leq \| {\psi}- {\phi}\|_1
                          \leq                2 \sqrt{\epsilon},
  $$
  as advertised.
\end{proof}

\begin{corollary}
\label{pomoc2}
Given an orthonormal basis $\{\ket{x} \}$,
let  $\| \sum_x p_x \proj{x} -  \sum_x q_x \proj{x} \|_1 \leq \epsilon$.
Define $\ket{\psi} = \sum_x \sqrt{p_x} \ket{x}$
and $\ket{\phi} =  \sum_x \sqrt{q_x} \ket{x}$. Then
\be
\| \psi - \phi \|_1 \leq  2 \sqrt{\epsilon}.
\ee
\end{corollary}

\begin{lemma}
\label{lemma:props}
  The following statements hold for density operators
  $\omega^{A}$, ${\omega'}^{A A'}$, $\sigma^{A}$, $\rho^{A'}$,
  $\Omega^{A_1 }$,
  and quantum operations $\cM', \cN': A A' B \rightarrow C$,
  $\cM, \cN: AB \rightarrow C$, $\cK, \cL:  A'B' \rightarrow C'$,
  and $\cM_i, \cN_i: A_i  B_i \rightarrow A_{i+1} C_{i+1}$.

\begin{enumerate}
\item If $\omega' \ext \omega$ then
      $\| \cM' - \cN' \|_{\omega'} \leq  \| \cM' - \cN' \|_{\omega}$.
\item $\| \cM - \cN \|_{\omega}
        \leq \| \cM - \cN \|_{\sigma} + 4 \sqrt{\|\omega - \sigma \|_1}$.
\item $\| \cM \otimes \cK - \cN \otimes \cL \|_{\omega \otimes \rho}
        \leq \| \cM - \cN \|_{\omega} + \| \cK - \cL \|_{\rho}$.
\item $\|\cM_k \circ \dots \circ \cM_1 - \cN_k \circ \dots \circ \cN_1 \|_{\Omega}
         \leq \sum_i \|\cM_i - \cN_i \|_{ (\cM_{i-1} \circ \dots \circ \cM_1)(\Omega)}$. 
\end{enumerate}
\end{lemma}
\begin{proof}
Straightforward.
\end{proof}

\medskip\noindent
Finally, $[n]$ denotes the set $\{1,\dots,n\}$ and if we have systems
$A_1$, $A_2$, \ldots, $A_n$, 
we use the shorthand $A^n = A_1 \dots A_n$.

\vfill\pagebreak

\section{Information processing resources}
\label{sec:resources}
In this section, the notion of a
information processing resource will be rigorously introduced.
Unless stated otherwise, we shall be using the QQ formalism 
(and occasionally the QP formalism) in order
to treat classical and quantum entities in a unified way.

\subsection{The distant labs paradigm}
\label{subsec:distantlabs}

\begin{figure}
\centerline{ {\scalebox{0.50}{\includegraphics{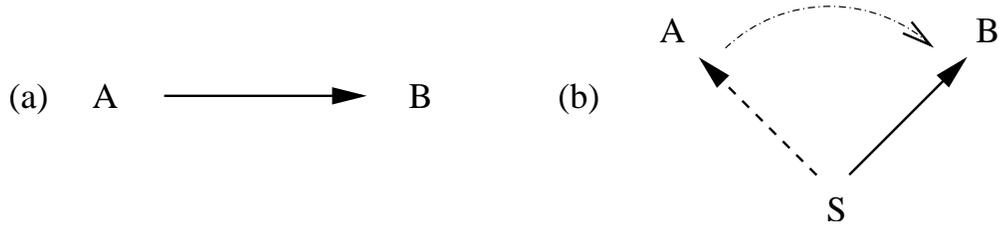}}}}
\caption{A channel (a) between Alice and Bob may be used in a source coding
problem (b) to convert a channel from the Source to Alice, into a channel
from the Source to Bob.
}
\label{fig:src}
\end{figure}

The communication scenarios we will be interested in involve two or more
separated parties.  Each  party is either active or passive. 
Active parties are allowed to perform arbitrary local operations in their lab for free,
while passive once are not allowed to perform any operations at all. 
Non-local operations (a.k.a. {channels}) and states connecting the parties are 
 the principal objects of our theory. They are valuable resources and are carefully 
accounted for.
In this paper, we consider the following parties:
\begin{itemize}
\item {\em Alice} ($A$) Alice is an active party, usually in the role of the sender.
\item {\em Bob} ($B$):  Bob is an active party, usually in the role of the receiver.
In this paper we consider only problems
involving communication from Alice to Bob.  This means we work with channels from Alice to Bob
(i.e. of the form $\cN:A'\ra B$) and arbitrary states $\rho^{AB}$
shared by Alice and Bob. More generally we have \emph{feedback channels}
$\cN:A' \ra A B$ with outputs on both sides. 
\item {\em Eve} ($E$): In the CP and QP formalisms, we purify noisy
channels and states by giving a share to the environment.  Thus, we
replace $\cN:A'\ra B$ with the isometry $U_\cN:A'\ra BE$ and replace
$\rho^{AB}$ with $\psi^{ABE}$. \footnote{In our paper, we think of Eve
as a passive environment, but other work, for example on private
communication \cite{Devetak03,BM04}, treats Eve as an active participant who
is trying to maximize her information.  In these settings, we introduce
 private environments for Alice and Bob $E_A$ and $E_B$, so that
they can perform noisy operations locally without leaking information
to Eve.} We consider a series of operations equivalent when they
differ only by a unitary rotation of the environment.
\item {\em Reference} ($R$):  Suppose Alice wants to send an ensemble of 
states $\{p_i,\ket{\alpha_i}^{A}\}$ to Bob with average density
matrix $\rho^A = \sum_i p_i \alpha_i^A$.  We would like to give a lower
bound on the average fidelity of this transmission in terms only of
$\rho$.  Such a bound can be accomplished (in the CP/QP formalisms) by
extending $\rho^A$ to a pure state $\ket{\phi}^{AR}\supseteq \rho^A$ and
finding the fidelity of the resulting state with the original state
when $A$ is sent through 
the channel and $R$ is left untouched \cite{BKN98}.  Here the reference
system $R$ is introduced to guarantee that transmitting system $A$
preserves its entanglement with an arbitrary external system.  Like
the environment, $R$ is always inaccessible and its properties are not
changed by local unitary rotations.  Indeed the only freedom in
choosing  $\ket{\phi}^{AR}$ is given by a local unitary rotation on $R$.
Both the Reference and Eve are passive.
\item {\em Source} ($S$)  In most coding problems Alice can choose how
she encodes the message, but cannot choose
the message that she wants to communicate to Bob; it can be thought of
as externally given.  Taking this a step further, we can identify the
source of the message as another protagonist ($S$), who begins a
communication protocol by telling Alice which message to
send to Bob. Alice's communication task
becomes to redirect the channel originating at the Source to Bob
(Fig. \ref{fig:src}). 
 Introducing $S$ is useful in cases when the Source does
more than simply send a state to Alice. For example, in distributed
compression, the Source distributes a bipartite state to
Alice and Bob. The Source is a passive party 
as it is  not allowed to code.
\end{itemize}

To each party corresponds a class of 
quantum or classical systems which they control or have access
to at different times. The systems corresponding to Alice 
are labeled by $A$ (for example, $A'$, $A_1$, $X_A$, etc.),
while Bob's systems are labeled by $B$.
When two classical systems, such as $X_A$ and $X_B$, have
the same principal label it means that they are instances of
the same random variable. In our example, $X_A$ is Alice's copy
and $X_B$ is Bob's copy of the random variable $X$.

%In addition to these two principal classes,
%the auxiliary class of \emph{reference} systems (labeled by $R$)
%need not belong to either party and
%plays a role in witnessing the quality of communication.
%For instance, $R$ may represent all or part of the outside world 
%that is initially entangled with the quantum system Alice 
%is transmitting (c.f. equation (\ref{lung})).

%When working in the QP formalism, the
%auxiliary class of \emph{environment} systems 
%(labeled by $E$) enables the purification of 
%all states, and isometric extensions of
%all quantum operations.

%The fourth system is the environment $E$ which
%is initially decoupled from and never interacts directly
%with the reference system $R$. It models noise 
%through interaction with the legitimate systems 
%$A$ and $B$. {\tt we may not need this here}.

\medskip
We turn to some important examples of quantum states and operations.
Let $A$, $B$, $A'$, $X_A$ and $X_B$ be $d$-dimensional systems
with respective distinguished bases
$\{ \ket{x}^{A} \}, \{ \ket{x}^{B} \}$, etc.
The standard maximally entangled state on $AB$ is given by 
$$
\ket{\Phi_d}^{AB} = \frac{1}{\sqrt{d}} \sum_{x = 1}^d 
                                        \ket{x}^A \ket{x}^B.
$$
The decohered, ``classical'', version of this state is 
$$
\bar{\Phi}_d^{X_A X_B} = \frac{1}{{d}} \sum_{x = 1}^d 
\proj{x}^{X_A} \otimes \proj{x}^{X_B},
$$
which may be viewed as two maximally correlated random variables
taking values on the set $[d] := \{ 1, \dots , d \}$.
The local restrictions of either of these
states is the maximally mixed state
$\tau_d := \frac{1}{d} \1_d$. (We write $\tau$ to remind
us that it is also known as the \emph{tracial state}.)
Define the identity quantum operation
$\id_d: A' \rightarrow B$ by the isometry 
$\sum_x  \ket{x}^{B} \bra{x}^{A'}$
(Note that this requires fixed bases of $A'$ and $B$!).
It represents a perfect quantum channel
between the systems $A'$ and $B$.
Its classical counterpart is the completely dephasing channel 
$\bar{\id}_d: X_{A'} \rightarrow X_B$, given in the Kraus representation
by $\{ \ket{x}^{X_B} \bra{x}^{X_{A'}}\}_{x\in[d]}$. It corresponds to a perfect
classical channel because it perfectly transmits density operators diagonal
in the preferred basis, i.e. random variables. 
The channel $\bar{\Delta}_d : X_{A'} \rightarrow
X_A X_B $ with 
Kraus representation  
$\{ \ket{x}^{X_B}  \ket{x}^{X_A} \bra{x}^{X_{A'}}\}_{x\in[d]}$
is a variation on $\bar{\id}_d$ in which Alice first makes a (classical)
copy of the data before sending it through the classical channel.
The two channels are essentially interchangeable.
Of considerable interest is the 
so-called \emph{coherent} channel $\Delta_d: A' \rightarrow AB$ 
\cite{Har03}, given by the
isometry $\sum_x  \ket{x}^{A}\ket{x}^{B} \bra{x}^{A'}$ which is  
a coherent version of the noiseless classical channel with \emph{feedback},
$\bar{\Delta}_d$. Here and in the following,
``coherent'' is meant to say that the operation preserves
coherent quantum superpositions.

The maximally entangled state 
$\ket{\Phi_d}^{AB}$ and perfect quantum channel $\id_d: A' \rightarrow B$ 
are locally basis covariant:
$(U \otimes U^*) \ket{\Phi_d}^{AB} = \ket{\Phi_d}^{AB}$
and $U^\dagger \circ \id_d \circ U = \id_d$ for
any unitary $U$. On the other hand, $\bar{\Phi}_d$, $\bar{\id}_d$,
$\bar{\Delta}_d$ and $\Delta_d$ are all locally basis-dependent.

\subsection{Finite resources}
\label{sec:finite-res}

In this subsection we introduce \emph{finite} or 
non-asymptotic resources. %They can be static or dynamic, or protected.
The central theme, which will carry over to the asymptotic setting,
is that of comparing two resources.
We introduce the notion of a \emph{protocol} 
in which resource $1$ is consumed in order to simulate 
resource $2$.  We then consider resource $1$ to be at least as strong
as resource $2$ (for any asymptotic communication task).

\begin{defi}[Static and dynamic resources]
  \label{defi:finite-resource}
  A finite \emph{static resource} 
  is a quantum state $\rho^{AB}$ shared between Alice and Bob.
  Let $\cN: A' \rightarrow AB$
  be a quantum operation which takes states living on 
  Alice's system $A'$ to a system $AB$ shared by Alice and Bob.
  The \emph{test state} $\omega^{A^\rel}$ lives on a subsystem
  ${A^\rel}$ of $A' = A^\abs A^\rel$.
  A  finite \emph{dynamic resource} is the ordered
  pair $(\cN : \omega)$.
  A static resource $\rho$ is a special kind of dynamic resource
  because of its equivalence to appending maps $\app{\rho}$.
\end{defi} 

States and channels can be used to perform information processing tasks
of interest. Hence the name ``resource''.
The operation $\cN$ comes with a test state
because it ``expects'' an extension of $\omega$ as input.
This will be formalized in  Definition \ref{d6}.
If $A^\rel = \emptyset$, we identify $(\cN : \omega)$ with the
\emph{proper} dynamic resource $\cN$. 
This is the usual 
notion of a quantum channel which can be used without restriction.
Note that $\app{\rho}$ is always a proper dynamic resource, as it has no inputs.
The dynamic resource $(\cN : \omega)$ is called  \emph{relative} if $A^\abs = \emptyset$.

\begin{defi}[Protected resources]
  Let $\cN: S \rightarrow AB$
  be a quantum operation which takes states living on 
  the Source system $S$ to a system $AB$ shared by Alice and Bob.
  The \emph{source state} $\omega^{S}$ lives on the system $S$.
  A  finite \emph{protected resource} is the ordered
  pair $(\cN : \omega)$.
\end{defi}

A {protected resource}  differs from a relative dynamic resource only in that
it originates at the Source. An example of a ``source coding'' problem is Schumacher compression.
There Alice expects a particular state from the Source, channeled through $\cN$.
%This example also illustrates why  such  resource is called protected.
%Alice's goal is to redirect the channel $\CN$ to Bob, and must 
%We will justify the name ``protected'' once we define how these resources are used in protocols. 
Information coming from the Source 
is supposed to be preserved (albeit redirected --- see Fig.\ref{fig:src}), 
and restrictions exist on the allowed operations.
Hence the adjective ``protected''.
This is  formalized in  Definition \ref{d6}.

We now unify the concepts of protected and unprotected (static and dynamic) resources.
%in order to make the exposition that follows more compact.

\begin{defi}[Generalized resources]
Let $\cN: A'S \rightarrow AB$ be 
a quantum operation which takes states living on 
the joint Alice-Source system $A'S$ to a system $AB$ shared by Alice and Bob.
Define $\omega^{A^\rel}$ and $\rho^S$ as above.
A finite \emph{generalized resource} is
the ordered pair $(\cN : [\omega \otimes \rho])$.
\end{defi}

In the next couple of paragraphs when we speak of resources we mean finite generalized resources.
We will often omit the system labels and absorb $\rho$ into $\omega$.

A resource $(\cN: \omega)$ is called \emph{pure} if 
$\cN$ is an isometry. It is called \emph{classical} 
if $\cN$ is a $\{ c \rightarrow c \}$ entity. 

Define a distance measure between 
two dynamic resources $(\cN: \omega)$ and $(\cN' : \omega)$ with
the same test state as
$$
 \| (\cN' : \omega) - (\cN : \omega) \| := \|\cN' - \cN \|_{\omega}.
$$
(If they have different test states then the distance is undefined.)
Define the tensor product of resources as
$$
(\cN_1 : \omega_1) \otimes (\cN_2 : \omega_2) :=
(\cN_1 \otimes \cN_2 : \omega_1 \otimes \omega_2). 
$$

\begin{defi}[Reduction]
\label{defred}
We are given two resources $(\cN:\omega)$ and $(\cN':\omega')$.
We say that $(\cN:\omega)$ \emph{reduces to} $(\cN':\omega')$ 
and write
$$
 (\cN : \omega) \reduction  (\cN' : \omega')
$$
if there exist encoding and decoding channels
$\cE$ and $\cD$ such that $\omega=\cE(\omega')$ and
%$\cN'$ acts on input $\rho$ according to 
$\cN'(\rho)=\cD\circ\cN\circ\cE(\rho)$. 
\end{defi}

The reduction has an operational significance. One can simulate
$(\cN':\omega')$ using the resource $(\cN:\omega)$ by means
of feeding some dummy input $\omega''$ along with the ``genuine'' input $\omega'$.
By definition $\omega''$ is of the form $\omega^{A^\rel} \otimes \rho^S$, 
and may thus be locally prepared.
The canonical example of a reduction is that from $(\cN:\omega)^{\ot 2}$ to $(\cN:\omega)$.
This natural reduction would cease to hold had we allowed source/test 
states to be generally correlated states between spatially separated parties.

\medskip
Resources as defined above are atomic primitives. 
If you have a resource $\cN$, you are allowed to apply the operation $\cN$ only once. 
This is why we speak of \emph{consuming} resources.
If you have a resource $\cN_1 \otimes \cN_2$, you have to apply the two
channels in parallel. You would not be able  
to use the output of $\cN_1$ as an input to $\cN_2$.
We extend our original definition in order to allow for such sequential use of resources.
% When formalizing the notion of ``having'' several resources,
%e.g., the choice from different channels, it would be too restrictive
%to model this by the tensor product, because it gives us just another
%resource, which the parties have to use in a sort of ``block code''.
%To allow for --- finite --- recursive depth (think, e.g., of feedback,
%where future channel uses depend on the past ones) in using the resources,
%we introduce the following:

\begin{defi}[Depth-$\ell$ resources]
  \label{defi:depth-l-res}
  A finite \emph{depth-$\ell$} {resource} is an unordered collection
  of  ``component'' resources 
  $$(\cN :  \omega)^{\ell} := 
      \bigl( (\cN_1 : \omega_1), \ldots,
             (\cN_{\ell} : \omega_{\ell}) \bigr).
  $$
\end{defi}

  What we previously called resources are now identified as depth-$1$ resources.
  To avoid notational confusion, for $\ell$ copies of the same resource,
  $\bigl( (\cN :  \omega),  \dots, (\cN : \omega) \bigr)$, we reserve
  the notation $(\cN : \omega)^{\times \ell}$.

  The definition of the distance measure
  naturally extends to the case of
  two depth-$\ell$ resources:
  $$
   \| (\cN' : \omega)^\ell - (\cN : \omega)^\ell \| 
       := \min_{\pi\in\cS_{\ell}, \omega_j = \omega_{\pi(j)}\forall j}
\sum_{j\in[\ell]} \|(\cN_j':\omega_j) - (\cN_{\pi(j)}:\omega_{\pi(j)})\|.
  $$
Here $\cS_\ell$ is the set of permutations on $\{1, \dots, \ell\}$ objects; we need
  to minimize over it to reflect the fact that we are free to use
  depth-$\ell$ resources in an arbitrary order.

  To \emph{combine} resources
  there is no good definition of a tensor product (which operations
  should we take the products of?), but we can take tensor {\em
  powers} of a resource:
  $$ \l((\cN : \omega)^{\ell}\r)^{\ot k} :=
\bigl((\cN_1 :\omega_1)^{\ot k}, \ldots, 
(\cN_\ell:\omega_\ell)^{\ot k}\bigr).$$

The way we combine a depth-$\ell$ and a depth-$\ell'$ resource is by
concatenation. From $(\cN : \omega)^{\ell}$ and $(\cN' : \omega')^{\ell'}$
we  obtain
  $$
     \bigl( (\cN_1 :\omega_1 ), \ldots, (\cN_\ell    :\omega_\ell    ),
                (\cN_1':\omega_1'), \ldots, (\cN_{\ell'}':\omega_{\ell'}') \bigr).
  $$

For resources with depth $>1$, 
$(\cN:\omega)^\ell=((\cN_1:\omega_1),\ldots,(\cN_\ell:\omega_\ell))$ and
$(\cN':\omega')^{\ell'} = ((\cN'_1:\omega'_1),\ldots,
(\cN'_{\ell'}:\omega'_{\ell'}))$,
we say that $(\cN:\omega)\reduction (\cN':\omega')$ if there exists an
injective function $f:[\ell']\ra[\ell]$ such that for all
$i\in[\ell']$, $(\cN_{f(i)}:\omega_{f(i)}) \reduction
(\cN'_i:\omega'_i)$.  In other words, for each $(\cN_i':\omega_i')$
there is a unique $(\cN_j:\omega_j)$ that reduces to $(\cN_i':\omega_i')$.
Note that this implies $\ell \geq \ell'$.

%Given a partition $\cS_1, \dots, \cS_{l_2}$ of the set $[l_1]$,
%a coarse graining of $(\cN :  \omega)^{l_1}$ is defined by 
%$$
%( \bigotimes_{j \in \cS_1} (\cN_j :  \omega_j), \dots,
%\bigotimes_{j \in \cS_{l_2}} (\cN_j :  \omega_j) ).
%$$

\medskip
Now we are in a position to define a \emph{protocol} as
a general way of simulating or \emph{creating} a depth-$1$ resource
while consuming a depth-$\ell$ resource.
At the same time we introduce the notions
of approximation that will be essential for the treatment of 
asymptotic resources in  Section \ref{secasy}. 

%Assume first that there are no protected components in the given 
%depth-$\ell$ resource. 
%The $\ell$ dynamic resources at Alice's disposal are applied in a certain order.
%Before each round Alice is allowed local processing of the quantum data in her possession.
%In this way she prepares the input state for the dynamic resource to be applied next.
%If this dynamic resource is not proper, Alice must ensure 
%that the state of the input is  \emph{approximately} compatible with
%the corresponding test state. As Bob is always on the receiving end, he may postpone
%the processing of his quantum data until the end of the protocol.
%The protocol is ``tested'' on an extension of the test states of the
%depth-$1$ resource we are trying to simulate.

%Protected resources require additional constraints.
%Ideally, we should insist that the consumed and created resources 
%have the same source states. Remember, the Source has a passive role and
%is not allowed to freely shape her input states like Alice can.
%For reasons that will be elucidated in
%Section \ref{secasy}, we only require that the source state for
%the created resource is a restriction of those for the consumed
%resources.

\begin{defi}[Protocol]
\label{d6}
A \emph{depth-$\ell$ protocol} ${\bf P}$ is a map taking a depth-$\ell$
resource to a depth-$1$ resource.  
Define the depth-$\ell$ resource $(\cN : \omega \otimes \rho)^\ell$
by the operations $\cN_i: A_i' S_i \rightarrow A_i B_i$, $A_i' = A_i^{\rel}A_i^{\abs}$
and
test/source states $\omega_i^{A_i^\rel} \otimes \rho_i^{S_i}$, $i = 1, \dots, \ell$.
Then ${\bf P}[(\cN : \omega  \otimes \rho)^\ell]$ is the
finite depth-$1$ resource $(\cP: \Omega^{A^\rel} \otimes \Theta^{S})$, where
%The protected aspect of the Source is embodied in the requirement that 
$\Theta^{S}$ is a restriction of $\bigotimes_i \rho_i^{S_i}$ to
a subsystem $S$ of $S^\ell$;
the quantum map $\cP: A' S \rightarrow AB$, $A' = A^{\rel}A^{\abs}$, is the following
composition of operations:
%which is constructed as follows:
\footnote{
  We use diverse notation to emphasize the role of the systems in question. 
  The primed systems, such as $A'_i$, are channel inputs. 
  Test systems like $A_i^\rel$ are always subsystems of the corresponding channel
  input $A'_i$. In case of operations  originating at the Source, the test
  system is the full  input system $S_i$. 
  The systems with no superscript, such as $B_i$, are channel outputs.  
  Furthermore, there
  are auxiliary systems, such as $A^\aux$.  Of course many of these
  systems can be null (i.e. one-dimensional).}
\begin{enumerate}
\item  select a permutation $\pi$ of the integers $\{1,\ldots,\ell\}$;
  \item perform local operations
    $\cE_0: A'\rightarrow A_0 A_0^\aux$;
  \item repeat, for $i = 1, \dots, \ell$,
    \begin{enumerate}
      \item $\!\!\!_i \,$  perform local isometries
        $\cE_i: A_{i-1} A^\aux \rightarrow A'_{i} A^\aux$;
      \item $\!\!\!_i \,$  apply the operation $\cN_{\pi(i)}$, 
mapping $A_i'S_i$ to $A_iB_i$;
    \end{enumerate}
  \item perform local operations
    $\cE_{\ell+1}:  A'_\ell A^\aux \rightarrow A$ and 
    $\cD: B^\ell \rightarrow B$. \footnote{recall, $B^\ell = B_1 \dots B_\ell$.}
\end{enumerate}
We allow the arbitrary permutation of the resources $\pi$ so that
depth-$\ell$ resources do not have to be used in a fixed order.
Denote by
$\cP_i
% := (\bf{P}^{(i)}[\rho, \cA^{i - 1}, (\cN : \omega )^{i - 1}])^{A_i}, 
$
the composition of all operations through 
step 3$ (a)_i$.
%(followed by restricting to the subsystem $A_i$).
Define $\hat{\cP}_i$ to be $\cP_i$ followed
by a restriction onto $A_i^\rel$.
The protocol ${\bf P}$ is called \emph{$\eta$-valid}
on the input finite resource $(\cN : \omega  \otimes \rho)^\ell$ if
the conditions
$$
\| \hat{\cP}_i(\xi) - \omega_{\pi(i)}^{A_i^\rel} \|_1 \leq
\eta 
$$
are met for all $i$ and 
for all extensions $\xi$ of $\Omega^{A^\rel} \otimes \Theta^S$.
Whenever the input resource is clear from the context, we will
just say that the protocol is $\eta$-valid.
\end{defi}

A protocol is thus defined to be the most general way one can use the available
resources to generate a new one. Each use of a resource $(\cN_i: \omega_i \otimes \rho_i)$
is preceded by Alice's  encoding layer (the operations $\cE_i$) which prepares an 
appropriate input based on feedback from the preceding layer and memory exemplified in 
the auxiliary system $A^{\aux}$. 
The $\eta$-validity condition ensures that each operation $\cN_i$ 
acts on a extension of a state close to $\omega_i \otimes \rho_i$.
The Source has a passive role and
is not allowed to freely shape her input states like Alice can.
Thus we require that the source state $\Theta^{S}$ for
the created resource is a restriction of the source state $\bigotimes_i \rho_i^{S_i}$ 
for the consumed resources.\footnote{
The simplest situation, which is seen in Fig. \ref{fig:src}, is when
the input states for
the consumed and created resource are \emph{identical}.
} 
%This is because, as we already explained, it is the same source that 
%gets processed throughout the protocol. 
In contrast, the test systems are virtual, and change from
$A_1^\rel \dots A_\ell^\rel$ to $A^\rel$.
%{ \tt Define $* \geq$ here, it's a $0$-valid protocol! In fact can always
%use the protocol name, and then the inequality!}

The  protocol $\bf{P}$ is completely characterized by
the ordered $\ell + 3$-tuple $(\pi,\cE_0, \dots \cE_{\ell + 1}, \cD)$.
Thus we may write
$$
{\bf P} = (\pi,\cE_0, \dots \cE_{\ell + 1}, \cD).
$$

The notion of a reduction from Definition \ref{defred}
provides a simple example of a  $0$-valid protocol.
% mapping depth-$1$ resources to depth-$1$ resources. 
If $ (\cN : \omega) \reduction (\cN' : \omega')$
then there exists a protocol $\bf{R}$ such that
${\bf R} [(\cN : \omega) ] =  (\cN' : \omega')$.
Another important example is given below.

\begin{defi}[Standard protocol]
Define the \emph{standard protocol} ${\bf S}$, which is a $0$-valid
elementary protocol 
on a depth-$\ell$ finite resource $(\cN : \omega)^\ell$, by 
$$
 {\bf S} [(\cN : \omega)^\ell]
    = \bigotimes_{i=1}^\ell (\cN_i : \omega_i).
$$
\end{defi}
This protocol takes a collection of resources and
``flattens'' them into a depth-$1$ tensor product. The standard protocol will
play a major role in the asymptotic theory of resources in Section \ref{secasy}.
Definition \ref{d6} does not account for the simulation of  
resources of arbitrary depth. It does allow the simulation of  the 
flattened version of any resource.
\par\medskip

We can define a tensor product of two protocols by their parallel execution.

\begin{defi}[Tensor product of protocols]
Given protocols ${\bf P}_1 = (\pi_1,\cE_{1,0}, \dots \cE_{1,\ell_1 + 1}, \cD_1)$ 
and ${\bf P}_2 = (\pi_2,\cE_{2,0}, \dots \cE_{2,\ell_2 + 1}, \cD_2)$ 
acting on two separate systems,
define ${\bf P}_1 \otimes {\bf P}_2$ by
$$
{\bf P}_1 \otimes {\bf P}_2 = 
( (\pi_1, \pi_2), \cE_{1,0}, \dots \cE_{1,\ell_1 + 1}, 
\cE_{2,0}, \dots \cE_{2,\ell_2 + 1}, 
\cD_1 \otimes \cD_2).
$$

\end{defi}

\begin{corollary}
If ${\bf P}_1$  and ${\bf P}_2$ are $\eta$-valid 
on $(\cN_1 : \omega_1)^{\ell_1}$ and $(\cN_2 : \omega_2)^{\ell_2}$,
respectively, then 
${\bf P}_1 \otimes {\bf P}_2$ is $\eta$-valid on
$((\cN_1 : \omega_1)^{\ell_1}, (\cN_2 : \omega_2)^{\ell_2})$.
\end{corollary}

The following three lemmas are  straightforward exercises in applying the
definitions and are given without proof.

\begin{lemma}
If $(\cN_1 : \omega_1)^{\ell_1} \reduction (\cN_2 : \omega_2)^{\ell_2}$ then
$$
{\bf S} [(\cN_1 : \omega_1)^{\ell_1}] \reduction {\bf S} [(\cN_2 : \omega_2)^{\ell_2}]. 
$$
\end{lemma}

\begin{lemma}
\label{svodi}
If ${\bf P}$ is an $\eta$-valid protocol for which
$$
\| {\bf P}[(\cN_2 : \omega_2)^{\ell_2}] -  
{\bf S} [(\cN_3 : \omega_3)^{\ell_3}] \|_1 \leq \epsilon
$$
and 
$$
(\cN_1 : \omega_1)^{\ell_1} \reduction (\cN_2 : \omega_2)^{\ell_2}
$$
$$
(\cN_3 : \omega_3)^{\ell_3} \reduction (\cN_4 : \omega_4)^{\ell_4}
$$
then there is an $\eta$-valid protocol ${\bf P'}$
such that
$$
\| {\bf P'}[(\cN_1 : \omega_1)^{\ell_1}] -  
{\bf S} [(\cN_4 : \omega_4)^{\ell_4}] \|_1 \leq \epsilon.
$$
\end{lemma}

\begin{lemma}
\label{tensor_protocol}
If, for $i = 1,2$, %${\bf P}_i$ is an $\eta_i$-valid protocol for which
$$
\| {\bf P}_i[(\cN_i : \omega_i)^{\ell_i}] -  
{\bf S} [(\cN'_i : \omega'_i)^{\ell'_i}] \|_1 \leq \epsilon_i,
$$
then 
$$
\left\| ({\bf P}_1 \otimes {\bf P}_2) [((\cN_1 : \omega_1)^{\ell_1}, (\cN_2 : \omega_2)^{\ell_2})] 
- {\bf S} [((\cN'_1 : \omega'_1)^{\ell'_1}, (\cN'_2 : \omega'_2)^{\ell'_2}) ] \right\|
 \leq \epsilon_1 + \epsilon_2.
$$
\end{lemma}

The next few lemmas help justify aspects of our definition of a
protocol such as $\eta$-validity and the flatness of outputs.
% fact that outputs are
%depth-1---
They will later be crucial in showing how protocols may be
composed.

\begin{lemma}[Protocol composition and validity]
\label{lemma:validity}
If some protocol ${\bf Q}$ is $\nu$-valid on
$({\bf P}[(\cN : \omega)^{\ell}])^{\times m}$,
and ${\bf P}$ is $\eta$-valid on $(\cN : \omega)^{\ell}$,
then the composition protocol ${\bf Q} \circ {\bf P}^{\times m}$,
defined by
$$
{\bf Q} \circ {\bf P}^{\times m}: ((\cN : \omega)^{\ell})^{\times m} \mapsto {\bf Q}[({\bf P}[(\cN : \omega)^{\ell}])^{\times m}],  
$$
is $\eta + 2 \sqrt{\nu}$-valid on $((\cN : \omega)^{\ell})^{\times m}$.
%If some protocol ${\bf Q}$ is $\eta$-valid on
%$( \{ {\bf P}[(\cN_1 : \omega_1)^{\ell_1}], 
%\dots,  {\bf P}_m[(\cN_m : \omega_m)^{\ell_m}]  )$ and for 
%$j = 1, \dots, m$, ${\bf P}_j$ is $\nu$-valid on 
%$(\cN_j : \omega_j)^{\ell_j}$, then 
%the protocol 
\end{lemma}
\begin{proof}
Let $(\cP: \Omega) = {\bf P}[(\cN : \omega)^{\ell}]$.
When the protocol ${\bf P}$ is applied to a purification
$\xi$ of $\Omega$, its $\eta$-validity is expressed as
$$
\| \hat{\cP}_i(\xi) - \omega_i \|_1 \leq \eta.
$$
In the protocol ${\bf Q}$, the resource $(\cP: \Omega)$ is 
applied to a state $\xi'$ which is, %not necessarily a purification of $\Omega$, but, 
according to   Lemma \ref{pomoc} and  the $\nu$-validity of ${\bf Q}$, 
$2 \sqrt{\nu}$-close to some purification $\xi$ of $\Omega$.
By the triangle inequality and monotonicity, 
$$
\| \hat{\cP}_i(\xi') - \omega_i \|_1 \leq \eta + 2 \sqrt{\nu}.
$$
This proves the claim.  

Note that $ {\bf P}^{\times m}$ by itself is not a well defined protocol 
because it would output a resource of depth $m$.
\end{proof}

\medskip

%First we explain the importance of $\eta$-validity.
In general we want our distance measures for states to satisfy the triangle
inequality, and to be nonincreasing under quantum operations.  These
properties guarantee that the error of a sequence of quantum
operations is no more than the sum of errors of each individual
operation (cf. part 4 of \lem{props} as well as \cite{BV93}).
This assumes that we are using the same distance measure
throughout the protocol. When working with relative resources
the distance measure is dependent on the test state in a continuous way. 
Thus for a protocol to map approximately correct inputs to approximately correct 
outputs the assumption of  $\eta$-validity is necessary.

\begin{lemma}[Continuity]
\label{lemma:continuity}
If some protocol 
${\bf P}$ 
is $\eta$-valid  on $ [(\cN : \omega)^\ell]$ and 
$$
\|(\cN :\omega)^\ell  - (\cN': \omega)^\ell \|_1 \leq \epsilon,
$$
then it is $( \epsilon + \eta + 4 \ell \sqrt{\eta})$-valid on 
$(\cN' : \omega)^\ell$ and 
$$
\| {\bf P} [(\cN : \omega)^\ell] - 
{\bf P} [(\cN' : \omega)^\ell] \| 
\leq \epsilon + 4 \ell \sqrt{\eta}.
$$
\end{lemma}
\begin{proof}
Let $(\cP: \Omega) = {\bf P} [(\cN : \omega)^\ell]$
and $(\cP': \Omega) = {\bf P} [(\cN' : \omega)^\ell]$.
By definition \ref{d6}, $\cP$ is of the form
$$
\cP =
  \cE_{\ell+1} \circ \cN_\ell \circ \cE_\ell \circ \dots \circ \cN_1 \circ \cE_1  \circ \cE_0,
%\circ \app{\rho} .
$$
and similarly for $\cP'$.
The $\eta$-validity condition reads, for all $i$ and for all extensions 
$\xi$ of $\Omega$,
$$
\| \hat{\cP}_i(\xi) - \omega_i \|_1 \leq \eta.
$$
By part 4 of \lem{props},
$$
\| \cP - \cP' \|_\xi \leq \sum_i \|\cN'_i - \cN_i\|_{\cP_i(\xi)}.
$$
By part 1 of \lem{props},
$$
\|\cN'_i - \cN_i\|_{\cP_i(\xi)} \leq \|\cN'_i - \cN_i
\|_{\hat{\cP}_i(\xi)}.
$$
By part 2 of \lem{props} and 
$\eta$-validity
$$
\| \cN'_i - \cN_i \|_{\hat{\cP}_i(\xi)}
\leq \| \cN'_i - \cN_i \|_{\omega_i} + 4 \sqrt{\eta}
$$
Hence
$$
\| \cP - \cP' \|_\xi \leq \epsilon + 4 \ell \sqrt{\eta},
$$
which is one of the statements of the lemma.
To estimate the validity of  ${\bf P}$  on 
$[(\cN' : \omega)^\ell]$, note that one obtains in the same way as above, 
for all $i$,
$$
\| \hat{\cP}_i - \hat{\cP}'_i \|_\xi \leq \epsilon + 4 \ell \sqrt{\eta}.
$$
Combining this with the $\eta$-validity condition via the
triangle inequality  finally gives
$$
\| \hat{\cP}'_i (\xi) - \omega_i \|_1
      \leq  \epsilon + \eta + 4 \ell \sqrt{\eta},
$$
concluding the proof.
\end{proof}

Recall that we can only simulate a depth-$\ell$ resource flattened
by the standard protocol. 
The following lemma states that the standard
protocol is in a sense sufficient to generate any other,
under some i.i.d.-like assumptions. Thus working with depth-$1$
resources is not overly restrictive.

\medskip

\begin{lemma}[Sliding]
\label{lemma:sliding}
If for some depth-$\ell$ finite resource
$(\cN : \omega)^\ell = ((\cN_1 : \omega_1), \dots, (\cN_\ell : \omega_\ell))$
and quantum operation $\cC$,
\be
\| (\cC : \bigotimes_i \omega_i  )  -  
     {\bf S}[(\cN : \omega)^\ell] \|_1 \leq \epsilon,
\label{cab''}
\ee
then for any integer $m \geq 1$ and for any  
$\eta$-valid protocol ${\bf P}$ on $(\cN :  \omega)^\ell$, 
there exists an\protect\\
$( (m+\ell-1) (\epsilon + 4 \sqrt{\ell \eta}) + \ell \eta)$-valid protocol 
${\bf K}$ on 
$(\cC:  \bigotimes_i \omega_i)^{\times (m + \ell - 1)}$,
such that
$$ 
\|{\bf K}[ (\cC:  \bigotimes_i \omega_i)^{\times (m + \ell - 1)}] - 
({\bf P}[ 
(\cN :  \omega)^\ell])^{\otimes {m}} \|
\leq (m+\ell-1) (\epsilon + 4 \sqrt{\ell \eta}).
$$
\end{lemma}
\begin{proof}
%Recall the sliding protocol ${\bf K}$ for which
%$$
%{\bf K}[({\bf S}[ (\cN: \omega)^\ell])^{\times (m + \ell - 1)}] = 
%({\bf P}[(\cN :  \omega)^\ell])^{\otimes {m}},
%$$
%and  ${\bf K}$ is $\ell \eta$-valid on 
%$({\bf S}[(\cN : \omega)^\ell])^{\times (m + \ell - 1)}$.
Let $( \cP : \Omega) = {\bf P}[(\cN :  \omega)^\ell]$
and $(\cM, \bigotimes_j \omega_j) = {\bf S}[ (\cN: \omega)^\ell]$.
Let ${\bf P} = (\pi, \cE_1, \dots, \cE_{\ell + 1}, \cD)$, absorbing
$\cE_0$ into $\cE_1$ and w.l.o.g. assuming that $\pi$ is 
the identity permutation $\id$.
We start by defining the sliding protocol $\bf{K}$
for which we show that
\be
{\bf K}[({\bf S}[ (\cN: \omega)^\ell])^{\times (m + \ell - 1)}] = 
({\bf P}[(\cN :  \omega)^\ell])^{\otimes {m}}.
\label{slicon}
\ee
In other words, the protocol ${\bf K}$ effects the map
$\cP =  \cD \circ \cN_\ell \dots  \circ \cN_1  \circ \cE_1$
on each of the $m$ realizations $\Omega^{(1)}, \dots, \Omega^{(m)}$ of 
the test/source state $\Omega$. In the $i$th round
of the protocol, $i = 1, \dots, m + \ell - 1$,  a realization $\cM_i$
of the map $\cM$ must be applied to the input $\bigotimes_j \omega_j$.
The structure of the protocol is shown in \fig{slide}, which is perhaps
more useful than the formal description below.
\begin{figure}
\centerline{ {\scalebox{0.50}{\includegraphics{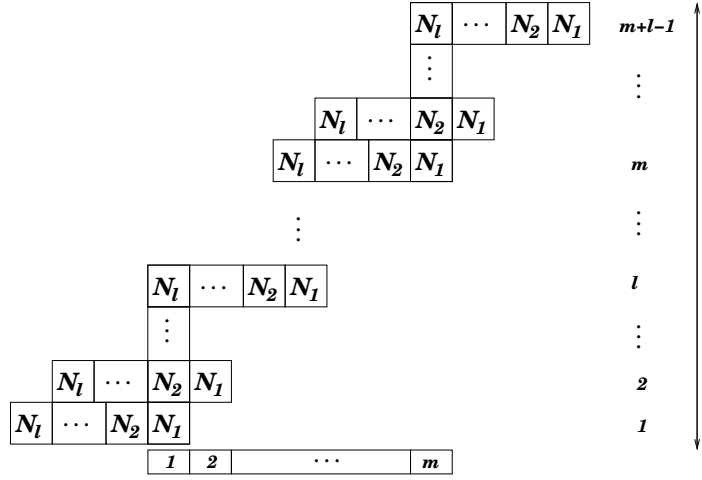}}}}
\caption{The sliding protocol {\bf K}.} 
%Each symbol $1, \dots, m$
%in the bottom flat rectangle denotes an input to the resource 
%${\bf P}[(\cN : \omega)^\ell]$ which is being simulated.
%Each of the rectangles containing the symbols $\cN_1, \dots, \cN_\ell$ denotes
%a single application of the map $\cN_1 \otimes , \dots, \otimes \cN_\ell$. 
%%An appropriate encoding layer between these rectangles is tacitly understood.
%The arrow of time is upward, and each channel is applied to the 
%appropriately encoded output of the channel directly below it. In this
%way the protocol ${\bf P}$ is executed on all $m$ inputs, while consuming...
%It has the exact same validity on $((\cN : \omega)^\ell)^{\times m}$ as
%${\bf P}^{\otimes m}$ does, namely $m \eta$.
%is valid. 
%YO what about dummy source inputs?! Strictly speaking the sliding
%protocol is not a protocol, because it does not preserve the source.
%So what do we do? Always require a sublinear amount of fake sources.
%Have $l - 1$ fake sources. Plan: define the fakeness of a protocol.
%It is $\frac {\ell-1}{m + \ell - 1}$.
\label{fig:slide}
\end{figure}
We proceed to decribe the elements of 
${\bf K} = (\id, \cF_0, \cF_1, \dots \cF_{m + \ell}, \cD')$. 
%In each round there are systems labeled by $1 \dots m + 2 \ell - 2$.
Let $\xi_i^{A_i'}$ %S_i?
be some  dummy locally prepared extension of $\omega_i^{A_i^{\rel}}$.
$\cF_0$ consists of Alice appending a number of such states, yielding
$$
\xi_\ell \ot \dots  \ot \xi_2 \ot \Omega^{(1)} \ot \dots \ot \Omega^{(m)}
\ot  \xi_1 \ot \dots  \ot \xi_1.
$$
For $1 \leq i \leq \ell+1$,
$$
\cF_i = I^{\otimes \ell-1} \otimes \cE_i \ot  \dots \ot  \cE_1 \ot  
I^{\otimes m +\ell -i -1}. 
$$
For $\ell+2 \leq i \leq m$,
$$
\cF_i = I^{\otimes i-2} \otimes \cE_{\ell + 1} \ot  \dots \ot  \cE_1 \ot  
I^{\otimes m + \ell -i -1}. 
$$
For $m+1 \leq i \leq m + \ell$,
$$
\cF_i = I^{\otimes i-2} \otimes \cE_{\ell + 1} \ot  \dots \ot  \cE_{i - m+1} \ot  
I^{\otimes \ell -1}. 
$$
For
$1 \leq i \leq m + \ell - 1$,
$$
\cM_i = I^{\otimes i-1} \otimes \cN_\ell \ot  \dots \ot  \cN_1 \ot  
I^{\otimes m +\ell -i -1}. 
$$
Finally Bob performs $\cD$ on the middle $m$ systems and traces out
the first and last $\ell-1$; in other words,
$$
\cD' =  \tr^{\otimes \ell-1} \ot \cD^{\otimes m} \ot \tr^{\otimes \ell-1}.
$$
The claim (\ref{slicon}) follows by inspection.
Observe that if $\cP$ is $\eta$-valid then
${\bf K}$ is  $\ell \eta$-valid.

Condition (\ref{cab''}) implies that 
\be
\| (\cC : \bigotimes_i \omega_i  )^{\times (m + \ell - 1)}  -  
     ({\bf S}[(\cN : \omega)^\ell])^{\times (m + \ell - 1)} \|_1 \leq (m + \ell - 1)
 \epsilon.
\label{cab'''}
\ee
The result follows from \lem{continuity}.
\end{proof}

\medskip

Relative resources are only guaranteed to work properly when applied to 
the corresponding test   state.  Here we show that using shared randomness,
some of the standard relative resources can be ``absolutized,''
removing the restriction to a particular input state.

\begin{lemma}
\label{lemma:absolutizy}
For a operation $\cN: A' \rightarrow AB$ which is either the
perfect quantum channel
$\id_d$, the coherent channel 
$\Delta_d$ or the perfect classical channel $\bar{\id}_d$, 
there exists a $0$-valid protocol ${\bf P}$ such that
$$
{\bf P}[\bar{\Phi}^{X_A  X_B}, (\cN : \tau_d^{A'})] = 
\cN \otimes \app{\bar{\Phi}^{X_A  X_B}},
$$
where $\dim X_A =  (\dim A')^2$, and $\tau_d^{A'}$ is
the maximally mixed state on $A'$.
\end{lemma}
\begin{proof}
Consider first the case where $\cN$ is either $\id_d$ or 
the coherent channel $\Delta_d$. 
The main observation is that there exists a set of 
unitary operations $\{U_x\}_{x \in [d^2]}$ (the
generalized Pauli, or discrete Weyl, operators) 
living on a $d$-dimensional Hilbert space
such that
\begin{enumerate}[(i)]
\item For any state $\rho$
\be
d^{-2} \sum_x U_x \rho U^\dagger_x = \tau_d, 
\label{randomo}
\ee
\item There exists a set of unitary operators $\{V_x\}_{x \in [d^2]}$
such that
\be
\Delta_d \circ U_x = [V_x \otimes U_x] \circ \Delta_d.
\ee
\end{enumerate}

Let Alice and Bob share the common randomness state
$$
\bar{\Phi}^{X_A  X_B} =  d^{-2} \sum_{x = 1}^{d^2} \proj{x}^{X_A} 
\otimes \proj{x}^{X_B},
$$
where $d:= \dim A'$.
Consider an arbitrary input state $\ket{\phi}^{R A'}$,
possibly entangled between Alice and a reference system $R$.
Alice performs the conditional unitary 
$\sum_x \proj{x}^{X_A} \otimes U_x^{A'}$, yielding a state
%$\Gamma^{X_A X_B R A'}$ 
whose restriction to $A'$ is precisely $\tau_d^{A'}$.
She then applies the operation $\cN$ (this is $0$-valid!),
which gives the state
$$
 d^{-2} \sum_{x = 1}^{d^2} \proj{x}^{X_A} 
\otimes \proj{x}^{X_B} \otimes (  \cN \circ U_x^{A'}) {\phi}^{R A'}.  
$$
In the case of the $\id_d$ channel, Bob simply applies the
conditional unitary $\sum_x \proj{x}^{X_B} \otimes (U^{-1}_x)^{B}$.
In the case of the $\Delta_d$ channel Alice must also perform
$$
\sum_x \proj{x}^{X_A} \otimes (V^{-1}_x)^{A}.
$$ 
In either case, the final state is 
\be
\bar{\Phi}^{X_A  X_B} \otimes \cN ({\phi}^{R A'}),
\label{eq:otdv}
\ee
as advertised.

The case where $\cN$ is the perfect classical channel $\bar{\id}_d$ is a
classical analogue of the above. The observation here is
that there exists a set of $d$ unitaries 
$\{U_x\}_{x \in [d]}$ (all the cyclic shifts of the basis vectors),
such that 
\begin{enumerate}[(i)]
\item (\ref{randomo}) holds for any state $\rho$ diagonal in the preferred basis.
\item 
$
U_x \circ  \bar{\id}_d  =  \bar{\id}_d \circ U_x \circ  \bar{\id}_d.
$
\end{enumerate}
Alice first applies a local $\bar{\id}_d$ on the $A'$ system 
(making the state of $A'$ input diagonal) before proceeding as above.
This concludes the proof.
\end{proof}

\medskip

In the above lemma, the final output of $\cN$ is
uncorrelated with the shared randomness that is used.  In the QQ
formalism, this is immediate from the tensor product  form of \eq{otdv}.
%between $\cN$ and $\app{\bar{\Phi}^{X_A  X_B}}$.  
Thus we say that the
shared randomness is (incoherently) {\em decoupled} from the rest of
the protocol. 

If we move to the QP formalism, so $\cN$ replaced by $U_\cN: A \rightarrow  BE$, 
this decoupling need not hold any more.
When $\cN=\bar{\id}_d$, the common randomness will remain coupled to 
the $E$ system for a particular input state ${\phi}^{R A'}$.
In a cryptographic setting this means that Eve has acquired information
about the key $\bar{\Phi}^{X_A  X_B}$.
%   If we condition on a particular
%message sent by Alice, then the randomness is no longer
%decoupled from the composite $BE$ system.  This is the problem of
%reusing the key in a one-time pad: if the message is not uniformly
%random, then information about the key leaks to Eve.
When $\cN$ is an isometry such as $\Delta_d$ or $\id_d$ then the shared
randomness is decoupled even from the environment. This stronger form
of decoupling is called {\em coherent decoupling}.  Below we extend 
 these notions of decoupling to arbitrary classical resources.\footnote{The notion
of an ``oblivious'' protocol for remotely preparing quantum states is
similar to coherent decoupling, but applies instead to quantum
messages \cite{LS02}.}

\begin{defi}[Incoherent decoupling of input resources]
\label{d7}
Consider some $\{c \rightarrow c\}$ entity  ${\cN_1}:  X'_1 \rightarrow Y_1$
with classical extension $\cC_{\cN_1}: X'_1 \rightarrow Y_1 X_1$.
This induces a modification of  the depth-$\ell$ resource 
$({\cN} : {\omega})^\ell$,
$$
({\cN'} : {\omega})^\ell = 
( ({\cC_{\cN_1}} : {\omega_1}), ({\cN_2} : {\omega_2}), \dots, 
({\cN_\ell} : {\omega_\ell})).
$$
For some protocol ${\bf P}$, define
$$
(\cP : \Omega) = \bf{P}[({\cN'} : {\omega})^\ell ].
$$ 
Assume that for all extensions $\xi$ of $\Omega$
\be
\| \sigma^{X_1 Q} - \sigma^{X_1} \otimes 
\sigma^{Q} \|_1 \leq \epsilon,
\label{otto}
\ee
where $\sigma^{X_1 Q} = \cP(\xi)$.
Then we say that the classical resource 
$({\cN_1} : {\omega_1})$
is $\epsilon-$\emph{incoherently decoupled} 
(or just $\epsilon-$\emph{decoupled}) with respect to 
the protocol ${\bf P}$ on $({\cN} : {\omega})^\ell$.
\end{defi}

%We describe separately how classical resources used in the input and
%the output of a protocol may be coherently decoupled.

\begin{defi}[Coherent decoupling of input resources]
\label{def:coh-decoupling-input}
Consider the setting of the previous definition.
%Again, consider a protocol ${\bf P}$ on 
%$((\bar{\cN} : \bar{\omega})^\ell,({\cN} : {\omega})^{\ell'})$,
%where $(\bar{\cN} : \bar{\omega})^\ell$ is a classical non-protected resource.  
Now adopt a QP view in which all operations except for the classical $\cN_1$ are
isometrically extended. Thus $({\cN'} : {\omega})^\ell$ is replaced by
$$
({\cN''} : {\omega})^\ell = 
( ({\cC_{\cN_1}} : {\omega_1}), (\cU_{\cN_2} : {\omega_2}), \dots, 
(\cU_{\cN_\ell} : {\omega_\ell}))
$$
 and ${\bf P} = (\pi, \cE_0,  \dots , \cD)$ is replaced by 
${\bf P}' = (\pi, \cU_{\cE_0},  \dots , \cU_{\cD})$.
Let
$$
(\cP' : \Omega) = {\bf P}'[({\cN''} : {\omega})^\ell ].
$$ 
Assume that for all extensions $\xi$ of $\Omega$
\be
\| \sigma^{X_1 Q E} - \sigma^{X_1} \otimes 
\sigma^{QE} \|_1 \leq \epsilon,
\label{otto2}
\ee
where $\sigma^{X_1 Q E} = \cP'(\xi)$.
Then we say that the classical resource  
$(\cN_1 : \omega_1 )$ is $\epsilon-$\emph{coherently decoupled} 
with respect to the protocol ${\bf P}$ on 
$({\cN} : {\omega})^{\ell}$.
\end{defi}

The above definitions naturally extend to the case where 
$({\cN_1} : {\omega_1})$ is replaced by a resource of arbitrary depth.
In this case each component resource must be $\epsilon$-decoupled.

\begin{defi}[Coherent decoupling of output resources]
\label{def:coh-decoupling-output}
Let ${\bf P}$ be a protocol mapping
$({\cN} : {\omega})^{\ell}$ to 
$({\cP_1}:{\Omega_1}) \otimes ({\cP_2}:{\Omega_2})$, 
where
$\cP_1: X'_1 \rightarrow Y_1$ is a $\{ c \rightarrow c \}$  entity
with classical extension ${\cC_\cP}: X_1' \rightarrow Y_1 X_1$.
Let $({\cN'} : {\omega})^{\ell}$ be the modification  of 
$({\cN} : {\omega})^{\ell}$ in which all operations are isometrically 
extended. Replace
${\bf P} = (\pi, \cE_0,  \dots , \cD)$ by 
${\bf P}' = (\pi, \bar{\Delta}^{X_1' \rightarrow X_1' X_1}, \cU_{\cE_0},  \dots , 
 \cU_{\cD} )$.
%Note that $\bf{P}'$ outputs the classical extension
%$({\cC_\cP}:{\Omega})$.
Let 
$$
(\cP' : \Omega) = \bf{P}'[({\cN'} : {\omega})^\ell ].
$$ 
Assume that for all extensions $\xi$ of $\Omega$
\be
\| \sigma^{X_1 Y_1 Q E} - \sigma^{X_1 Y_1} \otimes 
\sigma^{QE} \|_1 \leq \epsilon,
\label{otto3}
\ee
where $\sigma^{X_1 Y_1 Q E} = \cP'(\xi)$. %(note that $\sigma^{X Q E} = \cC_\cP(\xi)$). 
Then we say that the 
that the output classical resource  
$(\cP_1 : \Omega_1 )$ is $\epsilon-$\emph{coherently decoupled} 
with respect to the protocol ${\bf P}$ on 
$({\cN} : {\omega})^{\ell}$.
\end{defi}

One simple example of decoupling is when a protocol involves several
pure resources (i.e. isometries) and one noiseless classical resource.
In this case, decoupling the classical resource is rather easy, since
pure resources don't involve the environment.  However, it is possible
that the classical communication is correlated with the ancilla system
$Q$ that Alice and Bob are left with.  If $Q$ is merely discarded,
then the cbits will be incoherently decoupled.  To prove that coherent
decoupling is in fact possible, one has to carefully account for the
ancillas produced by the classical communication.  This was performed
in \cite{HL04}, which proved that classical messages sent through
isometric channels can always be coherently decoupled.

In this paper, we will instead focus on examples of decoupled
classical communication obtained through noisy channel coding.

\subsection{Asymptotic resources}
\label{secasy}

\begin{defi}[Asymptotic resources]
\label{def:asy-resource}
An  \emph{asymptotic resource} $\alpha$ is defined by 
a sequence of finite depth-$\ell$ resources $(\alpha_n)_{n = 1}^\infty$, 
where
$\alpha_n = (\cN_n : \omega_n )^\ell := 
((\cN_{n,1} : \omega_{n,1}), (\cN_{n,2} : \omega_{n,2}), \ldots,
 (\cN_{n,\ell} : \omega_{n,\ell})) $, such that
\begin{enumerate}
\item for all sufficiently large $n$
 \be \alpha_n \reduction \alpha_{n-1};
 \label{eq:asy-monotone}\ee
\item for any $\delta>0$, any integer $k$ 
and all sufficiently large $n$,
\be
\alpha_{\lfloor  n(1 + \delta) \rfloor} \stackrel{{}_{\scriptstyle *}}{\geq}
(\alpha_{\lfloor n/k \rfloor})^{\otimes k} \stackrel{{}_{\scriptstyle *}}{\geq}
\alpha_{ \lfloor  n(1 - \delta) \rfloor}.
 \label{eq:asy-quasi-iid}\ee
\end{enumerate}
\end{defi}
Denote the set of asymptotic resources by ${\cR}$. 

Given two resources $\alpha = (\alpha_n)_{n = 1}^\infty$ 
and $\beta = (\beta_n)_{n = 1}^\infty$, if 
$\alpha_n \reduction \beta_n$ for all sufficiently large $n$,
then we write $\alpha \reduction \beta$. 
%Denote the ``test state'' part of $\alpha_n$, by $\alpha^r_n:=  \omega_n$.
%For a, say, purely static resource, all but the first entry in $\alpha_n$ 
%are null. Furthermore,
%
%We shall use the following convention: if $\beta = (\cN_n)_n$, where
%all $\cN_n$ are proper dynamic resources and $\gamma = (\omega_n)_n$,
%where all $\omega_n$ are proper static resources, then $(\beta:
%\gamma) := (\cN_n : {\omega_n})_n$.  Note that
%typically $\omega_n$ is product state, so the resource $\gamma$
%reduces to the null resource $\emptyset$; however this is no problem
%as long as we are interested in $\gamma$ only as a test state for $\beta$. 
%If, for an asymptotic resource $\alpha = (\alpha_n)_n$, 
%each $\alpha_n$ is an (un)protected finite resource, 
%we call $\alpha$ an (un)protected asymptotic resource and denote it by 
%$\alpha^P$ ($\alpha^U$). 

Unless otherwise stated,
we shall henceforth abbreviate ``asymptotic resource'' to ``resource''.

\begin{defi}[I.i.d. resources]
We call a resource $\alpha$ 
\emph{independent and identically distributed  (i.i.d.)} if 
$\alpha_n = (\cN : \omega)^{\otimes n}$ for some depth-$1$ finite resource 
$(\cN : \omega)$.
We use the shorthand notation $\alpha =  \<\cN: \omega\>$.
\end{defi}

%\begin{lemma}
%Assume, that for some integers $k$, $n$, $n'$,
%\ben
%\cN_n    \reduction & (\cN^{\lfloor n/k \rfloor})^{\otimes k}
%&  \reduction \cN_{n'}, \\
%\omega_n   \reduction & (\omega^{\lfloor n/k \rfloor})^{\otimes k}
%& \reduction \omega_{n'}.
%\een
%If $\cE^{(i)}  \stackrel{\epsilon}{\sim} 
%\cN^{(i)} : {\omega^{(i)}}$ for $i = k, n, n'$, then
%\end{lemma}

We shall use the following notation for \emph{unit} asymptotic resources:
\begin{itemize}
\item ebit $[q \, q]:= \<\Phi_2 \>$
\item rbit $[c \, c] := \<\bar{\Phi}_2 \>$
\item qubit $[q \rightarrow q] := \< \id_2 \>$
\item cbit  $[c \rightarrow c] := \<\bar{\id}_2 \>$
\item cobit $[ q \rightarrow qq ] := \< \Delta_2 \>$
\end{itemize}
In this paper, we tend to use symbols for asymptotic resource
inequalities (e.g. ``$\<\cN\>\geq R \, \ctc$'') and words for finite protocols
(e.g. ``$\cN^{\ot n}$ can be used to send $\geq n(R-\delta_n)$ cbits
with error $\leq \epsilon_n$'').  However, there is no formal reason
that they cannot be used interchangeably.

We also can define versions of the dynamic resources with respect to
the standard ``reference'' state $\tau_2^{A'} = I_2^{A'}/2$: a
qubit in the maximally mixed state.  These are denoted as follows:
\begin{itemize}
\item $[q \rightarrow q : \tau] := \< \id_2 : \tau_2 \>$
\item $[c \rightarrow c : \tau] := \<\bar{\id}_2  : \tau_2\>$
\item $[ q \rightarrow qq  : \tau] := \< \Delta_2  : \tau_2 \>$
\end{itemize}

\begin{defi}[Addition]
The addition operation $+: \cR \times \cR \rightarrow \cR$ is defined
for $\alpha = (\alpha_n)_n$, 
$\alpha_n =  
((\cN_{n,1}  : \omega_{n,1}),  \dots, (\cN_{n,l}  : \omega_{n,l}) ) $,
and $\beta = (\beta_n)_n$, 
$\beta_n = 
((\cN'_{n,1}  : \omega'_{n,1}),  \dots, (\cN'_{n,l'}  : \omega'_{n,l'}) )$,
 as
$\alpha + \beta = (\gamma_n)_n$ with 
$$
\gamma_n = (\alpha_n, \beta_n):= 
((\cN_{n,1}  : \omega_{n,1}),  \dots, (\cN_{n,l}  : \omega_{n,l}),
(\cN'_{n,1}  : \omega'_{n,1}),  \dots, (\cN'_{n,l'}  : \omega'_{n,l'}) ).
$$
\end{defi}

Closure is trivially verified. It is also easy to see
that the operation $+$ is associative and commutative. 
Namely,
\begin{enumerate}
\item[(1)] ${\alpha}  + {\beta} = {\beta} + {\alpha}$ 
\item[(2)] $({\alpha}  + {\beta})  + \gamma =
{\alpha}  + ({\beta}  + \gamma)$ 
\end{enumerate}

\begin{defi}[Multiplication]
The multiplication operation 
$\cdot: \cR \times \mathbb{R}_+ \rightarrow \cR$ 
is defined for any positive real number $z$ and resource 
$\alpha = (\alpha_n)_n$
by $z \alpha  = (\alpha_{\lfloor z n \rfloor})_n$.
\end{defi}

We need to verify that $\cR$ is closed under
multiplication. Before we do so, it will be convenient
to introduce some notation.  Let $n$ be an integer
and $z_1, \dots, z_a$ be positive real numbers.
By $[z_1, z_2, \dots , z_a; n]$
we denote the set of numbers of the form 
$\lfloor z_{\pi(1)} \lfloor z_{\pi(2)}  z_{\pi(3)} 
\lfloor \dots \lfloor z_{\pi(a)} n \rfloor \dots 
\rfloor$, where $\pi$ is some permutation of $\{1,  \dots, a\}$.
There can be an arbitrary number of $\lfloor \, \rfloor$ brackets
as long as they all contain $n$. For instance,   
$\lfloor z \lfloor w  n   \rfloor  \rfloor$ and
$w \lfloor z n  \rfloor$ satisfy this requirement,  
while $\lfloor \lfloor zw   \rfloor n \rfloor$ does not.
It can be shown that for all $\delta > 0$ and all 
$n \geq N$, where $N = N(z_1, \dots, z_a, \delta)$, 
\be
b_{- \delta} \leq b_0 \leq b_{\delta}
\label{zwn}
\ee
holds for all $b_\nu \in  [z_1, z_2, \dots , z_a, (1 + \nu); n]$.

Define
$\beta := z \alpha$, so that $\beta_n = \alpha_{\lfloor zn \rfloor}$.
Condition 1 of Definition \ref{def:asy-resource} is trivially verified for $\beta$. 
For $\delta > 0$, all $k$ and all sufficiently large $n$
\ben
\alpha_{ \lfloor z\lfloor  n(1 + \delta)^3 \rfloor\rfloor} & \reduction &
\alpha_{\lfloor  \lfloor z \lfloor (1 + \delta)n \rfloor \rfloor(1 + \delta) \rfloor}\\
& \reduction &  (\alpha_{\lfloor \lfloor z \lfloor(1 + \delta)n \rfloor
\rfloor/k \rfloor})^{\otimes k}\\
& \reduction & (\alpha_{ \lfloor z \lfloor n/k \rfloor\rfloor})^{\otimes k}.
\een
The first and third inequality follow from (\ref{zwn}) and (\ref{eq:asy-monotone}),
and the second from (\ref{eq:asy-quasi-iid}).
Thus we get $\beta_{ \lfloor  n(1 + \delta)^3 \rfloor} \geq 
(\beta_{  \lfloor n/k \rfloor} )^{\otimes k}$.
Analogously it can be shown that  
$(\beta_{  \lfloor n/k \rfloor} )^{\otimes k} \geq 
\beta_{ \lfloor  n(1 - \delta)^3 \rfloor}$. Thus
$\beta$ satisfies condition 2 of Definition \ref{def:asy-resource}.

%Observe that 
%$R [q \, q] = ( \Phi_{D_n} )_n$, with $D_n = 2^{\lfloor n R \rfloor}$
%and
%$$
%\ket{\Phi_{D}}  =  \sum_x \frac{1}{\sqrt{D}} \ket{x} \otimes \ket{x}.
%$$
%A similar statement holds for $R [c \, c]$ with $\Phi_{D}$
%replaced by 
%$$
%\bar{\Phi}_{D} = \sum_x \frac{1}{{D}} \proj{x} \otimes \proj{x}.
%$$

\medskip

Our next goal is to define what it means to
simulate one (asymptotic) resource by another. 
This is the central definition of the paper.

\begin{defi}[Asymptotic resource inequality]
\label{def:asy-ineq}
A resource \emph{inequality} $\alpha \geq \beta$ holds
between two resources $\alpha = (\alpha_n)_n$
and $\beta = (\beta_n)_n$
if for any $\delta > 0$ there exists an integer $k$
such that for any $\epsilon>0$ there exists $N$ such that for all
$n\geq N$
there exists 
%an integer $n' \geq (1 - \delta) n$ and
an $\epsilon$-valid protocol ${\bf P}^{(n)}$ on 
$(\alpha_{\lfloor n/k \rfloor})^{\times k}$
for which
$$
\| {\bf P}^{(n)} [ (\alpha_{\lfloor n/k \rfloor})^{\times k}]
   - {\bf S}[\beta_{\lfloor (1- \delta)n \rfloor}] \|  \leq \epsilon.
$$
%where $n' = \lfloor (1- \delta)n \rfloor$.

$\alpha$ is called the input or \emph{consumed} resource, $\beta$ is
called the output or \emph{created} resource, $n$ is the {blocklength},
 $\delta$ is the inefficiency
and $\epsilon$ (which bounds both the
validity and the error) is called the accuracy or error.
\end{defi}

% If the dummy inputs are
%non-local, $o$-term resources are needed to effect the source coding.
%We should require $-$ notation.
%In this case have the difference of sources on the LHS.
%DO I REALLY NEED PROTECTION IN THIS CASE?

At first glance it may seem that we are demanding rather little from
asymptotic resource inequalities: we allow the depth of the input
resource to grow arbitrarily, while requiring only a depth-1 output.
This definition is nevertheless strong enough to
allow the sort of protocol manipulations we would like.
We show this in \thm{composability} using tools like the sliding
lemma.

Definition \ref{def:asy-ineq} is slightly inadequate for
source coding. There the data stream coming from the Source needs to be 
redirected in its entirety. In contrast, our definition allows a fraction 
$\delta$ of the Source-supplied data to get lost.  Alice and 
Bob can fix this problem by replacing this perishable data by fake data.
Section \ref{soconere} is dedicated to this issue.
 
\begin{corollary}
If $\alpha \reduction \beta$ then $\alpha \geq \beta$.
\end{corollary}

Resources that consist entirely of states and 
one-way channels  never require protocols with depth
$>1$. This fact will later be
useful in proving converses, i.e. statements about which
resource inequalities are impossible.

\begin{lemma}[Flattening]\label{lemma:flattening}
Suppose $\alpha\geq\beta$ and $\alpha$ is a \emph{one-way} resource,
meaning that it consists
entirely of static resources ($\app{\rho}$) and dynamic resources
which leave nothing on Alice's side (e.g. $\cN^{A'\ra BE}$).  Then for
any $\epsilon,\delta>0$ for sufficiently large $n$ there is an
$\epsilon$-valid protocol ${\bf P}^{(n)}$ on $\alpha_n$ such that
$$ \| {\bf P}^{(n)}[\alpha_n] - {\bf
S}[\beta_{\lfloor(1-\delta)n\rfloor}] \| \leq \epsilon.$$
\end{lemma}
\begin{proof}
To prove the lemma, it will suffice to convert a protocol on
$(\alpha_{\lfloor n/k\rfloor})^{\times k}$ to a protocol on
$(\alpha_{\lfloor n/k\rfloor})^{\otimes k}$.  
The lemma then follows from
 $\alpha_{\lfloor n(1+\delta)\rfloor} \reduction (\alpha_{\lfloor
n/k\rfloor})^{\otimes k}$ and  a suitable
redefinition of $n$ and $\delta$.

Since $\alpha$ is a one-way resource, any protocol that uses it can be
assumed to be of the following form:
\begin{enumerate}
\item First the Source applies all of its protected maps;
\item Alice applies all of the
appending maps;
\item Alice applies all of her encoding operations;
\item Alice applies all of the dynamic resources;
\item  Bob performs his decoding operation.
\end{enumerate}
The one-way nature of the protocol means that
Alice can apply the dynamic
resources last: they have no outputs on her side, so none of her
other operations can depend on them.
The protected and appending maps can be pushed to the beginning because they 
require no inputs from Alice.
Thus $(\alpha_{\lfloor n/k\rfloor})^{\times k}$ can be simulated using
$(\alpha_{\lfloor n/k\rfloor})^{\otimes k}$, completing the proof.
\end{proof}

\medskip

\begin{defi}[Asymptotic decoupling of input resources]
\label{d10}
Let the inequality $\alpha + \beta \geq \gamma$ 
hold, with a classical resource $\beta$.
Referring to Definition \ref{def:asy-ineq}, if  
$(\beta_{\lfloor n/k \rfloor})^{\times k}$
is $\epsilon-$(coherently) decoupled
with respect to ${\bf P}^{(n)}$ for 
each $\epsilon>0$ and all sufficiently large $n$,
then we say that $\beta$ is (coherently) decoupled
in the resource inequality.
\end{defi}

\begin{defi}[Asymptotic decoupling of output resources]
Let the resource inequality 
$
\alpha \geq \beta + \gamma
$
hold with $\beta$ a classical resource.
Referring to Definition \ref{def:asy-ineq}, if  $\beta_{n}$
is $\epsilon-$coherently decoupled
with respect to ${\bf P}^{(n)}$ for 
each $\epsilon>0$ and all sufficiently large $n$,
then we say that $\beta$ is coherently decoupled
in the resource inequality.
\end{defi}

The central purpose of our resource formalism is contained in the
following ``composability'' theorem, which states that resource
inequalities can be combined via concatenation and addition.  In other
words, the origin of a resource (like cbits) doesn't matter; whether
they were obtained via a quantum channel or a carrier pigeon, they can
be used equally well in any protocol that takes cbits as an input.  A
well-known example of composability in classical information theory is
Shannon's joint source-channel coding theorem which states that a
channel with capacity $\geq C$ can transmit any source with entropy
rate $\leq C$; the coding theorem is proved trivially by composing
noiseless source coding and noisy channel coding.

\begin{theorem}[Composability]
\label{thm:composability}
For resources in $\cR$:
\begin{enumerate}
\item   if $\alpha \geq  \beta$ and $\beta \geq \gamma$ then
$\alpha \geq \gamma$
\item  if $\alpha \geq \beta$ and $  \gamma \geq \varepsilon$ then
$\alpha +  \gamma \geq  \beta + \varepsilon $
\item  if $\alpha \geq  \beta$ then $z \alpha \geq z \beta$
\end{enumerate}
\end{theorem}
\begin{proof}
\begin{enumerate}\item Since $\alpha \geq \beta$ and $  \beta \geq \gamma$,
according to Definition \ref{def:asy-ineq}, $\forall \delta>0, \exists
k,k', \forall \epsilon>0, \exists N, \forall n\geq N$
\be
\| {\bf P}_1 [(\alpha_{\lfloor \lfloor n(1 -  \delta)^2/(mk')\rfloor /k \rfloor})^{\times k}]
- {\bf S}[\beta_{\lfloor \lfloor n(1 - \delta)^2/(mk')\rfloor (1 - \delta) \rfloor}] \| 
\leq \epsilon,
\label{eq:compose-a1}
\ee
\be
\| {\bf P}_2 [(\beta_{\lfloor \lfloor n(1 -  \delta)^4 /m  \rfloor /k' \rfloor})^{\times k'}]
- {\bf S}[\gamma_{\lfloor \lfloor n(1 -  \delta)^4 /m \rfloor  (1 - \delta) \rfloor}]
 \|  \leq \epsilon, 
\label{eq:compose-a2}
\ee
with $m \geq k' \ell/\delta$, where $\ell$ is the depth of $\beta$, and 
where ${\bf P}_1$ and ${\bf P}_2$ are both $\epsilon$-valid protocols. 
For sufficiently large $n$
\ben
(\gamma_{\lfloor \lfloor n(1 -  \delta)^4 /m \rfloor  (1 - \delta) \rfloor})^{\otimes m} 
& \reduction & 
(\gamma_{\lfloor \lfloor n(1 -  \delta)^6 \rfloor /m \rfloor})^{\otimes m}\\
& \reduction & 
\gamma_{ \lfloor n(1 -  \delta)^6 \rfloor (1 - \delta) \rfloor} \\
& \reduction & \gamma_{ \lfloor n(1 -  \delta)^8 \rfloor}.
\een
The first and third reductions follow from (\ref{zwn}) and
(\ref{eq:asy-monotone}), 
and the second from (\ref{eq:asy-quasi-iid}).  Together they imply the
existence of a 0-valid protocol $\bR_\gamma$ such that 
$$ \bR_\gamma[(\gamma_{\lfloor \lfloor n(1 -  \delta)^4 /m \rfloor
    (1 - \delta) \rfloor})^{\otimes m}]
= \bS[\gamma_{ \lfloor n(1 -  \delta)^8 \rfloor}].$$
Applying $\bR_\gamma$ and lemmas  \ref{tensor_protocol} and \ref{svodi} 
to (\ref{eq:compose-a2}):
% there  exists an $m \epsilon$-valid protocol $\tilde{\bf P}_2^{\otimes m}$
\be
\| \bR_\gamma \circ {\bf P}_2^{\otimes m}[
(\beta_{\lfloor \lfloor n(1 -  \delta)^4 /m  \rfloor /k' \rfloor})^{\times k'm}]
- {\bf S}[\gamma_{ \lfloor n(1 -  \delta)^8 \rfloor}] \|_1 \leq m \epsilon.
\label{eq:compose-a1a}
\ee
Define $K = \lfloor mkk'(1 + \delta) \rfloor$. Then
$\lfloor n/K \rfloor \geq  \lfloor n(1 - \delta)/ (mkk') \rfloor$,
which, combined with (\ref{zwn}) and (\ref{eq:asy-monotone}), gives
(for sufficiently large $n$)
\ben
\alpha_{\lfloor n/K \rfloor} & \reduction & 
\alpha_{\lfloor n(1 - \delta)/(mkk') \rfloor}\\
& \reduction & 
\alpha_{\lfloor \lfloor n(1 -  \delta)^2/(mk')\rfloor /k \rfloor}.
\een
Equations (\ref{zwn}) and (\ref{eq:asy-monotone}) also imply
\ben
\beta_{\lfloor \lfloor n(1 - \delta)^2/(mk')\rfloor (1 - \delta) \rfloor}
& \reduction & \beta_{\lfloor \lfloor n(1 - \delta)^4/m \rfloor /k' \rfloor}.
\een
Applying lemmas  \ref{tensor_protocol} and \ref{svodi} 
to (\ref{eq:compose-a1}), there 
 exists an $\epsilon$-valid protocol $\tilde{\bf P}_1^{\otimes k'}$ such that
\be
\label{eq:rulu}
\| \iota - 
{\bf S}[(\beta_{\lfloor \lfloor n(1 -  \delta)^4 /m  \rfloor /k' \rfloor})^{\times k'}] \|
 \leq k' \epsilon.
\ee
where
$
\iota = \tilde{\bf P}_1^{\otimes k'} 
[(\alpha_{\lfloor n/K \rfloor})^{\times kk'}].
$
By the Sliding Lemma \ref{lemma:sliding} and \eq{rulu}, there 
exists some $\epsilon'$-valid protocol ${\bf K}$
such that 
$$
\| {\bf K}[\iota^{\times  m + k' \ell - 1}] - 
{\bf P}^{\otimes m}_2 [(\beta_{\lfloor \lfloor n(1 - \delta)^4/m \rfloor /k' \rfloor})^{\times k'm}]
\|_1 \leq \epsilon',
$$
where
$$
\epsilon' = (m + k'\ell - 1)(k' \epsilon + 4 \sqrt{k' \ell \epsilon}) + k' \ell
\epsilon.
$$

Combining with (\ref{eq:compose-a1a}) and invoking Lemma \ref{lemma:validity}, the 
protocol  ${\bf K} \circ  (\tilde{\bf P}_1^{\otimes k'})^{\times m + k' \ell - 1}$
(which is $(\epsilon + 2 \sqrt{\epsilon'})$-valid) obeys
$$
\| {\bf K} \circ  (\tilde{\bf P}_1^{\otimes k'})^{\times m + k' \ell - 1}
[(\alpha_{\lfloor n/K \rfloor})^{\times kk' (m + k' \ell - 1)}] - 
 {\bf S}[\gamma_{ \lfloor n(1 -  \delta)^8 \rfloor}]
\|_1 \leq \epsilon' + m \epsilon.
$$
Since $ K \geq kk'(m + k' \ell - 1)$,
$$
(\alpha_{\lfloor n/K \rfloor})^{\times K} \reduction 
(\alpha_{\lfloor n/K \rfloor})^{\times kk'(m + k' \ell - 1)}. 
$$
Finally, by Lemma \ref{svodi},  there exists a
$(\epsilon + 2 \sqrt{\epsilon'})$-valid protocol ${\bf P}_3$
such that
$$
\| {\bf P}_3[(\alpha_{\lfloor n/K \rfloor})^{\times K}] -
 {\bf S}[\gamma_{ \lfloor n(1 -  \delta)^8 \rfloor}
 \| 
\leq \epsilon' + m \epsilon.
$$
Fixing $\delta$, which controls the inefficiency, 
since $k, k'$ and $m$ are functions of $\delta$, the
accuracy can be made arbitrarily small for a suitable choice of $\epsilon$.
Therefore $\alpha \geq \gamma$.
\item 
%We begin with the standard quantifiers from our definition of a
%resource inequality: 
Since $\alpha \geq \beta$ and $  \gamma \geq \varepsilon$,
according to Definition \ref{def:asy-ineq}, $\forall \delta>0, \exists
k,k', \forall \epsilon>0, \exists N, \forall n\geq N$
\be
\| {\bf P}_1 [(\alpha_{\lfloor \lfloor n/k'\rfloor /k \rfloor})^{\times k}]
- {\bf S}[\beta_{\lfloor \lfloor n / k' \rfloor(1 -  \delta) \rfloor }] \| 
 \leq \epsilon,
\label{osam}
\ee
\be
\| {\bf P}_2 [(\gamma_{\lfloor \lfloor n/k\rfloor /k' \rfloor})^{\times k'}]
- {\bf S}[\varepsilon_{\lfloor \lfloor n / k \rfloor(1 -  \delta) \rfloor }] \| 
 \leq \epsilon.
\ee
For sufficiently large $n$
\bea
(\beta_{\lfloor \lfloor n / k' \rfloor(1 -  \delta) \rfloor })^{\otimes k'} 
& \reduction & 
(\beta_{\lfloor \lfloor n (1 - \delta)^2 \rfloor /k'
  \rfloor})^{\otimes k'}  \non\\
& \reduction & 
\beta_{\lfloor \lfloor n (1 - \delta)^2 \rfloor (1 - \delta) \rfloor } \non\\
& \reduction & 
\beta_{\lfloor n (1 - \delta)^4 \rfloor }.
\label{eq:compose-reduct1}\eea
The first and third reductions follow from (\ref{zwn}) and (\ref{eq:asy-monotone}),
and the second from (\ref{eq:asy-quasi-iid}).  Thus there exists a
0-valid reduction $\bR_1$ mapping the LHS of \eq{compose-reduct1} to
the flattened version of the RHS. 
Combining $\bR_1$ with (\ref{osam}) via Lemmas  \ref{tensor_protocol}
and \ref{svodi}  
and (\ref{eq:asy-monotone}) gives
$$
\| \bR_1 \circ {\bf P}_1^{\otimes k'}[(\alpha_{\lfloor n/(kk')
    \rfloor})^{\times kk'} ] 
- {\bf S}[\beta_{\lfloor n (1 - \delta)^4 \rfloor }] \|_1 \leq k' \epsilon.
$$
Similarly there exists a reduction $\bR_2$ such that 
$$
\| \bR_2 \circ {\bf P}_2^{\otimes k}[(\alpha_{\lfloor n/(kk')
    \rfloor})^{\times kk'} ] 
- {\bf S}[\beta_{\lfloor n (1 - \delta)^4 \rfloor }] \|_1 \leq k \epsilon.
$$
Again invoking Lemma  \ref{tensor_protocol}, the $\epsilon$-valid
${\bf P} = {\bf P}_1^{\otimes k'} \otimes {\bf P}_2^{\otimes k}$ satisfies
\be
\| {\bf P}_3 [((\alpha+\gamma)_{\lfloor n/(kk') \rfloor})^{\times kk'}]
- {\bf S}[(\beta+\varepsilon)_{ \lfloor n(1 -  \delta)^4 \rfloor }] \|_1
 \leq (k + k') \epsilon.
\ee
Hence $\alpha +  \gamma \geq  \beta + \varepsilon$.

\item Immediate from the definitions.
\end{enumerate}
\end{proof}

It is worth noting that our definitions of resources and resource
inequalities were carefully chosen with the above theorem in mind; as
a result the proof exposes most of the important features of our
definitions.  (It is a useful exercise to try changing aspects of our
definitions to see where the above proof breaks down.)  
%By contrast,
%the remainder of this section will establish a number of details about
%the resource formalism that mostly depend only on
%\eqs{asy-monotone}{asy-quasi-iid} and not so much on the details of
%how we construct protocols and resource inequalities.

\begin{defi}[Equivalent resources]
Define an equivalence between resources $\alpha \equiv \beta$
iff $\alpha \geq  \beta$ and $\beta \geq \alpha$.
\end{defi}

\begin{eg}
It is easy to see that $R \, [q \, q] \equiv ( \Phi_{D'_n} )_n$ with 
$D'_n = {\lfloor 2^{n R} \rfloor}$. 
\end{eg}

\begin{lemma}
For resources in $\alpha,\beta \in {\cR}$ and $z,w \geq 0$:
\begin{enumerate}
\item $(zw) {\alpha} \equiv z(w {\alpha})$ 
\item $z({\alpha}  + {\beta}) = 
z {\alpha}  + z {\beta}$
\item $(z + w) {\alpha} \equiv  z {\alpha} + w {\alpha}$
%\item $1 \,  {\alpha} =  {\alpha}$.
\end{enumerate}
\end{lemma}
\begin{proof}
1. It suffices to show that 
$\alpha_{\lfloor zwn \rfloor} \reduction \alpha_{\lfloor z \lfloor w \lfloor n (1 - \delta)
 \rfloor \rfloor  \rfloor}$ and 
$\alpha_{\lfloor z \lfloor w  n  \rfloor \rfloor} \reduction  
\alpha_{\lfloor zw \lfloor n(1 - \delta)\rfloor \rfloor}$.
These follow from (\ref{zwn}) and (\ref{eq:asy-monotone}).

2. Immediate from the definitions.

3. Consider first the $\geq$ direction. 
From the first two parts of this lemma it suffices to prove the statement
when $z + w = 1$. Define $\beta = z \alpha + w \alpha$.
Fix $\delta < 0$. Let  $k = \lfloor zm \rfloor$ and 
$k' = m - \lfloor zm \rfloor$, 
where $m$ is chosen such that $k, k' \geq 1/\delta$. Clearly,
$(1 - \delta) k/m  \leq z \leq (1 + \delta) k/m $ and
$(1 - \delta) k'/m  \leq w \leq  (1 + \delta)k'/m $.
Hence, for sufficiently large $n$, 
\ben
(\alpha_{ \lfloor n/m \rfloor})^{\otimes k} & \reduction & 
\alpha_{\lfloor k  \lfloor n/m \rfloor (1 - \delta) \rfloor} \\
 & \reduction & \alpha_{\lfloor k/m \lfloor  n(1 - \delta)^2  \rfloor \rfloor} \\
 & \reduction &  \alpha_{\lfloor z \lfloor  n(1 - \delta)^3 \rfloor \rfloor}
\een
The first inequality follows from  (\ref{eq:asy-quasi-iid}),
and the last two from  (\ref{zwn}) and (\ref{eq:asy-monotone}).
Similarly it can be shown that
$
(\alpha_{ \lfloor n/m \rfloor})^{\otimes k'}   \reduction 
\alpha_{\lfloor w \lfloor  n(1 - \delta)^3 \rfloor \rfloor}.
$
%and
%$
%\alpha_{\lfloor (z + w) \lfloor  n(1 + \delta)^3 \rfloor \rfloor}   \reduction 
%(\alpha_{ \lfloor n/m \rfloor})^{\otimes (k + k')}. 
%$
Since 
$$
{\bf S}[(\alpha_{ \lfloor n/m \rfloor})^{\times m}] = 
{\bf S}[((\alpha_{ \lfloor n/m \rfloor})^{\otimes k}, 
(\alpha_{ \lfloor n/m \rfloor})^{\otimes k'})],
$$ 
by Lemma \ref{svodi} there
exists a $0$-valid protocol ${\bf P}$ such that
$$
{\bf P}[ (\alpha_{ \lfloor n/m \rfloor})^{\times m}] = 
{\bf S}[\beta_{\lfloor  n(1 - \delta)^3 \rfloor}].
$$
 Hence $\alpha \geq \beta$.
%Hence
%\ben
%\alpha_{\lfloor z \lfloor  n(1 + \delta)^3 \rfloor \rfloor} \otimes
%& \reduction & 
%(\alpha_{ \lfloor n/m \rfloor})^{\otimes k}\\
%\een
%\ben
%\alpha_{\lfloor z \lfloor  n(1 + \delta)^3 \rfloor \rfloor} & \reduction & 
%\alpha_{\lfloor k/m \lfloor  n(1 + \delta)^2  \rfloor \rfloor} \\
% & \reduction & \alpha_{\lfloor k n/m (1 + \delta) \rfloor} \\
% & \reduction &  (\alpha_{ \lfloor n/m \rfloor})^{\otimes k}
%\een

To prove the $\leq$ direction we observe that
% there exists a reduction ${\bf R}$ implementing 
\ben
\alpha_{\lfloor z \lfloor  n(1 + \delta)^3 \rfloor \rfloor} \otimes
\alpha_{\lfloor w \lfloor  n(1 + \delta)^3 \rfloor \rfloor}
& \reduction & 
(\alpha_{ \lfloor n/m \rfloor})^{\otimes k} \otimes
(\alpha_{ \lfloor n/m \rfloor})^{\otimes k'} \\
& = & (\alpha_{ \lfloor n/m \rfloor})^{\otimes m} \\
&  \reduction &  \alpha_{\lfloor  n(1 - \delta)^3  \rfloor}.
\een
Since 
$$
{\bf S}[\beta_{\lfloor  n(1 + \delta)^3 \rfloor}]
= 
{\bf S}[
\alpha_{\lfloor z \lfloor  n(1 + \delta)^3 \rfloor \rfloor} \otimes
\alpha_{\lfloor w \lfloor  n(1 + \delta)^3 \rfloor \rfloor}
],
$$
by Lemma \ref{svodi}, there exists a
$0$-valid protocol ${\bf P}$ such that
$$
{\bf P}[\beta_{\lfloor  n(1 + \delta)^3 \rfloor}] 
= {\bf S}[ \alpha_{\lfloor  n(1 - \delta)^3 \rfloor} ].
$$
Hence $\beta \geq \alpha$.
\end{proof}

\begin{defi}[Equivalence classes of resources]
Denote by $\tilde{\alpha}$ the equivalence class of $\alpha$,
i.e. the set of all $\alpha'$ such that $\alpha' \equiv \alpha$.
Define $\tilde{\cR}$ to be the set of
equivalence classes of resources in $\cR$.
Define the relation $\geq$ on $\tilde{\cR}$ by
$\tilde{\alpha} \geq \tilde{\beta}$ iff $\alpha' \geq \beta'$
for all $\alpha' \in \tilde{\alpha}$ and 
$\beta' \in \tilde{\beta}$.
Define the operation $+$ on $\tilde{\cR}$ such that
$\tilde{\alpha} + \tilde{\beta}$ is the union
of $\tilde{\alpha' + \beta'}$ over  all 
$\alpha' \in \tilde{\alpha}$ and $\beta' \in \tilde{\beta}$.
Define the operation $\cdot$ on $\tilde{\cR}$ such that
$z \tilde{\alpha}$ is the union
of $\tilde{z \alpha'}$ over  all 
$\alpha' \in \tilde{\alpha}$.
\end{defi}

\begin{lemma}
\label{ate}
For resources in $\cR$:
\begin{enumerate}
\item  $\tilde{\alpha} \geq \tilde{\beta}$ iff 
$\alpha \geq \beta$
\item $\tilde{\alpha} + \tilde{\beta} = \tilde{\alpha + \beta}$
\item $z \tilde{\alpha} = \tilde{z \alpha}$
\end{enumerate}
\end{lemma}
\begin{proof}
Regarding the first item: it suffices to show the
``if'' direction. Indeed, for any  $\alpha' \in \tilde{\alpha}$ and 
$\beta' \in \tilde{\beta}$
$$
\alpha' \geq \alpha \geq \beta \geq \beta',
$$
by \thm{composability}.
Regarding the second item: it suffices to show
that if  $\alpha' \equiv \alpha$, $\beta' \equiv \beta$
then $\alpha' + \beta' \equiv \alpha + \beta$.
This follows from \thm{composability}.
Similarly, for the  third item it suffices to show
that if  $\alpha' \equiv \alpha$ then
$z \alpha' \equiv z \alpha$, which is true by 
\thm{composability}.
\end{proof}

We now state a number of additional properties of $\tilde{\cR}$, each
of which can be easily verified.
\begin{proposition}
The relation $\geq$  forms a
partial order on the set $\tilde{\cR}$:
\begin{enumerate}
 \item $\tilde{\alpha} \geq \tilde{\alpha}$ (reflexivity) 
 \item if $\tilde{\alpha} \geq  \tilde{\beta}$ and 
$\tilde{\beta} \geq \tilde{\gamma}$ then
$\tilde{\alpha} \geq \tilde{\gamma}$  (transitivity) 
 \item  if $\tilde{\alpha} \geq  \tilde{\beta}$ and 
$\tilde{\beta} \geq \tilde{\alpha}$ 
 then $\tilde{\alpha} = \tilde{\beta}$ (antisymmetry) 
\end{enumerate}
\qed
\end{proposition}

\begin{proposition}
The following properties hold for the set $\tilde{\cR}$
with respect to $+$ and multiplication by positive real numbers.
\begin{enumerate}
\item $(zw) \tilde{\alpha} = z(w \tilde{\alpha})$ 
\item $(z + w) \tilde{\alpha} =  z \tilde{\alpha} + w \tilde{\alpha}$
\item $z( \tilde{\alpha}  + \tilde{\beta}) = 
z \tilde{\alpha}  + z \tilde{\beta}$
\item $1 \,  \tilde{\alpha} =  \tilde{\alpha}$
\end{enumerate}
\qed
\end{proposition}

\begin{proposition}
For equivalence classes in $\tilde{\cR}$:
\begin{enumerate}
\item if $\tilde{\alpha}_1 \geq  \tilde{\alpha}_2$ and 
$\tilde{\beta}_1 \geq \tilde{\beta}_2$ then
$\tilde{\alpha}_1 + \tilde{\beta}_1 \geq \tilde{\alpha}_2 + \tilde{\beta}_2 $
\item if $\tilde{\alpha} \geq \tilde{\beta}$ then 
$z \tilde{\alpha} \geq z \tilde{\beta}$
\end{enumerate}
\qed
\end{proposition}

Lemma \ref{ate} has essentially allowed us
to replace resources with their equivalence classes
and $\equiv$ with $=$. Henceforth we shall 
equate the two, and drop the $\sim$ superscript.

%{\tt loose this: }The one exception to this rule is when writing relative resources as
%$(\beta : \gamma)$ where $\beta$ is a proper dynamic resource and
%$\gamma$ is a proper static resource; in this case replacing $(\beta :
%\gamma)$ with its equivalence class is well-defined, but replacing
%$\beta$ and $\gamma$ with their equivalence classes wouldn't make sense.

%  SOURCE CODING  SOURCE CODING  SOURCE CODING  

\subsection{Source coding and improper resource inequalities}
\label{soconere}

In this subsection we will introduce \emph{improper} resource inequalities
as a means 
for overcoming the slight inadequacy of Definition \ref{def:asy-ineq}.
In this definition consumed  resources %on the left hand side of the RI
correspond to block length $n$ (or rather $k$ blocks of length ${\lfloor n/k \rfloor}$),
while created resources 
%Resources on the right hand side of the RI 
correspond to block length $\lfloor (1- \delta)n \rfloor$. 
In source coding we insist that created and consumed resources
are of the same blocklength. 
We will indicate this requirement with 
a superscript $s$ (for ``source coding'') above the resource $\geq$ sign.
Noting that there is no advantage in
breaking up a protected resource $\gamma = (\gamma_n)_n$ 
into a resource of depth $>1$,
we extend Definition \ref{def:asy-ineq} as follows.

\begin{defi}[Improper RI]
\label{def:asy-ineq2}
An \emph{ improper resource inequality} 
\be
\alpha + \gamma \ \sourceRI \ \beta + \gamma'
\label{eq:negii}
\ee
holds for general resources $\alpha = (\alpha_n)_n$ and $\beta = (\beta_n)_n$
and protected resources 
$\gamma = (\gamma_n)_n$ and  $\gamma' = (\gamma'_n)_n$,
if for any $\delta > 0$ there exists an integer $k$
such that for any $\epsilon>0$ there exists $N$ such that for all
$n\geq N$
there exists 
%an integer $n' \geq (1 - \delta) n$ and
an $\epsilon$-valid protocol ${\bf P}^{(n)}$ on 
$((\alpha_{\lfloor n/k \rfloor})^{\times k}, \gamma_{\lfloor (1- \delta/2)n \rfloor]})$
for which
\be
\| {\bf P}^{(n)} [ (\alpha_{\lfloor n/k \rfloor})^{\times k},
\gamma_{\lfloor (1- \delta/2)n \rfloor]}
   - {\bf S}[\beta_{\lfloor (1- \delta)n \rfloor}, \gamma'_{\lfloor (1- \delta/2)n \rfloor]}
 \|_1
\leq \epsilon.
\label{eq:prota}
\ee
%where $n' = \lfloor (1- \delta)n \rfloor$.
\end{defi}
While the unprotected resources $\alpha$ and $\beta$ appear as 
in Definition \ref{def:asy-ineq}, the protocol consumes slightly less
of the protected resource $\gamma$ and creates slightly more
of its ``partner'' protected resource $\gamma'$.

\medskip

A simple example of a source coding resource inequality is
the one illustrated in figure \ref{fig:src}.
A channel  between Alice and Bob may be used in a source coding
problem  to convert the channel from the Source to Alice into a channel
from the Source to Bob.
$$
\< \id^{A' \rightarrow B}: \rho^{A'} \> 
+ \< \id^{S \rightarrow \hat{A}}: \rho^{S} \>
\ \sourceRI \
\< \id^{S \rightarrow \hat{B}}: \rho^{S} \>. 
$$
In contrast, the proper RI (from Definition~\ref{def:asy-ineq})
$$
\< \id^{A' \rightarrow B}: \rho^{A'} \> 
+ \< \id^{S \rightarrow \hat{A}}: \rho^{S} \>
\geq 
\< \id^{S \rightarrow \hat{B}}: \rho^{S} \>. 
$$
allows a fraction $\delta$ of the Source-supplied data to get lost.

%This way of writing it reflects the fact that
%the \emph{redirection} of protected resources (in this case from Alice to Bob)
%is the information processing task Alice and Bob are trying to accomplish.
%Does this mean that we never need to write ``unnatural'' resource inequalities of the
%form
%\be
%\alpha + \gamma  + \varepsilon' \geq \beta + \varepsilon + \gamma' \,?
%\label{eq:pozii}
%\ee
%Unfortunately not.
The problem with Definition \ref{def:asy-ineq2} is that
composition of protocols via the sliding
lemma will always introduce a small inefficiency $\delta$. 
Thus improper resource inequalities cannot be composed.
In general we will have to switch back and forth between
proper and improper resource inequalities. To prove an
improper resource inequality we typically prove its proper version first, 
and then convert it to the improper version.
Rules for doing this appear in the next section as 
Lemmas \ref{lemma:nosorog} and \ref{lemma:sopm}.

\newpage

\vfill\pagebreak

\section{General resource inequalities}
\label{sec:general-inequalities}

In this section, we present several  resource inequalities and theorems that will
be useful for manipulating and combining other resource inequalities. 

\begin{lemma}
\label{lemma:relatif}
%$>=*$?
\medskip
The following resource inequalities hold:
\begin{enumerate}
\item 
$
\< \cN^{A' \rightarrow AB} \> \geq \< \cN^{A' \rightarrow AB} : \omega^{A'} \> 
%\geq \< \cN^{A' \rightarrow AB}(\omega^{A'}) \>,
$
\item 
$
\< \cN^{S \rightarrow AB} : \rho^{S} \> \geq \< \cN^{S \rightarrow AB}(\rho^{S}) \>,
$
\item
$\< \cN^{S \ra A' B'} : \rho^S \> + 
\< \cM^{A' \ra AB} : \tr_{B'} \{ \cN^{S \ra A'B'}(\rho^S) \}  \> 
\geq \< \cN^{S \ra A'B'} \circ  \cM^{A' \ra AB} : \rho^S \>$, 
% \omega^{A'} with $ \omega^{A'}  =  \tr_{B'} \{ \cN^{S \ra A'B'}(\rho^S) \}$,
%Let $\cN: {S \rightarrow AB}$ be a quantum operation
%and $\rho^{S}$ a source state. 
\item $\< \rho^{A'B'} \> + \< \cN^{A' \rightarrow AB} : \rho^{A'} \> \geq
\< \cN^{A' \rightarrow AB}(\rho^{A'B'}) \>$,
\item 
If $(\cN_1 : \omega_1) \reduction (\cN_2 : \omega_2)$
then $ \< \cN_1 : \omega_1 \> \geq  \< \cN_2 : \omega_2 \>$.

\end{enumerate}
\end{lemma}
\begin{proof}
Immediate from definitions.
\end{proof}

%Let $\beta$ and $\beta'$ be proper dynamic resources,
%and $\gamma$ and $\gamma'$ static test resources.
%the following  resource inequalities hold:
%\begin{enumerate}
%\item  \< \cN : \omega \> \geq  \< \cN : \omega \>
%$\beta   \geq   (\beta : \gamma)  $
%\item $ (\beta : \gamma)  + \gamma  \geq  \beta(\gamma) $
%\item if $\gamma \ext \gamma'$ then 
%$(\beta : \gamma') \geq (\beta : \gamma) $ 
%\item
%$ \beta : \gamma  + \beta' : (\beta \gamma) 
% \geq   (\beta' \circ  \beta) : \gamma.$
%\end{enumerate}
%\end{lemma}
%\begin{proof}
%Immediate from definitions.
%\end{proof}

\begin{lemma}[Closure]\label{lemma:closure}
Given $z_0 >0$ and $\alpha, \beta \in \cR$,
if 
$$
z  \, \alpha \geq \beta 
$$ 
for every $z > z_0$ then 
$$
z_0 \,\alpha  \geq \beta.
$$
% The same holds for $z_0 = 0$
%if $z \alpha \geq  \gamma$ for some $z$.
\end{lemma}
\begin{proof}
The statement is equivalent to
$$
z_0 \alpha \geq (1 - \delta) \, \beta, \,\,\, \forall \delta > 0, 
$$
which by Definition \ref{def:asy-ineq} implies the statement for $\delta = 0$.
\end{proof}

\medskip

The case of $z_0=0$ is special and corresponds to the use of a
sublinear amount of a resource.
\begin{defi} [Sublinear  $o$ terms]
We write
$$
\alpha + o  \, \gamma \geq \beta 
$$
if for every $z>0$
$$
\alpha + z  \, \gamma \geq \beta.
$$
\end{defi}

At the other extreme we might consider a setting 
in which we are allowed an arbitrary rate of some resource.
%, typically when proving converse
%theorems. 
\begin{defi} [$\infty$ terms]
We write
$$\alpha + \infty \, \gamma \geq \beta$$
if there exists an $z$ for which
$$\alpha + z  \, \gamma \geq \beta.$$
%if for any $\delta>0$, there exists $k$ such that for any $\epsilon>0$
%there exists $n_1,n_2$ and a $\epsilon$-valid protocol $\bP$
%satisfying
%$$\l\| \bP[(\alpha_{\lfloor n_1/k\rfloor} + \gamma_{\lfloor
%n_2/k\rfloor})^{\times k}] - \beta_{\lfloor(1-\delta)n\rfloor}
%\r\|_1 \leq \epsilon.$$
\end{defi}
%This means that we can use an amount of $\gamma$
%that increases arbitrarily quickly with $n$.
%Note that $\infty\gamma$ cannot be defined as a resource, since it
%violates \eq{asy-quasi-iid}.

Note that ``$\infty\gamma$'' does not actually mean that our protocols may
use an arbitrary amount of the resource $\gamma$; more precisely, they
may, in the asymptotic limit, use an arbitrary but finite rate.

Let us focus on sublinear terms.
In general we cannot neglect sublinear resources.
In entanglement
dilution, for instance, they are both necessary \cite{HL02,HW02} and
sufficient \cite{LP99}.  This situation only occurs when
the sublinear resources cannot be generated from the other resources being 
consumed in the protocol.

\begin{lemma} [Removal of $o$  terms]
\label{lemma:noo}
For $\alpha, \beta , \gamma \in \cR$,
if 
\ben
\alpha + o  \, \gamma &  \geq &  \beta \\ 
z \alpha &\geq & \gamma 
\een
for some real $z > 0$,
then 
$$
\alpha \geq \beta.
$$
\end{lemma}
\begin{proof}
For any $z' > 0$ 
$$
(1 + z'z)  \, \alpha  \geq \alpha + z'  \, \gamma \geq \beta.
$$
The lemma follows by the Closure Lemma (\ref{lemma:closure}).
\end{proof}

\medskip

One place that sublinear resources often appear is as catalysts,
meaning they are used to enable a protocol without themselves being
consumed.  Repeating the protocol many times reduces the cost of the
catalyst to sublinear:

\begin{lemma}[Cancellation]
\label{lemma:cancel}
For $\alpha, \beta , \gamma \in {\cR}$, if
$$
 \alpha +  \gamma \geq \beta +  \gamma, \quad\text{ then }\quad
                                             \alpha + o  \, \gamma \geq  \beta.
$$
\end{lemma}
\begin{proof}
Combine $N$ copies of the inequality (using part 1 of \thm{composability}) to
obtain  
$$
\gamma + N  \, \alpha \geq \gamma + N  \, \beta.
$$
Divide by $N$:
$$N^{-1} \, \gamma  + \alpha \geq N^{-1} \, \gamma + \beta \geq \beta.
$$
As $N^{-1}$ is arbitrarily small, the result follows.
\end{proof}

\medskip
This cancellation result motivates us to extend the set ${\cal R}$ of
all resources into the negative domain: we will in the future also
call expressions $\alpha-\beta$ ``resources''. The rules of arithmentic
will be clear, including the one implicit in the above
Lemma, $\alpha-\alpha = o \alpha$. We only need to define
what the inequality sign means. Also that is straightforward, by
declaring, for $\alpha,\alpha',\beta,\beta'\in{\cal R}$,
\be
  \label{eq:negative-RI}
  \alpha-\beta \ \geq \ \alpha'-\beta' \quad:\Longleftrightarrow\quad
                                         \alpha+\beta' + o \,\beta  \ \geq \ \alpha'+\beta.
\ee
Allowing negative terms is mostly
for notational convenience, but it often also helps to concisely
state a resource inequality.

\medskip
Often we will find it useful to use shared randomness as a catalyst.
The condition for this to be possible is that the randomness be
incoherently decoupled.
\begin{lemma}[Recycling common randomness]
\label{lemma:rcr}
If $\alpha$ and $\beta$ are resources for which
$$
\alpha + z \, [c \, c] \geq \beta, 
$$
and the $[c \, c]$ is incoherently decoupled in
the above RI, then 
$$
\alpha +  o \, [c \, c] \geq \beta.
$$
\end{lemma}
\begin{proof}
Since $[c \, c]$ is asymptotically independent
of the $\beta$ resource, by definitions \ref{d7} and 
\ref{d10} 
it follows that
$$
\alpha + z \, [c \, c] \geq \beta + z \, [c \, c].
$$
An application of the Cancellation Lemma \ref{lemma:cancel}
yields the desired result.
\end{proof}

\begin{corollary}
\label{cor:purim}
If $\alpha \geq [c \, c]$ and $\beta$ is pure then
$$
\alpha + z \, [c \, c] \geq \beta 
$$
can always be derandomized to
$$
\alpha \geq \beta. 
$$
\end{corollary}
\begin{proof}
It suffices to notice that for a  pure  output resource $\beta$, 
equation (\ref{otto}) is automatically satisfied. 
\end{proof}

%{\tt Is this really so?}
\medskip
The following theorem tells us that in proving channel coding theorems
one only needs to consider the case where the input state is maximally
mixed.  A similar result was shown in \cite{BKN98}, though with quite
different techniques and formalism.

\begin{theorem} [Absolutization]
\label{thm:absolutize}
The following resource inequalities hold:
\begin{enumerate}
\item $\qtqtau  = [q \rightarrow q]$
\item $\coftau  =  [q \rightarrow qq] $
\item $\ctctau  =  [c \rightarrow c] $
\end{enumerate}
\end{theorem}
\begin{proof}
The lemma is a direct consequence of \lem{absolutizy}.
By part 1 of \lem{relatif}, it suffices to show the $\geq$ direction.
We shall only prove item 1.; the proofs of 2. and 3. are identical.
By \lem{absolutizy}, we know that
$$
{[q \rightarrow q  : \tau] }  + 2  \, [c \, c] \geq  [q \rightarrow q]
+ 2 \,  [c \, c].
$$
By the cancellation lemma,
$$
{[q \rightarrow q  : \tau] }  + o \,  [c \, c] \geq  [q \rightarrow q].
$$
Since 
$$ 
{[q \rightarrow q : \tau] } \geq  [c \, c],
$$
by \lem{noo} the $o$ term can be dropped, and we are done.
\end{proof}

\medskip

In section \ref{soconere} we showed how to write source coding problems 
as improper resource inequalities. We need to be able to move 
between proper and improper resource inequalities
in order to take advantage of composability properties of proper resources inequalities. 

\begin{lemma}[Faking the Source]
\label{lemma:nosorog} 
If for some resources $\alpha$ and $\beta$
$$
\alpha + \< \cN^{S \rightarrow AB} : \rho^{S} \> \geq 
\beta
$$
and $\beta$ does not refer to the Source system $S$  
then the protected resource 
$\< \cN^{S \rightarrow AB} : \rho^{S} \>$ 
may be ``faked'' by Alice and Bob alone:
$$
\alpha + \< \cN^{S \rightarrow AB}(\rho^{S}) \> \geq 
\beta.
$$
\end{lemma}

\begin{proof} Obvious.

%takes $(\cN_1 : \omega^{S_1}) \otimes (  \cN_2 :  \omega^{S_2})$
%to $(\cN'_2 : \omega^{S_2})$, replacing the protected resource $(\cN_1 : \omega^{S_1})$ by 
%the static resource $\cN_1(\omega^{S_1})$ has no effect.
\end{proof}

\begin{lemma}[Improper and proper resource inequalities]
\label{lemma:sopm}
Let $\< \cN^{S \rightarrow AB} : \omega^S \>$ and
$\< \cM^{S \rightarrow AB} : \omega^S \>$ be 
two i.i.d. protected resources, and $\alpha$ and $\beta$ be arbitrary resources in 
$\cR$. 
\begin{enumerate}
\item[(1)] If 
\be
\alpha  +  \< \cM : \omega \> \ \sourceRI \ \beta + \< \cN : \omega \>
\ee
then 
\be
\alpha + \< \cM : \omega \> \geq \beta + \< \cN : \omega \>.
\label{eq:balos}
\ee
\item[(2)] Conversely, if \eq{balos} holds then
$$
\alpha  +   \< \cM : \omega \> +  o \, \< \cM(\omega) \>  \ \sourceRI \
\beta + \< \cN : \omega \> 
$$
\end{enumerate}
\end{lemma}
\begin{proof}
Item (1) is immediate from definitions \ref{def:asy-ineq}  and \ref{def:asy-ineq2}.
Item (2) needs also the following observation (cf. \lem{nosorog}):
if 
$$
{\bf P}[(\cN_1^{S_1 \rightarrow AB} : \omega_1^{S_1}), 
(  \cN_2^{S_2 \rightarrow AB} :  \omega_2^{S_2})] = 
({\cM}_2^{S_2 \rightarrow AB} : \omega_2^{S_2}) 
$$
then
$$
{\bf P}[\{\cN_1(\omega_1^{S_1})\}^{AB}, ( \cN_2^{S_2 \rightarrow AB} :  \omega_2^{S_2})]
 = (\cM_2^{S_2 \rightarrow AB} : \omega_2^{S_2}).
$$
%I.e. there is no significance to a Source system not appearing in the output of 
%a protocol.
In other words, sources originating at $S_2$ don't care
if we can ``fake'' data coming from an independent source $S_1$.

\end{proof}

\medskip

Finally, we note how convex combinations of static resources can be thought
of as states conditioned on classical variables.
\begin{theorem}
\label{thm:alfav}
Consider the static i.i.d. resource 
$\alpha = \< \sigma \>$, where
$$
\sigma^{A X_A B X_B} = 
\sum_x p_x \, \proj{x}^{ X_A} \otimes \proj{x}^{ X_B} 
 \otimes  \rho^{AB}_x.
$$
In other words, Alice and Bob share a bipartite state chosen from
an ensemble and both parties have the classical information identifying the state.
Denote $\alpha_x = \<\rho_x \>$.
Then
$$
\alpha \geq \sum_x p_x \alpha_x.
$$
\end{theorem}
\begin{proof}
We will show that for all $\epsilon, \delta > 0$ and sufficiently
large $n$, $\sigma^{\otimes n}$ can be transformed into a state
$\epsilon$-close to $\omega_{ \lfloor n( 1 - \delta) \rfloor}$, where
$$
\omega_n = \bigotimes_x \rho_x^{\otimes \lfloor p_x n \rfloor}.
$$
Recall the notion of the typical set $\cT^n_{p,\delta}$.
For any $x^n \in \cT^n_{p,\delta}$,
$$
| n_x - p_x n| \leq \delta n,
$$
where $n_x$ is the number of occurrences of the symbol $x$ in 
$x^n$. In addition, $p^{\otimes n}(\cT^n_{p,\delta}) \geq 1 - \epsilon$ 
for any $\epsilon, \delta > 0 $ and sufficiently
large $n$.
Then
$$
\left\| \sigma^{\otimes n} - \sum_{x^n \in \cT^n_{p,\delta}} p^{\ot n}(x^n)
\proj{x^n}^{ X_A} \otimes \proj{x^n}^{ X_B} \, \otimes \rho_{x^n} \right\|_1
\leq \epsilon.
$$
%The state that we need to approximately create is 
%$\sum_x p_x \alpha_x = (\omega_n)_n$ with  
%$$
%\omega_n = \bigotimes_x \rho_x^{\otimes \lfloor p_x n \rfloor}.
%$$
For any $x^n \in \cT^n_{p,\delta}$ there is, clearly, a unitary 
$U_{x^n}^{A} \otimes U_{x^n}^{B}$ that maps $\rho_{x^n}$ to 
$\omega_{([1 - \delta] n - 1)} \otimes \hat{\rho}_{x^n}$ 
exactly for some state $\hat{\rho}_{x^n}$. Performing 
$$
\left(\sum_{x^n} \proj{x^n}^{ X_A} \otimes U_{x^n}^{A}\right) \otimes 
\left(\sum_{x^n} \proj{x^n}^{ X_B} \otimes U_{x^n}^{B}\right) 
$$
and tracing out subsystems thus brings $\sigma^{\otimes n}$
$\epsilon$-close to $\omega_{ \lfloor n( 1 - \delta) \rfloor}$.
%$\omega_{([1 - \delta] n - 1)}$.
Hence the claim.
\end{proof}

\medskip

In fact, the above result could be strengthened to the equality
\be \alpha = \sum_x p_x \, \alpha_x + H(X_A)_{\sigma} \, [c \, c], \ee
but we will not need this fact, so omit the proof.  However, we will
show how a similar statement to \thm{alfav} can be made about
source coding.

%We state a generalization of \thm{alfav} without proof.

\begin{theorem} 
\label{thm:betav}
Consider a source state of the form
$$
\rho^{X_S S} = 
\sum_x p_x \, \proj{x}^{ X_S} \otimes  \omega^{S}_x.
$$
Then
$$
\sum_x p_x  \, \<\id^{A' \rightarrow B}: \omega_x^{A'}\>
 +  \< \bar{\Delta}^{ X_S \rightarrow X_A X_B} 
\otimes {\id}^{ S \rightarrow A} : \rho^{X_S S}   \> \geq 
 \<\bar{\Delta}^{ X_S \rightarrow X_A X_B} 
\otimes {\id}^{ S \rightarrow B} : \rho^{ X_S S}   \>. 
$$
%\qed
\end{theorem}
%An important special case is when $\cN = \id^{S \rightarrow B}$.

\begin{proof}
The proof is very similar to that of the previous theorem and
is hence omitted.
\end{proof}

\begin{corollary}
 \label{betav2}
In the setting of the above theorem, let $\bar{\cN}^{X_S \rightarrow Y_A}$
be a $\{ c \rightarrow c \}$ entity and let
$$
\bar{\cN}^{X_S \rightarrow Y_A}(\rho^{X_S S}) = 
\sum_y q_y \, \proj{y}^{Y_A} \otimes \sigma_y^{S}.
$$
Define ${\bar{\cN}'}^{X_S \rightarrow Y_A Y_B} = 
\bar{\Delta}^{Y_A \rightarrow Y_A Y_B} \circ \bar{\cN}^{X_S \rightarrow Y_A}$.
Then
$$
\sum_y q_y  \, \<\id^{A' \rightarrow B}: \sigma_y^{A'}\>
 +  \< {\bar{\cN}'}^{ X_S \rightarrow Y_A Y_B} 
\otimes {\id}^{ S \rightarrow A} : \rho^{X_S S}   \> \geq 
 \<{\bar{\cN}'}^{ X_S \rightarrow Y_A Y_B} 
\otimes {\id}^{ S \rightarrow B} : \rho^{ X_S S}   \>.
$$
\qed
\end{corollary}

\vfill\pagebreak

\section{Known coding theorems and converses expressed as resource
inequalities} 
\label{sec:known}

There have been a number of quantum and classical coding theorems 
discovered to date, typically along with so-called converse theorems
which prove that the coding theorems cannot be improved upon. The
theory of resource inequalities has been 
developed to provide an underlying unifying principle. 
This direction was initially suggested in \cite{DW03a}.

We shall state theorems such as Schumacher compression, the classical reverse 
Shannon theorem,  the instrument compression theorem, 
the classical-quantum Slepian-Wolf theorem,
the HSW theorem, and common randomness concentration  as resource inequalities.
Then we will show how some of these can be used as building
blocks, yielding transparent and concise proofs of other important results.

We shall work within the QQ formalism.

\paragraph{Schumacher compression.}
The quantum source compression theorem was proven by Schumacher in \cite{JS94,Sch95}.
Given a quantum state $\rho^{A'}$, define $\sigma^{B} := \id^{ A'
\rightarrow B}(\rho^{A'})$.  Then the 
following RI holds:  
\be
(H(B)_\sigma + \delta) [q \rightarrow q] \geq  
\<\id^{ A' \rightarrow B}  : \rho^{A'} \>, 
\label{eq:schu}
\ee
if and only if $\delta\geq 0$.

Note that this formulation simultaneously
expresses both the coding theorem and the converse theorem.

The Source version of this theorem states that
\be
(H(B)_\sigma + \delta) [q \rightarrow q]
+ \< \id^{S \rightarrow {A}}: \rho^{S} \>
 \ \sourceRI \  
\< \id^{S \rightarrow {B}}: \rho^{S} \>, 
\label{eq:schu2}
\ee
if and only if $\delta\geq 0$.

\paragraph{Entanglement concentration.} The problem 
of entanglement concentration was solved in \cite{BBPS96}, and is,
in a certain sense, a static counterpart to Schumacher's 
compression theorem.
Entanglement concentration can be thought of as a coding theorem which
says that given a pure bipartite quantum state $\ket{\phi}^{AB}$ 
the following RI holds: 
$$
\< \phi^{A B} \> \geq H(B)_\phi \,[q \, q].
$$

The reverse direction is known as \emph{entanglement dilution} \cite{BBPS96},
and thanks to Lo and Popescu \cite{LP99} it is known that
$$
H(B)_\phi \,[q \, q] + o\, [c \ra c] \geq \< \phi^{AB} \>.
$$

Were it not for the $o\,[c\ra c]$ term, we would have the equality
$\<\phi^{AB}\>=H(B)_\phi \, \qq$.  However, it turns out
that the $o [c \ra c]$ term cannot be avoided \cite{HL02,HW02}.  This
means that the strongest equality we can state has a sublinear amount
of classical communication on both sides:
\be
H(B)_\phi \,[q \, q] + o\,\ctc = \< \phi^{AB}\> + o\,\ctc
.\label{eq:BBPS-equality}\ee

Note how \eq{BBPS-equality} states the converse in a form that is in
some ways stronger than \eq{schu}, since it implies the transformation
is not only optimal, but also asymptotically reversible.  We can also
state a converse when unlimited classical communication is allowed:
$$ \< \phi^{A B} \> + \infty\,\ctc 
\geq (H(B)_\phi -\delta) \,[q \, q]$$
iff $\delta\geq 0$; and similarly for entanglement dilution.

\paragraph{Shannon compression.} Shannon's classical compression
theorem was proven in \cite{Shannon48}.  Given a classical state
${\rho}^{X_A}$ and defining
$${\sigma}^{X_B} = \bar{\id}^{ X_A \rightarrow X_B}({\rho}^{X_A}),$$
Shannon's theorem says that
\be
(H(X_B)_{{\sigma}} + \delta) [c \rightarrow c] \geq  
\<\bar{\id}^{ X_A \rightarrow X_B} : {\rho}^{X_A} \>, 
\label{shan}
\ee
if and only if $\delta\geq 0$.
The Source version of this theorem reads
\be
(H(X_B)_{{\sigma}} + \delta) [c \rightarrow c]
+ \<\bar{\id}^{ X_S \rightarrow X_A} : {\rho}^{X_A} \>
 \ \sourceRI \ 
\<\bar{\id}^{ X_S \rightarrow X_B} : {\rho}^{X_A} \>
\label{shan2}
\ee
if and only if $\delta\geq 0$.

\paragraph{Common randomness concentration.}
This is the classical analogue of entanglement concentration,
and a static counterpart to Shannon's compression theorem.
It states that, if Alice and Bob have a copy of the same random
variable $X$, embodied in the classical bipartite state
$$
\rho^{X_A X_B} = \sum_x p_x \proj{x}^{X_A} \otimes \proj{x}^{X_B},
$$
then
\be
\< \rho^{X_A X_B} \> \geq H(X_B)_\rho \, [c \, c].
\label{eq:crc}
\ee
Incidentally, common randomness dilution can do without the $o [c\ra
  c]$ term: 
$$
H(X_B)_\rho \, [c \, c] \geq \< \rho^{X_A X_B} \>.
$$

Thus we obtain a simple resource equality:
$$H(X_B)_\rho \,\cc = \<\rho^{X_AX_B}\>.$$

\paragraph{Classical reverse Shannon theorem (CRST).} This theorem
was proven in \cite{BSST01,Winter:02a}, and it generalizes
Shannon's compression theorem to compress probability distributions
of classical states instead of pure classical states.  
Given a classical channel $\bar{\cN}: X_{A'} \rightarrow Y_B$ and a
classical state ${\rho}^{X_{A'}}$, the CRST states that
\be
I(X_A; Y_B)_\sigma [c \rightarrow c] + H(X_A |Y_B)_\sigma [c \, c] \geq  
\<\bar{\cN} : \rho^{X_{A'}} \>, 
\label{eq:crst}
\ee
where 
$$
{\sigma}^{X_A Y_B} = {\cC_{\bar{\cN}}}^{X_{A'} \rightarrow Y_{B} X_{A}}  
(\rho^{X_{A'}}).
$$

We can also express this in the Source formalism,
$$
I(X_A; Y_B)_\sigma [c \rightarrow c] + H(X_A |Y_B)_\sigma [c \, c]
                                     + \<\bar{\id}^{X_S \ra X_{A'}} : \rho^{X_S} \>
        \ \sourceRI \ \<\bar{\cN}^{X_S \ra X_B} : \rho^{X_S} \>.
$$
%(Note that here, as well as in Schumacher and Shannon compression, the
%Source formulation was never necessary, since the ``improper'' resource
%involved was perfectly well-formed.)

Moreover, given a modified classical channel $\bar{\cN}': 
X_{A'} \rightarrow Y_A Y_B$ which also provides Alice with a copy
of the channel output,
$$
\bar{\cN}' =  \bar{\Delta}^{Y_B  \rightarrow Y_A Y_B}   \circ \bar{\cN}, 
$$
 the following stronger RI also holds:
\be
I(X_A; Y_B)_\sigma [c \rightarrow c] + H(X_A| Y_B)_\sigma [c \, c] \geq  
\<\bar{\cN}'  : \rho^{X_{A'}} \>, 
\label{eq:crst2}
\ee
In fact, this latter RI can be reversed to obtain the equality
\be
I(X_A; Y_B)_\sigma [c \rightarrow c] + H(X_A| Y_B)_\sigma [c \, c] =
\<\bar{\cN}'  : \rho^{X_{A'}} \>.
\ee

However, in the case without feedback, the best we can do is a
tradeoff curve between cbits and rbits, with \eq{crst} representing
the case of unlimited randomness consumption.  The full tradeoff
will be given by an RI of the following form
$$a \,\ctc + b \,\cc \geq \<\bar{\cN}  : \rho^{X_{A'}} \>$$
where $(a,b)$ range over some convex set $CR(\bar{\cN})$.  It can be
shown \cite{Wyner75, BW05} that $(a,b)\in CR(\bar{\cN})$ iff there
exist channels 
$\bar{\cN}_1:X_{A'}\ra W_{C'}, \bar{\cN}_2:W_{C'}\ra Y_B$ such that 
$\bar{\cN}=\bar{\cN}_2 \circ \bar{\cN}_1$
and $a \geq I(X_A;W_C)_\omega, b \geq I(X_AY_B;
W_C)_\omega$, where 
$$\omega^{X_AW_CY_B} := 
{\cC_{\bar{\cN}_2}}^{W_{C'} \rightarrow Y_{B} W_{C}}
\circ  {\cC_{\bar{\cN}_1}}^{X_{A'} \rightarrow W_{C'} X_{A}} 
(\rho^{X_{A'}}).
$$

\paragraph{Classical compression with quantum side information.}
This problem  was solved in \cite{DW02, WinterPhD}, 
and is a generalization of Shannon's classical compression
theorem in which Bob has quantum side information 
about the source. 
Suppose Alice and Bob are given an ensemble  
$$
{\rho}^{X_A B} = \sum_x p_x \proj{x}^{X_A} \otimes \rho_x^{B}, 
$$ 
and Alice wants to communicate $X_A$ to Bob, which would give them the
state 
$$
{\sigma}^{X_B B} := \bar{\id}^{X_A \rightarrow X_B}({\rho}^{X_A B}).
$$
To formalize this situation, we use the Source as one of the protagonists in the
protocol, so that the coding theorem redirects a channel from the Source
to Alice and Bob $\<\bar{\id}^{X_S\ra X_A}\ot \id^{S\ra B}:\rho^{X_S
S}\>$ to a channel from the Source entirely to Bob.  The coding
theorem is then
\be
(H(X_B | B)_{{\sigma}}+\delta) [c \rightarrow c] 
+  \< \bar{\id}^{X_S\ra X_A}\ot \id^{S\ra B}:\rho^{X_SS}\>
\ \sourceRI \  
\<\bar{\id}^{X_S \rightarrow X_B} \ot \id^{S\ra B}:\rho^{X_SS}\>,
\label{eq:cqsw}
\ee
which holds iff $\delta\geq 0$.  
%This formulation ensures that we work
%with well-defined resources instead of using the improper 
%$\<\bar{\id}^{X_A\ra X_B}:\rho^{X_AB}\>$. Of course, with
%definition~\ref{def:asy-ineq2}, we may shorthand the inequality
%$$
%  (H(X_B | B)_{{\sigma}}+\delta) [c \rightarrow c]
%      \ \sourceRI \  \<\bar{\id}^{X_A\ra X_B}:\rho^{X_AB}\>.
%$$
%\footnote{Resources are defined only relative to states
%on $A$ and $S$. Otherwise
%\eqs{asy-monotone}{asy-quasi-iid} would be violated by
%examples like $\<\bar{\id}^{X_A\ra X_B}:\rho^{X_AB}\>$.}

Of course, with no extra resource cost Alice could keep a copy of
$X_A$.

\paragraph{Instrument compression theorem.}
This theorem was proven in \cite{Winter01}, 
and is a generalization of the CRST.
Given a remote instrument ${{\mathbb {T}}}: 
A' \rightarrow {A} X_B$,
and a quantum state $\rho^{A'}$, the 
following RI holds:
\be
I(R; X_B)_\sigma [c \rightarrow c] + 
H(X_B| R)_\sigma [c \, c] \geq \< {{\mathbb {T}}} : \rho^{A'} \>,
\label{eq:ict}
\ee
where 
$$
{\sigma}^{R A X_B} = {{\mathbb {T}}}(\psi^{R A'})
$$
and $\proj{\psi}^{R A'} \ext \rho^{A'}$.
Moreover, given a modified remote instrument
 which also provides Alice with a copy of the instrument output,
$$
{{\mathbb {T}}}'  =  \bar{\Delta}^{X_B  \rightarrow X_A X_B} \circ {{\mathbb {T}}}, 
$$
the  RI still holds:
\be
I(R; X_B)_\sigma [c \rightarrow c] + 
H(X_B| R)_\sigma [c \, c] \geq \< {{\mathbb {T}}}' : \rho^{A'} \>.
\label{eq:ict2}
\ee
Only this latter RI is known to be optimal (up to a trivial
substitution of $\ctc$ for $\cc$); indeed 
\be
a \, [c \rightarrow c] + b \, [c \, c] \geq \< {{\mathbb {T}}}' : \rho^{A'} \>.
\ee
iff $a\geq I(R;X_B)_\sigma$ and $a+b \geq H(X_B)_\sigma$.

By contrast, only the communication rate of \eq{ict} is known to be
optimal; examples are known in which less randomness is necessary.

\paragraph{Teleportation and super-dense coding.} 
Teleportation \cite{BBCJPW98} and super-dense coding \cite{BW92} 
are finite protocols, and we have discussed them already in the
introduction.
In a somewhat weaker form they may be written as resource 
inequalities. Teleportation (TP):
\be
 2\,[c \rightarrow c] + [q \, q] \geq [q \rightarrow q].
\label{eq:tp}
\ee
Super-dense coding (SD):
\be
[q \rightarrow q] + [q \, q] \geq 2  \,[c \rightarrow c].
\label{eq:sd}
\ee
Finally, entanglement distribution:
\be
[q \rightarrow q] \geq  [q \, q].
\ee
All of these protocols are optimal (we neglect the precise
statements), but composing them with each other (e.g. trying to
reverse teleportation  by using super-dense coding) is wasteful.  By
replacing classical communication with coherent classical
communication (below), the protocols become reversible.

\paragraph{Coherent classical communication identity.} In \cite{Har03}
two more resource inequalities involving unit resources were
discovered. Coherent versions of teleportation and 
super-dense coding, respectively:
\ben
[q \rightarrow q] + [q \, q] & \geq & 2  \,[q \rightarrow qq], \\
2\,[q \rightarrow qq] + [q \, q] & \geq & [q \rightarrow q] + 2\,[q \, q].
\een
The $[q \, q]$ term on the left hand side of the second inequality
may be canceled completely by \lem{cancel}, \lem{noo} 
and the fact that $[q \rightarrow qq] \geq [q \, q]$.
This brings us to the \emph{coherent communication identity}
\be
[q \rightarrow qq] = \frac{1}{2} ([q \rightarrow q] + [q \, q]),
\label{eq:ccc}
\ee
which will turn out to be an important tool for constructing new protocols.

\paragraph{Holevo-Schumacher-Westmoreland (HSW) theorem.}
The direct part of this theorem was proven in \cite{Holevo98, SW97} and
the converse in \cite{Holevo73}. Together they say that 
given a quantum channel ${\cN}: A' \rightarrow B$, 
for any ensemble  
$$
{\rho}^{X_A A'} = \sum_x p_x \proj{x}^{X_A} \otimes \rho_x^{A'} 
$$ 
the following RI holds:  
\be
\<{\cN} : \rho^{A'}\>  \geq
(I(X_A; B)_\sigma -\delta) [c \rightarrow c], 
\label{hsw}
\ee
iff $\delta\geq 0$, where 
$$
{\sigma}^{X_A B} = {\cN}^{A'\ra B}(\rho^{X_A A'}).
$$

%In fact, it was proven in \cite{Holevo98, SW97} that
%This is because by the Ahlswede-Winter lemma
%$$
%\< \cN : \rho^{A'} \> + z [c c]  \geq I(X_A;B) 
%([c \rightarrow c] : [c c]),
%$$
%for some $z > 0$, and the $z [c \, c]$ term 
% was incoherently decoupled in the RI.
%From lemma \ref{rcr} and theorem \ref{absolutize}
%follows an alternative version of (\ref{hsw}): 
%\be
%\<{\cN} : \rho^{A'} \>  + o [c c] \geq
%I(X_A; B)_\sigma [c \rightarrow c].
%\label{hsw2}
%\ee

\paragraph{Shannon's noisy channel coding theorem}
This theorem was proven in \cite{Shannon48} and today can be understood
as a special 
case of the HSW theorem.
One version of the theorem says that
given a classical channel $\bar{\cN}: X_{A'} \rightarrow Y_B$
and any classical state ${\rho}^{X_{A'}}$ the 
following RI holds:  
\be
\<\bar{\cN} \>  \geq
(I(X_A; Y_B)_\sigma-\delta) [c \rightarrow c], 
\label{shan2a}
\ee
iff $\delta\geq 0$ and where 
\be
{\sigma}^{X_A Y_B} := 
 {\cC_{\bar{\cN}}}^{X_{A'} \rightarrow Y_{B} X_{A}}
%\bar{\cN} \circ \bar{\Delta}^{X_{A'} \rightarrow X_{A'}X_{A}} 
(\rho^{X_{A'}}).
\label{eq:shannon-sigma-def}\ee
If we optimize over all input states, then we find that
\be \<\bar{\cN}\> \geq \,C \ctc \ee
iff there exists an input ${\rho}^{X_{A'}}$ such that $C\leq
I(X_A;Y_B)_\sigma$, with $\sigma$ given by \eq{shannon-sigma-def}.

%Again, the alternative RI holds:
%\be
%\<\bar{\cN}  : \rho^{X_A} \>  \geq
%I(R; X_B)_\sigma [c \rightarrow c]. 
%\label{shan2b}
%\ee
%One could combine this with (\ref{crst}) into an identity like 
%\be
%\<\bar{\cN}  : \rho^{X_A} \>   = 
%I(R; X_B)_\sigma [c \rightarrow c] + O {[c \, c]}  . 
%\label{shan12}
%\ee

\paragraph{Entanglement-assisted capacity (EAC) theorem.}
This theorem was  proven in \cite{BSST01,Holevo01a,HsDeWi05}.
The direct coding part of the theorem says that, 
given a quantum channel ${\cN}: A' \rightarrow B$, 
for any quantum state $\rho^{A'}$
the following RI holds:  
\be
 \< \cN : \rho^{A'}\> + H(R)_\sigma [q \, q]
\geq I(R; B)_\sigma \,[c  \rightarrow c],
\label{eq:eac},
\ee
where 
$$
{\sigma}^{R B} = {\cN}(\psi^{R A'})
$$
for an arbitrary $\psi$ satisfying  $\proj{\psi}^{R A'} \ext
\rho^{A'}$.

The only converse proven in \cite{BSST01,Holevo01a} was for the case of
infinite entanglement: they found that $\<\cN\> + \infty\qq \geq
C\ctc$ iff $C \leq I(R; B)_\sigma$ for some appropriate $\sigma$.
\mscite{Shor04} gave a full solution to the tradeoff problem for
entanglement-assisted classical communication which we will present an
alternate converse  for in \sect{NCE-toff}.

%Furthermore, in the appendix we show a stronger RI:
%\be
% \< \cN :  \rho^{A'} \> + H(R)_\sigma [q \, q]
%\geq I(R; B)_\sigma \,[c  \rightarrow c].
%\label{eac2}
%\ee

\paragraph{Quantum capacity (LSD) theorem.}
This theorem was conjectured in \cite{Sch96,SN96},
a heuristic (but not universally accepted) proof given
by Lloyd~\cite{Lloyd96}
and finally proven by Shor~\cite{Shor02} and
with an independent method by Devetak~\cite{Devetak03}.
The direct coding part of the theorem says that, 
given a quantum channel ${\cN}: A' \rightarrow B$, 
for any quantum state $\rho^{A'}$
the following RI holds:  
\be
 \< \cN \> \geq I(R\,\rangle B)_\sigma  \,[q  \rightarrow q],
\label{eq:lsd}
\ee
%iff $\delta\geq 0$ and 
where 
$$
{\sigma}^{R B} = {\cN}(\psi^{R A'})
$$
for any $\psi^{RA'}$ satisfying $\proj{\psi}^{R A'} \ext \rho^{A'}$.
%Furthermore, in \cite{Devetak03} it was shown that
%$$
% \< \cN : \rho^{A'} \> + z [ c \, c]
%\geq 
%I(R\,\rangle B)_\sigma \, [q  \rightarrow q : \tau],
%$$
%for some $z > 0$, and the $z [ c \, c]$ term 
% was incoherently decoupled in the RI.
%By lemmas \ref{rcr} and \ref{noo} 
%with (\ref{hsw2}) and 
%$[c \rightarrow c] \geq [c \, c]$,
%$$
% \< \cN  : \rho^{A'}\> + o [c \,c ] \geq I(R\,\rangle B)_\sigma \,
%[q  \rightarrow q : \tau] .
%$$
%Invoking theorem \ref{absolutize}, we have an alternative
%version of (\ref{lsd}):
%\be
% \< \cN  : \rho^{A'}\> + o [c \,c ] 
%\geq I(R\,\rangle B)_\sigma \,[q  \rightarrow q].
%\label{lsd2}
%\ee

\paragraph{Noisy super-dense (NSD) coding theorem.} 
This theorem was  proven in \cite{HHHLT01}.
The direct coding part of the theorem says that, 
given a bipartite quantum state $\rho^{A B}$, 
the following RI holds:  
\be
   \< \rho^{AB} \> + 
H(A)_\rho \, [q \rightarrow q]  
\geq  I(A; B)_\rho \,[c \rightarrow c].
\label{eq:nsd}
\ee
A converse was proven in \cite{HHHLT01} only for the case when an
infinite amount of $\<\rho^{AB}\>$ is supplied, but we will return to
this 
problem and provide a full trade-off curve in \sect{nSD-toff}.

\paragraph{Entanglement distillation.}
The direct coding theorem for one-way entanglement distillation
is embodied in the \emph{hashing inequality}, proved in~\cite{DW03b,DW03c}:
given a bipartite quantum state $\rho^{A B}$, 
\be
\<\rho^{AB}\> + I(A;E)_\psi \, [c \rightarrow c] 
                      \geq I(A\,\rangle B)_\psi \,[q \, q],
\label{eq:hashing}
\ee
where 
$\proj{\psi}^{A B E} \ext \rho^{A B}$.

Again, the converse was previously only known for the case when an
unlimited amount of classical communication was
available \cite{Sch96,SN96, DW03b,DW03c}.  In \sect{distill-toff} we
will give an expression for the full trade-off curve.

\paragraph{State merging.}
The state merging RI was proved in \cite{HOW05}
\be
 \< U^{S\ra AB} : \rho^{S} \> +  I(A;E)_\psi \, [c \ra c] 
+ H(A | B)_\psi
 [q \, q]
\ \sourceRI \ 
 \< \id^{S\ra B} : \rho^{S} \>,
\label{eq:merging}
\ee
where $U^{S\ra AB}$ is an isometry,
$\rho^{AB} =  U^{S\ra AB}(\rho^{S})$ and $\psi^{ABE}$ is defined as above. 
It holds irrespectively of the sign of $H(A|B)$.
It implies entanglement distillation via Lemmas \ref{lemma:sopm} 
and  \ref{lemma:nosorog}.
Conversely, the protocol \cite{DW03b} implementing 
(\ref{eq:hashing}) may be easily modified (replacing  Eve with the reference system)
to give (\ref{eq:merging}) for $H(A | B)_\psi < 0$.

Lemma (\ref{lemma:sopm}) says that proper and improper resource inequalities 
are equivalent up to $o$ terms. In this vein, we may equivalently 
write (\ref{eq:merging}) as
\be
 I(A;E)_\psi \, [c \ra c]
+ H(A | B)_\psi [q \, q]
 \ \sourceRI \
 \< \id^{S\ra B} : \rho^{S} \> - \< U^{S\ra AB} : \rho^{S} \>,
\ee
 reflecting the fact that
the \emph{redirection} of protected resources (in this case from Alice to Bob)
is the information processing task Alice and Bob are trying to accomplish.
Taking this a step further, one may be inclined to disregard the Source
altogether and define
$$
 \< \id^{A \rightarrow B'}: \rho^{AB} \>  := 
\< U^{S\ra AB} : \rho^{S} \>  - \< \id^{S\ra B} : \rho^{S} \>,
$$
in analogy to the Source-free version of Schumacher compression (\ref{eq:schu})
(strictly speaking, our current formalism does not permit this). 
Curiously, $ \< \id^{A \rightarrow B'}: \rho^{AB} \> $ on the right hand
side of a RI can be an asset or liability, depending on
whether $H(A|B)_\psi$ is negative or positive.

\paragraph {Noisy teleportation.} This RI was 
discovered in \cite{DHW03}. 
Given a bipartite quantum state $\rho^{A B}$,
\ben
\label{eq:ntp}
\<\rho^{AB}\> + I(A;B)_\rho \, [c \rightarrow c] 
 \geq  I(A\,\rangle B)_\rho \,[q \rightarrow q].
\een 
Indeed, letting
$\proj{\psi}^{A B E} \ext \rho^{A B}$,
\ben
\<\rho^{AB}\> + I(A;B)_\psi \, [c \rightarrow c] 
& = & \<\rho^{AB} \> + I(A;E)_\psi \, [c \rightarrow c] + 
2 I(A\,\rangle B)_\psi [c \rightarrow c] \\
& \geq &  I(A\,\rangle B)_\psi \,[q \, q]  + 
2 I(A\,\rangle B)_\psi [c \rightarrow c] \\
& \geq & I(A\,\rangle B)_\psi \,[q \rightarrow q].
\een 
The first inequality follows from \eq{hashing} and
the second from teleportation.

\paragraph{Quantum  compression with classical side 
information} 
%-quantum communication trade-off
%           for remote state preparation.} 
Suppose Alice is given the 
ensemble 
$$
{\rho}^{X_{A} A} = \sum_x p_x \proj{x}^{X_{A}} \otimes \rho_x^{A}, 
$$
and she wants Bob to end up with the quantum part $A$ \cite{HJW02}.
The resources at their disposal are $[c \ra c]$ and $[q \ra q]$.
%This is quantum compression with classical side information.
As in the classical compression with quantum side information
problem above, we first give $\rho^{X_A A}$ to the Source (and
rename it $\rho^{X_SS}$).
%this is written as a source coding problem.
For any classical channel $\bar{\cN}: X_{S} \rightarrow Y_B$,
the following RI holds \cite{HJW02}:
\be
\label{eq:cqrsp}
\< \bar{\id}^{X_S\ra X_A}\ot \id^{S\ra A}:\rho^{X_SS}\> +
 H(B|Y_B)_\sigma [q \rightarrow q] + I(X_A;Y_B)_\sigma [c \rightarrow c]
\ \sourceRI \
\< \id^{S\ra B}:\rho^{S}\> 
%\< \bar{\id}^{X_S\ra X_A}\ot \id^{S\ra B}:\rho^{X_SS} \>-
,
\ee
where
$$
\sigma^{X_A Y_B B} = 
%((\bar{\cN}^{X_{A'}\ra Y_B}\circ\bar{\Delta}^{X_{S}\ra X_{A'}X_{A}})
 ({\cC_{\bar{\cN}}}^{X_{S} \rightarrow Y_{B} X_{A}}
\otimes \id^{S \rightarrow B})
 \rho^{X_{S}  S}.
$$
Conversely, if $a\qtq + b \ctc $ is $\geq$ to the right hand side of
\eq{cqrsp} 
then there exists a classical channel $\bar{\cN}: X_A \rightarrow Y_B$
with corresponding state $\sigma$ such that $a\geq H(B|Y_B)_\sigma$
and $b\geq I(X_A;Y_B)_\sigma$.

We shall now show how the proof from \cite{HJW02}
may be written very succinctly
in terms of the resource calculus.
Define $\bar{\cN}' = \bar{\Delta}^{Y_{B} \rightarrow Y_{A} Y_{B}}
\circ \bar{\cN}$. By the Classical Reverse Shannon Theorem
\eq{crst2}  
$$
I(X_A; Y_B)_\sigma [c \rightarrow c] + H(X_A| Y_B)_\sigma [c \, c] \geq  
\<{\bar{\cN}'}^{X_{A'} \ra Y_A Y_B}  : \rho^{X_{A'}} \>. 
$$
Combining with part 3 of \lem{relatif} gives
$$
\< \bar{\id}^{X_S\ra X_A}\ot \id^{S\ra A}:\rho^{X_SS}\>
+ I(X_A; Y_B)_\sigma [c \rightarrow c] + H(X_A| Y_B)_\sigma [c \, c]
\geq  
\< {\bar{\cN}'}^{X_S\ra Y_A Y_B}\ot \id^{S\ra A}:\rho^{X_SS}\>
$$
On the other hand, combining Schumacher compression \eq{schu} with 
Corollary \ref{betav2} %and \lem{sopm}  
gives
%Schumacher compression gives, for a source $\omega_y^{A}$
%$$
% H(B)[q \rightarrow q] \geq  \<\id_{ A \rightarrow B}  : \omega_y^{A} \>. 
%$$
$$
H(B| Y_B)_\sigma [q \rightarrow q] 
+ \< {\bar{\cN}'}^{X_S\ra Y_A Y_B}\ot \id^{S\ra A}:\rho^{X_SS}\>
\geq
\< {\bar{\cN}'}^{X_S\ra Y_A Y_B}\ot \id^{S\ra B}:\rho^{X_SS}\>.
$$
Adding the two equations gives
\be
\begin{split}
& \< \bar{\id}^{X_S\ra X_A}\ot \id^{S\ra A}:\rho^{X_SS}\> +
H(B|Y_B)_\sigma [q \rightarrow q] + I(X_A;Y_B)_\sigma [c \rightarrow c]
+ H(X_A| Y_B)_\sigma [c \, c] \\
& \geq 
\< {\bar{\cN}'}^{X_S\ra Y_A Y_B}\ot \id^{S\ra B}:\rho^{X_SS}\> \\
& \geq \< \id^{S\ra B}:\rho^{S}\>.
\end{split}
\ee
The last line is by part 4 of  \lem{relatif}.
Derandomizing via \cor{purim}
gives
\be
%\begin{split}
 H(B|Y_B)_\sigma [q \rightarrow q] + I(X_A;Y_B)_\sigma [c \rightarrow c] 
 + \< \bar{\id}^{X_S\ra X_A}\ot \id^{S\ra A}:\rho^{X_SS}\>
\geq \< \id^{S\ra B}:\rho^{S}\>.
%\end{split}
\ee
Invoking \lem{sopm}  and  \lem{noo}
yields the desired result \eq{cqrsp}.

%The result of \cite{HJW02} may be also viewed as a statement about 
%remote state preparation. Suppose we are given a classical state
%$\rho^{X_{A''}}$ and a remote $\{c \rightarrow q \}$ map
%$\cE' : X_{A''} \rightarrow B$, $\cE' = \id^{A' \rightarrow B} \circ 
%\cE$, where $\cE$ has Kraus representation 
%$\{ \ket{\phi_x}^{A^*A'} \bra{x}^{X_{A''}} \}$.  Then 
%for any classical channel $\bar{\cN}: X_{A} \rightarrow Y_B$,
%the following RI holds:
%\be
%\label{eq:cqrsp2}
% H(B|Y_B)_\sigma [q \rightarrow q] + I(X_A;Y_B)_\sigma [c \rightarrow c]
%\geq \<\cE' : \rho^{X_{A''}} \>,
%\ee
%where $\sigma^{X_A Y_B B}$ is defined as above and
%$$
%\rho^{X_{A'}  A^* A'} = (\cE \circ 
%\bar{\Delta}^{X_{A''} \rightarrow X_{A''} X_{A'}})\rho^{X_{A''}}  .
%$$
%This follows from adding (\eq{cqrsp}) to
%\ben
%\<\id^{A' \rightarrow B} : \rho^{X_{A'}  A'} \> & \geq & 
%\<\id^{A' \rightarrow B} : \rho^{X_{A'} A^*  A'} \>  + 
%\< (\cE \circ 
%\bar{\Delta}^{X_{A''} \rightarrow X_{A''} X_{A'}}) : \rho^{X_{A''}}\> \\
%& \geq &   \<\cE' : \rho^{X_{A''}} \>.
%\een
%The first inequality follows from part 3 of \lem{relatif} and
%the locality of the map $\cE$.
%The second is an application of  part 4 of \lem{relatif}.

\paragraph{Common randomness distillation.}
This theorem was originally proven in \cite{DW03a}.
Given an ensemble 
$$
\rho^{X_A B} = \sum_x p_x \proj{x}^{X_A} \otimes \rho_x^{B}, 
$$ 
the
following RI holds:
\be
\< \rho^{X_A B} \> + H(X_A|B)_\rho [c \rightarrow c] 
 \geq  H(X_A)_\rho [c \, c].
\label{eq:crd}
\ee
Our formalism makes transparent the intimate
relation between \eq{crd} and the problem of 
classical compression with quantum side
information \eq{cqsw}.
\be
\begin{split}
& \< \bar{\id}^{X_S \ra X_A} \otimes {\id}^{S \ra B}:  \rho^{X_S S} \> 
+ \, H(X_A|B)  [c \rightarrow c]  \\
& \geq \< \bar{\Delta}^{X_S \rightarrow X_A X_B}  \otimes {\id}^{S \ra B} : \rho^{X_S S} \>
\\
& \geq \< \bar{\Delta}^{X_S \rightarrow X_A X_B} : \rho^{X_S} \> \\
& \geq \< \bar{\Delta}^{X_S \rightarrow X_A X_B}(\rho^{X_S}) \> \\
& \geq H(X_A)\, [c \, c].
\end{split}
\ee
The first inequality is by \eq{cqsw} and \lem{sopm};
the second and third are  by  parts 5 and 2, respectively, of \lem{relatif}.
The last inequality is common randomness concentration \eq{crc}.
%Second, check out secrecy?
By \lem{nosorog}, $\< \bar{\id}^{X_S \ra X_A} \otimes {\id}^{S \ra B}:  \rho^{X_S S} \> $
can be replaced by 
$$
\< \rho^{X_A B} \> = \< \bar{\id}^{X_S \ra X_A} \otimes {\id}^{S \ra B} (  \rho^{X_S S}) \>,
$$
proving \eq{crd}.
\vfill\pagebreak

\section{A family of quantum protocols.} 
\label{sec:family}

\subsection{The family tree.}
A large class of problems in quantum Shannon theory involves
transforming a noisy resource, such as a channel or bipartite state,
into a noiseless one (such as cbits, ebits or qubits), perhaps by 
consuming some other noiseless resource.  In the prequel to this
paper \cite{DHW03} we gave a unified treatment of four such protocols
that were already known together with three new such protocols.
This section and the next one are devoted to a detailed
treatment of these results.
This is now possible because of the rigorous theory
of resource inequalities developed above.
All of the RIs presented in this section
involve a single noisy resource.
The ``static'' members of the family 
involve a noisy bipartite state $\rho^{AB}$,
% is denoted by $\{q \, q\}$,
while the ``dynamic'' members 
involve  a general quantum channel 
$\cN: {A'} \rightarrow B$. 
%is denoted by $\{q \rightarrow q\}$.
In the former case one may define a class of purifications
$\proj{\psi}^{ABE} \ext  \rho^{AB}$.
In the latter case one may define a class of pure states $\ket{\psi}^{RBE}$,
which corresponds to the outcome of sending half of
some $\ket{\phi}^{RA'}$ through the channel's isometric extension
$U_\cN : {A'} \rightarrow BE$, $U_\cN \ext \cN$.
%channel's Stinespring \cite{stinespring}
%extension  
%$U_\cN: {A} \rightarrow BE$ 
%($\cN$, mapping states on $A$ to states on $B$,
%is obtained as the isometry $U_\cN$ followed by the partial trace over $E$.)

Recall the identities, for a tripartite pure state $\ket{\psi}^{ABE}$,
\ben
  \frac{1}{2} I(A;B)_\psi + \frac{1}{2} I(A;E)_\psi &= & H(A)_\psi, \\
  \frac{1}{2} I(A;B)_\psi - \frac{1}{2} I(A;E)_\psi &= & I (A\,\rangle B)_\psi.
\een
Henceforth, all entropic quantities will be defined with respect
to $\ket{\psi}^{RBE}$ or $\ket{\psi}^{ABE}$, depending on the context,
so we shall drop the $\psi$ subscript.

The two ``parent'' resource inequalities were introduced
in \cite{DHW03}.
The ``mother'' RI reads
\begin{equation}
  \< \rho \>  + \frac{1}{2} I(A;E) \, [q \rightarrow q] 
                          \geq \frac{1}{2} I(A;B)\,[q \, q]. 
\label{eq:mama}
\end{equation}
%The entropic quantities implicitly refer to any
%$\ket{\psi}^{ABE}$ associated with the noisy resource $\rho^{AB}$.
There exists a  dual ``father'' RI, related to the mother by
interchanging  dynamic and static resources, and the $A$ and $R$ systems:
\begin{equation}
  \frac{1}{2} I(R;E) \, [q \, q] + \< \cN \>
                          \geq \frac{1}{2} I(R;B)\,  [q  \rightarrow q]. 
\label{eq:papa}
\end{equation}
%The entropic quantities implicitly refer to any
%$\ket{\psi}^{RBE}$ associated with the noisy resource $\cN$.
The main observation of \cite{DHW03} was that these parent RIs
may be combined with the unit RIs corresponding
to teleportation, super-dense coding and entanglement distribution
to recover several previously known ``children'' protocols.

\par
Each parent has her or his own children.
%(like the Brady Bunch \cite{bradybunch}),
Let us consider the mother first; she has three children.
The first is a variation of 
the  hashing inequality \eq{hashing},
which follows from the mother and teleportation.
\ben
\<\rho\> +  I(A;E) \, [c \rightarrow c] + \frac{1}{2} I(A;E) [q \, q]
&  \geq & \<\rho\> + \frac{1}{2} I(A;E) [q \rightarrow q]\\
&  \geq &  \frac{1}{2} I(A;E) [q \, q] \\
& = & I(A\,\rangle B) \,[q \, q] + \frac{1}{2} I(A;E) [q \, q].
\een
By the cancellation lemma,
\be
\<\rho\> +   I(A;E) \, [c \rightarrow c] + o[q \, q] 
 \geq I(A\,\rangle B) \,[q \, q].
  \label{eq:hashing2}
\ee
This is slightly weaker than \eq{hashing} itself.
Further combining with teleportation gives a variation
on  noisy teleportation \eq{ntp}:
\be
\<\rho\> + I(A;B) \, [c \rightarrow c] + o[q \, q] 
 \geq  I(A\,\rangle B) \,[q \rightarrow q].
\label{eq:ntp2}
\ee 
The third child is noisy super-dense coding
(\eq{nsd}), obtained by combining the mother with super-dense coding:
%and is obtained from (M) + $\frac{1}{2} I(A;B)$ (SD):
\ben
 H(A) \, [q \rightarrow q] +  \< \rho \> 
& = & \frac{1}{2} I(A;B)  \, [q \rightarrow q] +
\frac{1}{2} I(A;E)  \, [q \rightarrow q] +
 \< \rho \> \\
& \geq & \frac{1}{2} I(A;B) [q \rightarrow q] + 
\frac{1}{2} I(A;B) [q \, q]\\
& \geq & I(A; B) \,[c \rightarrow c].
  \label{eq3}
\een
%\par
%\par
The father happens to have only two children (that we know of).
One of them is the entanglement-assisted classical capacity
RI (\ref{eq:eac}), obtained by combining the father with super-dense
coding
\ben
 H(R) \, [q \, q] +  \< \cN \> 
& = & \frac{1}{2} I(R;B)  \, [q \, q] +
\frac{1}{2} I(R;E)  \, [q \, q] +
 \< \cN \> \\
& \geq & \frac{1}{2} I(R;B) [q \, q] + 
\frac{1}{2} I(R;B) [q \rightarrow q]\\
& \geq & I(R; B) \,[c \rightarrow c].
  \label{eq13}
\een
The second is a variation on the quantum channel capacity result 
\eq{lsd}. It is obtained by combining the father with entanglement
distribution. 
\ben
 \frac{1}{2} I(R;E) \, [q \, q] + \< \cN \>
   & \geq & \frac{1}{2} I(R;B)\,  [q  \rightarrow q] \\
 & =  & \frac{1}{2} I(R;E)\,  [q  \rightarrow q] 
+ \frac{1}{2} I(R \, \> B)\,  [q  \rightarrow q] \\
  & =  & \frac{1}{2} I(R;E)\,  [q \, q] 
+  \frac{1}{2} I(R \, \> B)\,  [q  \rightarrow q].
\een
Hence, by the cancellation lemma
\be
   \< \cN \> + o[q \, q]   \geq I(R\,\rangle B) \,[q  \rightarrow q].
  \label{eq5}
\ee
In the following subsection we give a rigorous proof
of the parent RIs using so-called coherification rules.
%Alas, we do not know how to get rid of the $o$ term without
%invoking further results. For instance, the
%original proof of the hashing inequality and 
%the HSW theorem allow
%us to get rid of the $o$ term,
%by lemma \ref{noo}.
%Quite possibly the original proof \cite{Lloyd96,Shor02,Devetak03} 
%is needed.

\subsection{Constructing the parent protocols using coherification rules.}
\label{sec:ccc}

Having demonstrated the power of the parent resource inequalities,
we now address the question of constructing protocols implementing them.
The lessons learned in \cite{Devetak03,DW03b,DW03c} regarding
making protocols coherent and the observations
of \cite{Har03} (in particular the 
coherent communication identity \eq{ccc}),
lead us to two general rules regarding making
classical communication coherent.  When coherently-decoupled cbits are
in the input to a protocol, Rule I (``input'') says that replacing them with
cobits not only performs the protocol, but also has the side 
effect of generating entanglement.  Rule O (``output'') is simpler; it
says that if a protocol outputs coherently-decoupled cbits, then it
can be modified to instead output cobits.  Using these rules, we can
give simple proofs of the parent protocols by making coherent
previously known protocols.

Below, we give formal statements of rules I and O, deferring their
proofs till the end of the section.  We shall be working in the CP picture.

\begin{theorem}[Rule I]
If for resources $\alpha, \beta \in {\cR}$
$$
\alpha + R \, [c \rightarrow c : \tau] \geq \beta 
$$
and the classical resource $ R \, [c \rightarrow c : \tau]$ is 
coherently decoupled  then
$$
\alpha + \frac{R}{2} \,[q \rightarrow q] \geq \beta + \frac{R}{2} \,[q \, q].
$$
\end{theorem}
There is also an incoherent version of Rule I  which
is easy to prove (cf. \lem{rcr}):
\begin{proposition}[Incoherent Rule I]
If for resources $\alpha, \beta \in {\cR}$
$$
\alpha + R \, [c \rightarrow c : \tau] \geq \beta 
$$
and the classical resource $ R \, [c \rightarrow c : \tau]$ is 
incoherently decoupled  then
$$
\alpha +  R \, [c \rightarrow c : \tau] \geq \beta + R\,[c \, c].
$$
\end{proposition}

\begin{theorem}[Rule O]
If for resources $\alpha, \beta \in {\cR}$
$$
\alpha  \geq \beta + R \,[c \rightarrow c]
$$
and the classical resource $R \,[c \rightarrow c] $ is 
coherently decoupled then 
$$
\alpha  \geq \beta + \frac{R}{2} \, [q \, q] +  
\frac{R}{2} \, [q \rightarrow q].
$$
\end{theorem}

\begin{corollary}
The mother inequality \eq{mama} is obtained 
from the hashing inequality \eq{hashing} by applying rule I. 
It can be readily checked that the classical message in
\cite{DW03b,DW03c}'s protocol is coherently decoupled and is uniformly
random (so the protocol is 0-valid).
\end{corollary}

\begin{corollary}
The father inequality \eq{papa} follows from the EAC inequality 
\eq{eac} by applying rule O. In \cite{HsDeWi05}
it was shown explicitly that the conditions of rule O hold 
for the protocol implementing the EAC inequality exhibited therein.
These conditions also hold for the original protocol of  \cite{BSST01}.

\end{corollary}

\begin{corollary}
The mother inequality also follows from the NSD inequality \eq{nsd} by
applying rule O. The proof is almost the 
same as for the previous corollary. It is easy to see that the
conditions of rule O hold for the protocol from \cite{HHHLT01}.
\end{corollary}

\medskip

We now give the proofs of rules I and O.

\medskip

\begin{proof}{\bf \!\!(of rule I)}
In what follows we shall fix $\epsilon$ and consider a sufficiently
large  blocklength $n$ so that the protocol ${\bf P}_n$ is $\epsilon$-valid,
$\epsilon$-decoupled and accurate to within $\epsilon$.
%We also drop the index $n$ for readability.
Whenever the resource inequality features
$[c \rightarrow c]$ in the input this means that Alice performs
a von Neumann measurement on some subsystem $A_1$ of dimension $D$,
with $\log D = \lfloor n(R + \delta) \rfloor$.
\footnote{If the protocol has
depth $>1$, then in the $i$th round a measurement
is performed on some $A_{1,i}$ of dimension $D_i$ 
such that $\sum_i D_i =  \lfloor n(R + \delta) \rfloor$.
In the analysis below we simply refer to $D$.}
The outcome of this measurement is sent to Bob  
who at the end of the protocol
performs an isometry depending on the received information.
Before Alice's von Neumann measurement, the joint state of  $A_1$ and 
the remaining quantum system $Q$ is
$$
\sum_x \sqrt{p_x}\ket{x}^{A_1} \ket{\phi_x}^{Q},
$$
where 
\be
\left\| \sum_x p_x \proj{x}^{A_1} - \tau_D^{A_1} \right\|_1 \leq \epsilon, 
\label{eq:wqr} 
\ee
and $\tau_D$ is the $D$ dimensional maximally mixed state.
At the end of the protocol Bob
performs some isometry $U_x$ on $Q$, leaving it $\eps$-decoupled
from $x$:
\be
 \left\|\sum_x p_x \proj{x}^{A_1} \otimes \theta_x^{Q} -
\sum_x p_x \proj{x}^{A_1} \otimes \bar{\theta}^{Q} \right\|_1 
%= \sum_x p_x \frac{1}{2} \| \theta_x - \bar{\theta}  \|_1 
\leq \epsilon.
\label{eq:wqr2}
\ee
where $\ket{\theta_{x}} = U_{x} \ket{\phi_{x}}$ and
$\bar{\theta} = \sum_x p_x \theta_x$.
Combining \eq{wqr} and \eq{wqr2} gives
\be
 \left\|\sum_x p_x \proj{x}^{A_1}  \otimes \theta_x^Q -
\tau^{A_1}  \otimes \bar{\theta}^Q \right\|_1 
%= \sum_x p_x \frac{1}{2} \| \theta_x - \bar{\theta}  \|_1 
\leq 2 \epsilon.
\label{eq:wqr3}
\ee
If  Alice refrains from the measurement and
instead sends $A_1$ through a \emph{coherent} channel, %(\ref{map}),
the resulting state is
$$
\sum_x \sqrt{p_x} \ket{x}^{A_1} \ket{x}^{B_1} \ket{\phi_x}^{Q}.
$$
Bob now performs the \emph{controlled} unitary 
$\sum_x \proj{x}^{B_1} \otimes  U_x^{B_1}$, giving rise to
$$
\ket{\Upsilon}^{A_1B_1Q} = 
\sum_x \sqrt{p_x} \ket{x}^{A_1} \ket{x}^{B_1} \otimes \ket{\theta_x}^Q.
$$
\eq{wqr2} may be written as
$$
\| \Upsilon^{A_1Q} - \tau_D^{A_1} \otimes {\Upsilon}^{Q} \|_1 \leq 2 \epsilon.
$$
Invoking  Lemma \ref{pomoc}, there exists
an isometry  $V:B_1 \rightarrow B_2 B_3$ on Bob's side
taking $\Upsilon$ to $\Upsilon'$
such that
$$
\|{\Upsilon'}^{A_1B_2B_3Q} - {\Phi_D}^{A_1 B_2} \otimes  
{\xi}^{B_3Q} \|_1 \leq 2\sqrt{2 \epsilon},
$$
for some purification ${\xi}^{B_3Q}$ of ${\Upsilon}^{Q}$.
Tracing out subsystems gives
$$
\|{\Upsilon'}^{A_1B_2} - {\Phi_D}^{A_1 B_2} \| 
\leq 2\sqrt{2 \epsilon}.
$$
Thus, the total effect of replacing $[c \rightarrow c : \tau]$ by
$[q \rightarrow qq: \tau]$ is the generation of a state close to
$\Phi_D$.
This mapping preserves the $\epsilon$-validity of the original
protocol (with respect to the inputs of $\alpha$) 
since all of Alice's reduced density operators are the same. It also preserves the
$\epsilon$-accuracy of the protocol concerning the $\beta$ resource, as
the final state of $Q$ is the same.
We have thus shown
$$
\alpha + R \,[q \rightarrow qq ] \geq \beta + R \,[q \, q].
$$
Equation (\ref{eq:ccc}) and lemmas \ref{lemma:noo} and \ref{lemma:cancel}
give the desired result
$$
\alpha + \frac{R}{2} \,[q \rightarrow q] \geq \beta + \frac{R}{2} \,[q \, q].
$$
\end{proof}

\medskip

\begin{proof}{\bf \!\!(of rule O)}
Again we  fix $\epsilon$ and consider a sufficiently
large blocklength $n$ so that the protocol ${\bf P}_n$ is $\epsilon$-valid, $\epsilon$-decoupled and
accurate to within $\epsilon$.
Now the roles of Alice and Bob are somewhat interchanged.
Assume that the message $x$ being sent is uniformly
distributed over a set of size $D$, $\log D = \lfloor n(R + \delta) \rfloor$.
Alice performs a unitary operation depending on $x$. 
At the end of the protocol Bob performs a von Neumann measurement on
some subsystem $B_1$ of dimension $D$, yielding  outcome $x'$ with
some probability $p_{x'|x}$.
%which succeeds in reproducing the message with probability $\geq 1 - \epsilon$.
%Alice makes a copy of her message in some system $A_1$.
%Namely, if  we denote by $p_{x'|x}$ the probability of outcome $x'$ given 
%Alice's message was $x$ then, for sufficiently large $n$,
By the $\epsilon$-accuracy of the protocol

\be
%\frac{1}{D} \sum_x p_{x|x} \geq 1 - \epsilon.
\left\| \frac{1}{D} \sum_{x x'} p_{x'|x} \proj{x} \otimes
 \proj{{x'}} - 
\frac{1}{D} \sum_{x} \proj{x} \otimes  \proj{x} \right\|
\leq  \epsilon.
\label{eq:nichi}
\ee
Before Bob's measurement, the state of $B_1$ and 
the remaining quantum system $Q$  conditioned on Alice's message being
$x$ is
$$
\sum_{x'} \sqrt{p_{x'|x}} \ket{x'}^{B_1} \ket{\phi_{xx'}}^{Q}.
$$
Based on the outcome $x'$ of his measurement, Bob
performs some unitary $U_{x'}$ on $Q$, 
yielding the state 
$$
\ket{\Upsilon_{x}}^{B_1 Q} =  \sum_{x'} \sqrt{p_{x'|x}}  \ket{{x'},x}^{B_1Q},
$$
where $ \ket{{x',x}}^{B_1Q} =  \ket{x'}^{B_1} \ket{\theta_{xx'}}^{Q}$
and $\ket{\theta_{xx'}} = U_{x'} \ket{\phi_{xx'}}$.
The decoupling condition says that the state
$$
\sigma^{A_1B_1Q} = \frac{1}{D} \sum_x \proj{x}^{A_1} \otimes 
\Upsilon_x^{B_1 Q}
$$
satisfies
\be
\| \sigma^{A_1B_1Q} - \sigma^{A_1B_1} \otimes \sigma^{Q} \|_1 \leq \epsilon.
\label{eq:dekl}
\ee

The above protocol may be modified to implement \emph{coherent}
communication in lieu of ordinary classical communication. Given a subsystem $A_1$ in the
state $\ket{x}^{A_1}$, Alice  encodes via 
\emph{controlled} unitary operations, eventually yielding 
$$
\ket{x}^{A_1} \sum_{x'} \sqrt{p_{x'|x}} \ket{x'}^{B_1} \ket{\phi_{xx'}}^{Q}.
$$
Bob refrains from measuring $B_1$ and instead 
performs the \emph{controlled} unitary 
$\sum_{x'} \proj{x'}^{B_1} \otimes  U_{x'}^{Q}$, giving rise to
$
\ket{x}^{A_1}\ket{\Upsilon_x}^{B_1Q}$.
Applying the protocol on the purification $\ket{\Phi_D}^{RA_1}$ yields
$$\ket{\Upsilon}^{RA_1B_1Q} := \frac{1}{\sqrt{D}}
%D^{-\smfrac{1}{2}}
\sum_x
\ket{x}^R\ket{x}^{A_1}\ket{\Upsilon_x}^{B_1Q}.$$
\eq{nichi} may be rewritten as
$$
\left\| \frac{1}{D} \sum_{x x'} p_{x'|x} \proj{x}^R \otimes \proj{x}^{A_1} \otimes
 \proj{{x'},x}^{B_1 Q} - 
\frac{1}{D} \sum_{x} \proj{x}^R \otimes \proj{x}^{A_1} \otimes  \proj{{x},x}^{B_1 Q} \right\|
\leq  \epsilon.
$$
From this and Corollary \ref{pomoc2} we get
$$
\| \Upsilon^{RA_1B_1Q} -  \Gamma^{RA_1B_1Q} \|_1 \leq 2 \sqrt{ \epsilon},
$$
where
$$ 
\ket{\Gamma}^{RA_1B_1Q} :=\frac{1}{\sqrt{D}}
%D^{-\smfrac{1}{2}}
\sum_x
\ket{x}^R\ket{x}^{A_1}\ket{x}^{B_1} \ket{\theta_{xx}}^{B_1Q}.
$$
Since ${\Gamma}^{RB_1} = \Phi_D^{RB_1}$,
$$
\| \Upsilon^{RB_1} - \Phi_D^{RB_1} \|_1 \leq 2 \sqrt{ \epsilon}.
$$
By \eq{dekl}
$$
\| \Upsilon^{RB_1Q} - \Upsilon^{RB_1} \otimes \Upsilon^{Q}\|_1 \leq \epsilon.
$$
Combining the two gives
$$
\| \Upsilon^{RB_1Q} - \Phi_D^{RB_1} \otimes \Upsilon^{Q}\|_1 \leq \epsilon + 
2 \sqrt{ \epsilon}.
$$
Define the GHZ state
$$
\ket{\Phi_{GHZ}}^{RA_1B_1} = \frac{1}{\sqrt{D}}
%D^{-\smfrac{1}{2}}
\sum_x \ket{x}^R\ket{x}^{A_1}\ket{x}^{B_1},
$$
so that
$$
\Phi_{GHZ}^{R A_1 B_1} = \Delta^{A_1 \rightarrow A_1 B_1} (\Phi_D^{RA_1}).
$$
Invoking Lemma \ref{pomoc}, there exists
an isometry  $V:A_1 \rightarrow A_2 A_3$ on Alice's side
taking $\Upsilon$ to $\Upsilon'$
such that
$$
\|{\Upsilon'}^{RB_1A_2A_3Q} - {\Phi_{GHZ}}^{R B_1 A_2} \otimes  
{\xi}^{A_3Q} \|_1 \leq 2\sqrt{ \epsilon + 2 \sqrt{ \epsilon}},
$$
for some purification ${\xi}^{A_3Q}$ of ${\Upsilon}^{Q}$.
Tracing out subsystems gives
$$
\|{\Upsilon'}^{RB_1A_2} - {\Phi_{GHZ}}^{R B_1 A_2} \|_1
\leq 2\sqrt{ \epsilon}.
$$
Thus we have successfully replaced $[c \rightarrow c]$ by $[q \rightarrow qq]$.
This mapping preserves the $\epsilon$-validity of the original
protocol (with respect to the inputs of $\alpha$) 
since all of Alice's reduced density operators are the same. It also preserves the
$\epsilon$-accuracy of the protocol concerning the $\beta$ resource, as
the final state of $Q$ is the same.
We have thus shown that
$$
\alpha \geq \beta + R \,[q \rightarrow q q : \tau].
$$
Using \thm{absolutize} and \eq{ccc}
gives the desired result
$$
\alpha  \geq \beta + \frac{R}{2} \, [q \, q] +  \frac{R}{2} \, [q \rightarrow q].
$$
\end{proof}

\vfill\pagebreak

\section{Two dimensional trade-offs for the family}
\label{sec:trade-off}

It is natural to ask about the optimality of our family of resource
inequalities.  In this section we show that they indeed give rise to
optimal two dimensional capacity regions, the boundaries of which are
referred to as trade-off curves.  To each family member corresponds a
theorem identifying the operationally defined capacity region
$C(\rho^{AB})$ ($C(\cN)$) with a formula $\tilde{C}(\rho^{AB})$
($\tilde{C}(\cN)$) given in terms of entropic quantities evaluated on
states associated with the given noisy resource $\rho^{AB}$ ($\cN$).
Each such theorem consists of two parts: the
\emph{direct coding theorem} which establishes $\tilde{C} \subseteq C$
and the \emph{converse} which establishes $C \subseteq \tilde{C} $.

\subsection{Grandmother protocol}

To prove the trade-offs involving static resources, we will first need
to extend the mother protocol \peq{mama} to a ``grandmother'' RI
by combining it with instrument compression \peq{ict2}.

%The other pertains to the dynamic trade-offs; we give a sort
%of grandfather protocol that subsumes direct coding for
%the two dynamic trade-offs. 

\begin{theorem}[Grandmother]
Given a static resource $\rho^{AB}$, 
for any remote instrument ${\bf T}: A \rightarrow A' X_B$,
the following RI holds
\begin{equation}
  \frac{1}{2} I({A'};EE'|X_B)_\sigma  \, [q \rightarrow q] 
 + I(X_B; BE)_\sigma   [c \rightarrow c] +
\< \rho^{AB} \> \geq \frac{1}{2} I({A'};B|X_B)_\sigma   \,[q \, q]. 
\label{eq:granny}
\end{equation}
In the above,
%$$
%\bbT' = \Delta^{X_B \rightarrow X_A X_B} \circ \bbT;
%$$
the state $\sigma^{X_B A' B E E'}$ is defined by
$$
\sigma^{X_B A' B E E'} = {\tilde{\bbT}}^{A\ra A'E'X_B} (\psi^{ABE}),
$$
where $\proj{\psi}^{ABE} \ext \rho^{AB}$ and
$\tilde{\bbT}: A \rightarrow {A'} E' X_B$ is 
a QP extension of ${\bbT}$.
\end{theorem}
\begin{proof}
By the instrument compression RI (\ref{eq:ict2}),
\ben
\< \rho^{AB} \> + I(X_B; BE)_\sigma [c \rightarrow c] + 
H(X| BE)_\sigma [c \, c] 
& \geq  & \< \rho^{AB} \> + \< \bar{\Delta}^{X_B \rightarrow X_A X_B} \circ  
{\bbT} : \rho^{A} \>\\
& \geq  & \< \bar{\Delta}^{X_B \rightarrow X_A X_B} (\sigma^{X_B A}) \>. 
\label{parte1}
\een
%The state $\sigma$ $T'(\psi)^{ABEE'X_A X_B}$ is of the form
%$$
%\sum_x p_x \proj{x}^{X_A} \otimes \proj{x}^{X_B} \otimes 
%\phi_x^{ABEE'}.
%$$
%By the mother inequality, for all $x$,
%$$
% \< \phi_x^{AB} \> + \frac{1}{2} I(A;EE')_{\phi_x} \, [q \rightarrow q] 
%   \geq \frac{1}{2} I(A;B)_{\phi_x} \,[q \, q]. 
%$$
On the other hand, by \thm{alfav} and the 
mother inequality (\ref{eq:mama}), 
%and noting that
%$I(A;EE'|X) = \sum_x p_x I(A;EE')_{\phi_x}$, etc.,  
$$ 
\< \bar{\Delta}^{X_B \rightarrow X_A X_B} (\sigma^{X_B A'}) \>
  + \frac{1}{2} I(A';EE'|X_B)_\sigma \, [q \rightarrow q] 
\geq \frac{1}{2} I(A'; B| X_B)_\sigma \,[q \, q].
$$
The grandmother RI is obtained by adding the above RIs,
followed by a derandomization via \cor{purim}. 
%Combining the above with (\ref{parte1}), 
%$$
%\< \psi^{AB} \> +  I(X; BE) [c \rightarrow c] + H(X| BE) [c c] 
%+ \frac{1}{2} I(A;EE'|X) \, [q \rightarrow q] 
%\geq \frac{1}{2} I(A; B| X) \,[q \, q].
%$$
%Finally, by the derandomization lemma
%$$
%\< \psi^{AB} \> +  I(X; BE) [c \rightarrow c] 
%+ \frac{1}{2} I(A;EE'|X) \, [q \rightarrow q] 
%\geq \frac{1}{2} I(A; B| X) \,[q \, q],
%$$
%and we are done. 
\end{proof}

\begin{corollary}
\label{cor:granny2}
In the above theorem, one may consider the special case
where ${\bbT}: A \rightarrow A' X_B$ corresponds to
some ensemble of operations $(p_x,  \cE_x)$, 
$\cE_x: A \rightarrow A'$, via the identification
$$
{\bbT}: \rho^{A} \mapsto 
\sum_x p_x \proj{x}^{X_B} \otimes \cE_x(\rho^{A}). 
$$
Then the $[c \rightarrow c]$ term from \eq{granny}
vanishes identically.
\qed
\end{corollary}

%Note, there is a much simpler
%version of the grandmother: a sort
%of average mother (mother + time sharing), 
%also obtained as a special of the 
%grandmother where $T = (p_x,  \cE_x)$ is
%merely an ensemble of operations. Then $I(X; BE) = 0$.

\subsection{Trade-off for noisy super-dense coding}
\label{sec:nSD-toff}

Now that we are comfortable with the
various formalisms, the formulae will reflect the
QP formalism, whereas the language  will be more 
in the CQ spirit.

\begin{figure}
\centerline{ {\scalebox{0.50}{\includegraphics{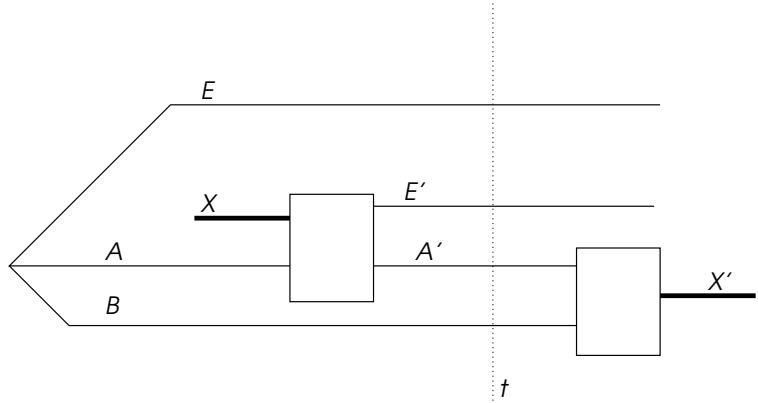}}}}
\caption{A general protocol for noisy super-dense coding.}
\label{fig:nsdfig}
\end{figure}

Given a bipartite state $\rho^{AB}$, 
the noisy super-dense coding capacity region 
$C_{\rm NSD}(\rho^{AB})$ is the two-dimensional region in the 
$(Q,R)$ plane with $Q \geq 0$ and $R \geq 0$ satisfying the RI
\begin{equation}
  \< \rho^{AB} \>  + Q \,[q \rightarrow q] \geq R \,[c \rightarrow c].
\label{nsddef}
\end{equation}

\begin{theorem}
The capacity region $C_{\rm NSD}(\rho^{AB})$ is given by
$$
C_{\rm NSD}(\rho^{AB}) = \tilde{C}_{\rm NSD} (\rho^{AB}):=
\bar{\bigcup_{n=1}^\infty \frac{1}{n}
\tilde{C}_{\rm NSD}^{(1)}( (\rho^{AB})^{\otimes n})},
$$
where the $\bar{S}$ means the closure of a set $S$ and
$\tilde{C}_{\rm NSD}^{(1)}(\rho^{AB})$
is the set of all $R \geq 0$, $Q \geq 0$ such that
$$
 R \leq  Q + \max_\sigma \left\{ I(A' \, \> BX)_\sigma : 
H(A'|X)_\sigma \leq Q \right\}.
$$
In the above, $\sigma$ is of the form
\be
\sigma^{XA'B} = \sum_x p_x \proj{x}^{X} \otimes \cE_x^{A\ra
A'}(\rho^{AB}). 
\label{eq:nsdsig} \ee
for some ensemble of operations $(p_x,  \cE_x)$, 
$\cE_x: A \rightarrow A'$.
\end{theorem}
\begin{proof}
We first prove the converse. Fix $n,R,Q, \delta, \epsilon$,
and use the Flattening Lemma (\ref{lemma:flattening}) so that we can
assume that $k = 1$.
The resources available are
\begin{itemize}
\item The state $(\rho^{AB})^{\otimes n}$
shared between Alice and Bob. Let it be
contained in the system $A^n B^n$, of total dimension
$d^n$, which we shall call $AB$ for short.
\item A perfect quantum channel $\id: A' \rightarrow A'$,
$ \dim A' = 2^{n Q}$,
from Alice to Bob
(after which $A'$ belongs to Bob despite the notation!).
 \end{itemize}
The resource to be simulated is the perfect
classical channel of size $D = 2^{n(R - \delta)}$
on any source, in particular on the
random variable $X$ corresponding to
the uniform distribution $\tau_{D}$.

In the protocol (see \fig{nsdfig}), 
Alice performs a $\{ c q \rightarrow q \}$ 
encoding $(\cE_x: A \rightarrow A')_x$, 
depending on the source random variable, 
and then sends the $A'$ system through the perfect quantum channel.
After time $t$ Bob performs a POVM 
$\Lambda: A'B \rightarrow X'$, 
on the system $A'B$, yielding the random variable $X'$.
The protocol ends at time $t_f$.
Unless otherwise stated, the entropic quantities below
refer to the state of the system at time $t$. 
%The classical ones are timeless.

Since at time $t_f$
the state of the system $XX'$ is supposed to be $\epsilon$-close to 
$\bar{\Phi}_{D}$,
\lem{fano} implies
$$
I(X;X')_{t_f} \geq n (R - \delta) - \eta(\epsilon) - K \epsilon n R.
$$ 
%where $d$ is the dimension of the single shot $AB$ Hilbert space.
By the Holevo bound \cite{Holevo73},
$$
I(X;X')_{t_f} \leq I(X; A'B).
$$
Recall from \eq{trip-entropy} the identity 
$$
I(X; A'B) = H(A') + I(A'\,\rangle BX) - I(A';B) + I(X;B). 
$$
Since $I(A';B) \geq 0$, and in our protocol $I(X;B) = 0$,
this becomes
$$
I(X; A'B) \leq H(A') + I(A'\,\rangle BX). 
$$
Observing that 
$$
nQ \geq H(A') \geq H(A'|X),
$$
these all add up to
$$
R \leq Q + \frac{1}{n} I(A'\,\rangle BX) + \delta + KR \epsilon
+ \frac{\eta(\epsilon)}{n}.
$$
As these are true for any $\epsilon, \delta > 0$ and 
sufficiently large $n$, the converse holds.

Regarding the direct coding theorem, it suffices to
demonstrate the RI
$$
 \< \rho^{AB} \>  + H(A'|X)_\sigma \,[q \rightarrow q] 
\geq I(A'; B|X)_\sigma  \,[c \rightarrow c].
$$
This, in turn, follows from linearly combining
\cor{granny2} with super-dense coding \peq{sd}
much in the same way the noisy super-dense coding RI \peq{nsd}
follows from the mother \peq{mama}.
\end{proof}

\subsection{Trade-off for quantum communication assisted
            entanglement distillation}

Given a bipartite state $\rho^{AB}$, 
the quantum communication assisted entanglement distillation  
capacity region ( or ``mother'' capacity region for short) 
$C_{\rm M}(\rho^{AB})$ is the set of
$(Q,E)$ with $Q \geq 0$ and $E \geq 0$ satisfying the RI
\begin{equation}
  \< \rho^{AB} \>  + Q \,[q \rightarrow q] \geq E \,[q \, q].
\label{eq:momdef}
\end{equation}
(This RI is trivially false for $Q<0$ and trivially true for $Q\geq 0$ and
$E\geq 0$.)
\begin{theorem}
The capacity region $C_{\rm M}(\rho^{AB})$ is given by
$$
C_{\rm M}(\rho^{AB}) = \tilde{C}_{\rm M} (\rho^{AB}):=
\bar{\bigcup_{n=1}^\infty \frac{1}{n}
\tilde{C}_{\rm M}^{(1)}( (\rho^{AB})^{\otimes n})},
$$
where
$\tilde{C}_{\rm M}^{(1)}(\rho^{AB})$
is the set of all $Q \geq 0$, $E \geq 0$ such that
\be
E \leq  Q + \max_\sigma \left\{ I(A' \, \> BX)_\sigma: 
\frac{1}{2} I(A';EE'|X)_\sigma \leq Q \right\}.
\label{eq:mom-toff}\ee
In the above, $\sigma$ is the QP version of \eq{nsdsig},
namely
\be
\sigma^{XA'BEE'} = 
\sum_x p_x \proj{x}^{X} \otimes U_x^{A\ra A'E'}(\psi^{ABE}).
\label{eq:nsdsig-QP}
\ee
for some ensemble of isometries $(p_x, U_x)$, 
$U_x: A \rightarrow A'E'$, and purification
$\proj{\psi}^{ABE} \ext \rho^{AB}$.
\end{theorem}
\begin{proof}
We first prove the converse, which in this case follows
from the converse for the noisy super-dense coding trade-off.
The main observation is that super-dense coding (\eq{sd}) induces
an invertible linear map $f$ 
between the $(Q,E)$ and $(Q,R)$ planes corresponding
to the mother capacity region and that of noisy super-dense coding,
respectively, defined by
$$
f: (Q, E) \mapsto (Q + E, 2E).
$$
By adding superdense coding (i.e. $E\qq+E\qtq \geq 2E\ctc$) to the
mother \peq{momdef}, we find
\be
f( C_{\rm M}) \subseteq C_{\rm NSD}.
\ee
On the other hand, by inspecting the definitions of $\tilde{C}_{\rm
NSD}$ and $\tilde{C}_{\rm M}$, we can verify
\be 
\tilde{C}_{\rm NSD} =  f(\tilde{C}_{\rm M}).
\ee
The converse for the noisy super-dense coding trade-off 
is written as $ C_{\rm NSD} \subseteq \tilde{C}_{\rm NSD}$.
As $f$ is a bijection, putting everything together we have
$$
{C}_{\rm M} \subseteq f^{-1} (C_{\rm NSD})
\subseteq  f^{-1} (\tilde{C}_{\rm NSD}) = \tilde{C}_{\rm M},
$$
which is the converse for the mother trade-off.

The direct coding theorem follows immediately from 
\cor{granny2}.
\end{proof}

\subsection{Trade-off for noisy teleportation}
\label{ntp-toff}

Given a bipartite state $\rho^{AB}$, 
the noisy super-dense coding capacity region 
$C_{\rm NTP}(\rho^{AB})$ is a two-dimensional region in the 
$(R,Q)$ plane with $R \geq 0$ and $Q \geq 0$ satisfying the RI
\begin{equation}
  \< \rho^{AB} \>  + R \,[c \rightarrow c] \geq Q \,[q \rightarrow q].
\label{eq:ntpdef}
\end{equation}

\begin{theorem}
The capacity region $C_{\rm NTP}(\rho^{AB})$ is given by
$$
C_{\rm NTP}(\rho^{AB}) = \tilde{C}_{\rm NTP} (\rho^{AB}):=
\bar{\bigcup_{n=1}^\infty \frac{1}{n}
\tilde{C}_{\rm NTP}^{(1)}( (\rho^{AB})^{\otimes n})},
$$
where $\tilde{C}_{\rm NTP}^{(1)}(\rho^{AB})$
is the set of all $R \geq 0$, $Q \geq 0$ such that
\be
Q \leq  \max_\sigma \left\{ I(A' \, \> BX)_\sigma : 
 I(A'; B| X)_\sigma + I(X; BE)_\sigma \leq R \right\}.
\label{eq:ntp-toff}\ee
In the above, $\sigma$ is of the form
\be
\sigma^{XA'BE} =  {\bbT}(\psi^{ABE}),
\label{eq:ntpsig}
\ee
for some instrument ${\bbT}: A \rightarrow A'X$
and purification $\proj{\psi}^{ABE} \ext \rho^{AB}$.
\end{theorem}
\begin{proof}
We first prove the converse. Fix $n,Q,R, \delta, \epsilon$,
and use the Flattening Lemma so we can assume that the depth is one.
The resources available are
\begin{itemize}
\item The state $(\rho^{AB})^{\otimes n}$
shared between Alice and Bob. Let it be
contained in the system $A^n B^n$, which we shall call $AB$ for
short. 
\item A perfect classical channel of size $2^{n R}$.
 \end{itemize}
The resource to be simulated is
the perfect quantum channel
 $\id_D: A_1 \rightarrow B_1$,
$D = \dim A_1 = 2^{n (Q - \delta)}$,
from Alice to Bob,
on any source, in particular on the maximally entangled
state $\Phi^{A' A_1}$. 

\begin{figure}
\centerline{ {\scalebox{0.50}{\includegraphics{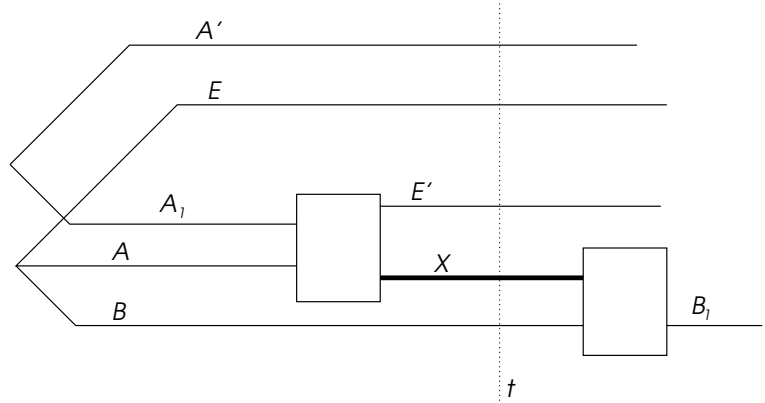}}}}
\caption{A general protocol for noisy teleportation.}
\label{fig:ntpfig}
\end{figure}

In the protocol (see \fig{ntpfig}), 
Alice performs a 
POVM $\Lambda: AA_1 \rightarrow  X$ on the system $AA_1$,
and  sends the outcome random variable $X$ 
through the classical channel.
After time $t$ Bob performs a $\{ cq \rightarrow q \}$ 
decoding quantum operation $\cD: XB \rightarrow B_1$.
The protocol ends at time $t_f$.
Unless otherwise stated, the entropic quantities below
refer to the time $t$. 

Our first observation is that 
performing the POVM $\Lambda$ induces 
an instrument 
${\bbT}: A \rightarrow A' X$,\footnote{  
Indeed, first a pure ancilla 
$A' A_1$ was appended, then another pure ancilla $X$ was appended,
the system  $A A'A_1 X$ was rotated to 
$A' E' X$, and finally $X$ was measured and $E'$ was traced out.}
so that the state of the system $XA'BE$ at time $t$ is indeed of the form
of \eq{ntpsig}.

Since at time $t_f$ the state of the system $A'B_1$ 
is supposed to be $\epsilon$-close to 
${\Phi}_{D}$,
\lem{fano} implies
$$
I(A' \> B_1)_{t_f} \geq n (Q - \delta) - \eta(\epsilon) - K \epsilon n Q.
$$ 
By the data processing inequality,
$$
I(A' \> B_1)_{t_f} \leq I(A' \> BX).
$$
Thus
\be
Q \leq  \frac{1}{n} I(A' \,\rangle BX) + \delta + KQ \epsilon
+ \frac{\eta(\epsilon)}{n}.
\label{eq:NTP-Q-bound}\ee
To bound $R$, start with the identity
$$
I(X; A'B E) = H(A')  + I(A'\,\rangle BEX) - I(A';BE) + I(X;BE). 
$$
Since $I(A';BE) = 0$, $H(A') \geq H(A'|X) $ and 
$I(A'\,\rangle BEX) \geq I(A'\,\rangle BX)$,
this becomes
$$
I(X; A'B E) \geq  I(A'; B|X) + I(X;BE).
$$
Combining this with 
$$
nR  \geq   H(X)  \geq   I(X; A'BE)
$$
gives the desired
\be
R \geq \frac{1}{n} [ I(A' ;B |X) + I(X;BE)].
\label{eq:NTP-R-bound}\ee
As \eqs{NTP-Q-bound}{NTP-R-bound} are true for any $\epsilon, \delta >
0$ and sufficiently large $n$, the converse holds.

Regarding the direct coding theorem, it suffices to
demonstrate the RI
\be
 \< \rho^{AB} \>  + (I(A'; B|X)_\sigma + I(X;BE)_\sigma)
 \,[c \rightarrow c]
\geq  I(A'\, \> BX)_\sigma \,[q \rightarrow q]. 
\label{dctntp}
\ee
Linearly combining
the grandmother RI (\eq{granny}) with teleportation (\eq{tp}),
much in the same way the variation on the noisy teleportation
RI (\eq{ntp2}) was obtained  from the mother (\eq{mama}), we have
$$
 \< \rho^{AB} \>  + (I(A'; B|X)_\sigma + I(X;BE)_\sigma)
 \,[c \rightarrow c] + o [q \, q]
\geq  I(A'\, \> BX)_\sigma \,[q \rightarrow q]. 
$$
Equation (\ref{dctntp}) follows by invoking \lem{noo}
and \eq{hashing}.
\end{proof}

\subsection{Trade-off for classical communication assisted
            entanglement distillation}
\label{sec:distill-toff}

Given a bipartite state $\rho^{AB}$, 
the classical communication assisted entanglement distillation  
capacity region (or ``entanglement distillation'' capacity region for short) 
$C_{\rm ED}(\rho^{AB})$ is the two-dimensional region in the 
$(R,E)$ plane with $R \geq 0$ and $E \geq 0$ satisfying the RI
\begin{equation}
  \< \rho^{AB} \>  + R \,[c \rightarrow c] \geq E \,[q \, q].
  \label{eq:eddef}
\end{equation}

\begin{theorem}
The capacity region $C_{\rm ED}(\rho^{AB})$ is given by
$$
C_{\rm ED}(\rho^{AB}) = \tilde{C}_{\rm ED} (\rho^{AB}):=
\bar{\bigcup_{n=1}^\infty \frac{1}{n}
\tilde{C}_{\rm ED}^{(1)}( (\rho^{AB})^{\otimes n})},
$$
where
$\tilde{C}_{\rm ED}^{(1)}(\rho^{AB})$
is the set of all $R \geq 0$, $E \geq 0$ such that
\be
E \leq  \max_\sigma \left\{ I(A' \, \> BX)_\sigma: 
I(A'; EE'|X)_\sigma + I(X; BE)_\sigma \leq R \right\},
\label{eq:ed-toff}\ee
In the above, $\sigma$ is the
fully QP version of \eq{ntpsig},
namely
\be
\sigma^{XA'BEE'} =  {\bbT}'(\psi^{ABE}),
\label{edsig}
\ee
for some instrument ${\bbT}: A \rightarrow A'E'X$
with pure quantum output
and purification $\proj{\psi}^{ABE} \ext \rho^{AB}$.
\end{theorem}
\begin{proof}
We first prove the converse, which in this case follows
from the converse for the noisy teleportation trade-off.
The argument very much parallels that of the converse for the 
mother trade-off.
The main observation is that teleportation (\eq{tp}) induces
an invertible linear map $g$ 
between the $(R,E)$ and $(R,Q)$ planes corresponding
to the entanglement distillation capacity region
and that of noisy teleportation,
respectively, defined by
$$
g: (R, E) \mapsto (R + 2E, E).
$$
By applying TP to \eq{eddef}, we find
\be
g( C_{\rm ED}) \subseteq C_{\rm NTP}.
\ee
On the other hand, from the definitions of $\tilde{C}_{\rm ED}$ and
$\tilde{C}_{\rm NTP}$ (\eqs{ed-toff}{ntp-toff}), we have
\be 
\tilde{C}_{\rm ED} =  g(\tilde{C}_{\rm NTP}).
\ee
The converse for the noisy teleportation trade-off 
is written as $ C_{\rm NTP} \subseteq \tilde{C}_{\rm NTP}$.
As $g$ is a bijection, putting everything together we have
$$
{C}_{\rm ED} \subseteq g^{-1} (C_{\rm NTP})
\subseteq  g^{-1} (\tilde{C}_{\rm NTP}) = \tilde{C}_{\rm ED},
$$
which is the converse for the entanglement distillation trade-off.

Regarding the direct coding theorem, it suffices to
demonstrate the RI
\be
 \< \rho^{AB} \>  + (I(A'; EE'|X)_\sigma + I(X;BE)_\sigma)
 \,[c \rightarrow c]
\geq  I(A'\, \> BX)_\sigma \,[q \, q]. 
\label{eq:dcted}
\ee
Linearly combining
the grandmother RI (\eq{granny}) with teleportation (\ref{eq:tp}),
much in the same way the variation on the hashing
RI (\eq{hashing2}) was obtained  from the mother (\eq{mama}), we have
$$
 \< \rho^{AB} \>  + (I(A'; EE'|X)_\sigma + I(X;BE)_\sigma)
 \,[c \rightarrow c] + o [q \, q]
\geq  I(A'\, \> BX)_\sigma \,[q \rightarrow q]. 
$$
\eq{dcted} follows by invoking \lem{noo}
and \eq{hashing}.
\end{proof}

\subsection{Trade-off for entanglement assisted
            quantum communication}
\label{sec:father-toff}

Given a noisy quantum channel $\cN: A' \rightarrow B$, 
the entanglement  assisted quantum communication
capacity region ( or ``father'' capacity region for short) 
$C_{\rm F}(\cN)$ is the region of
$(E,Q)$ plane with $E \geq 0$ and $Q \geq 0$ satisfying the RI
\begin{equation}
  \< \cN \>  + E \,[q \, q] \geq Q \,[q \rightarrow  q].
  \label{daddef}
\end{equation}

\begin{theorem}\label{thm:father-toff}
The capacity region $C_{\rm F}(\cN)$ is given by
$$
C_{\rm F}(\cN) = \tilde{C}_{\rm F} (\cN):=
\bar{\bigcup_{n=1}^\infty \frac{1}{n}
\tilde{C}_{\rm F}^{(1)}( \cN^{\otimes n})},
$$
where
$\tilde{C}_{\rm F}^{(1)}(\cN)$
is the set of all $E \geq 0$, $Q \geq 0$ such that
\ben
Q & \leq &  E + I(A \, \> B)_\sigma \\
Q & \leq &  \frac{1}{2} I(A;B)_\sigma.
\een
In the above, $\sigma$ is of the form 
$$
\sigma^{ABE} =  U_\cN \circ  \cE (\phi^{A A''}),
$$
for some pure input state $\ket{\phi^{A A''}}$,
encoding operation $\cE: A'' \rightarrow A'$,
and where $U_\cN: A' \rightarrow BE$ is an isometric extension of $\cN$.

%WE WANT THE CAPACITY REGION TO LOOKLIKE THIS:
%\ben
%Q & \leq &  E + I(R \, \> BX)_\sigma \\
%Q & \leq &  \frac{1}{2} I(R;B|X)_\sigma.
%\een
%In the above, $\sigma$ is of the form 
%$$
%\sigma^{XRBE} = \sum_x p_x \proj{x}^X \otimes U(\phi_x^{RA'})
%$$
%for some pure input ensemble $( p_x, \ket{\phi_x}^{RA'} )_x$,
%and where $U: A' \rightarrow BE$ is an isometric extension of $\cN$.

\end{theorem}

This tradeoff region includes two well-known limit points.  When
$E=0$, the quantum capacity of $\cN$ is
$I(A\>B)$ \cite{Lloyd96,Shor02,Devetak03}, and for $E>0$, entanglement
distribution ($\qtq\geq\qq$) means it should still be bounded by
$I(A\>B) + E$.  On the other hand, when given unlimited entanglement,
the classical capacity is $I(A;B)$ \cite{BSST01} and thus the quantum
capacity is never greater than $\half I(A;B)$ no matter how much
entanglement is available.  These bounds meet when $E=\half I(A;E)$
and $Q=\half I(A;E)$, the point corresponding to the father protocol.
Thus, the goal of our proof is to show that the father protocol is
optimal.

\begin{proof}
We first prove the converse. Fix $n,E,Q, \delta, \epsilon$,
and use the Flattening Lemma to reduce the depth to one.
The resources available are
\begin{itemize}
\item The channel  $\cN^{\otimes n}: {A'}^n \rightarrow {B^n}$ 
from Alice to Bob. We shall shorten ${A'}^n$ to $A'$ and 
${B}^n$ to $B$.
\item  The maximally entangled state $\Phi^{T_A T_B}$, 
$\dim T_A = \dim T_B = 2^{n E }$,
shared between Alice and Bob.
 \end{itemize}
The resource to be simulated is
the perfect quantum channel
 $\id_D: A_1 \rightarrow B_1$,
$D = \dim A_1 = 2^{n (Q - \delta)}$,
from Alice to Bob,
on any source, in particular on the maximally entangled
state $\Phi^{R A_1}$. 

\begin{figure}
\centerline{ {\scalebox{0.50}{\includegraphics{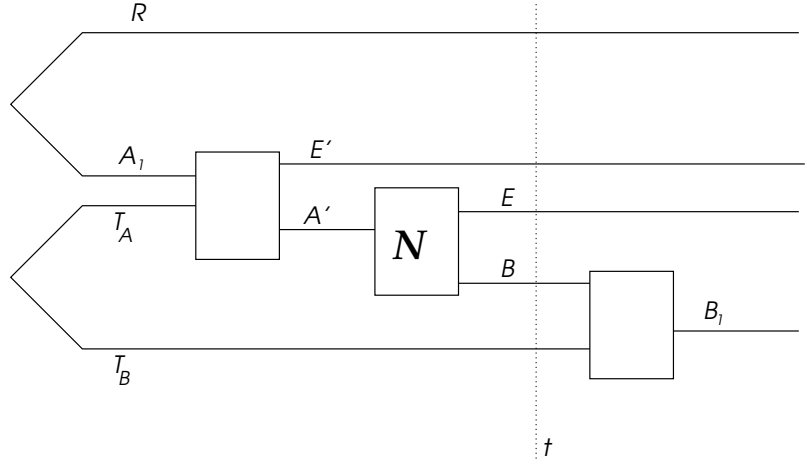}}}}
\caption{A general protocol for entanglement assisted
quantum communication.}
\label{fig:dadfig}
\end{figure}

In the protocol (see \fig{dadfig}), 
Alice performs a general
encoding map $\cE: A_1 T_A \rightarrow A' E'$
and sends the system $A'$ through the noisy channel
$\cN:{A' \rightarrow B}$.
After time $t$ Bob performs a decoding 
operation $\cD: B T_B \rightarrow B_1$.
The protocol ends at time $t_f$.
Unless otherwise stated, the entropic quantities below
refer to the time $t$.

Define $A := R T_B$ and $A'' := A_1 T_A$.
Since at time $t_f$ the state of the system $RB_1$ is supposed 
to be $\epsilon$-close to ${\Phi}_{D}$,
\lem{fano} implies
$$
I(R \, \> B_1)_{t_f} \geq n (Q - \delta) - \eta (\epsilon) - 
K \epsilon n Q.
$$ 
By the data processing inequality,
$$
I(R \,\> B_1)_{t_f} \leq I(R \, \> BT_B).
$$
Together with the inequality 
$$ 
I(R \,\rangle B T_B) \leq
I(R T_B\,\rangle B) + H(T_B),
$$
since $E = H(T_B)$, the above implies
$$
Q  \leq  E + \frac{1}{n} I(A \> B) 
+ \delta + KQ \epsilon + \frac{\eta(\epsilon)}{n}. 
$$
Combining this  with
$$
H(A) = H(R) + H(T_B) = nQ + nE.
\label{sumy}
$$
gives
$$
Q \leq \frac{1}{2n} I(A;B) + 
\delta/2 + KQ \epsilon/2 + \frac{\eta(\epsilon)}{2n}.
$$
As these are true for any $\epsilon, \delta > 0$ and 
sufficiently large $n$, the converse holds.

Regarding the direct coding theorem, it follows 
directly form the father RI
$$
 \< \cN \>  + \half I(A;E)_\sigma \,[q \, q] \geq 
\half I(A;B)_\sigma \,[q \rightarrow  q].
$$

%Notice, however, that due to the existence of $E'$,
%the state  that is fed into the channel is generally mixed.
%So we have proved the converse with respect to mixed input states
%$\phi^{A A'}$. It is easy to show, however, that there
%is always a pure input state that does better.
\end{proof}

\subsection{Trade-off for entanglement assisted
            classical communication}
\label{sec:NCE-toff}

The result of this subsection was first proved by Shor 
in \cite{Shor04}. Here we state it for completeness, and 
give an independent proof of the converse.
An alternative proof of the direct coding theorem was sketched 
in \cite{DS03} and is pursued in \cite{DHLS05} to
unify this result with the father trade-off.

Given a noisy quantum channel $\cN: A' \rightarrow B$, 
the entanglement assisted classical communication
capacity region (or ``entanglement assisted'' capacity region 
for short)
$C_{\rm EA}(\cN)$ is the set of all points
$(E,R)$ with $E \geq 0$ and $R \geq 0$ satisfying the RI
\begin{equation}
  \< \cN \>  + E \,[q \, q] \geq R \,[c \rightarrow  c].
  \label{cl-daddef}
\end{equation}

\begin{theorem}
The capacity region $C_{\rm EA}(\cN)$ is given by
$$
C_{\rm EA}(\cN) = \tilde{C}_{\rm EA} (\cN):=
\bar{\bigcup_{n=1}^\infty \frac{1}{n}
\tilde{C}_{\rm EA}^{(1)}( \cN^{\otimes n})},
$$
where
$\tilde{C}_{\rm EA}^{(1)}(\cN)$
is the set of all $E \geq 0$, $R \geq 0$ such that
\be
R \leq  \max_\sigma \left\{ I(AX;B)_\sigma
    : E \geq H(A|X)_\sigma  \right\}.
\label{eq:eac-toff}
\ee
In the above, $\sigma$ is of the form 
\be
\sigma^{XAB} = \sum_x p_x \proj{x}^X \otimes \cN(\phi_x^{AA'}),
\label{sigea}
\ee
for some pure input ensemble $( p_x, \ket{\phi_x}^{AA'} )_x$.
\end{theorem}
\begin{proof}
We first prove the converse. Fix $n,E,Q, \delta, \epsilon$,
and again use the flattening lemma to reduce depth to one.
The resources available are
\begin{itemize}
\item The channel  $\cN^{\otimes n}: {A'}^n \rightarrow {B^n}$ 
from Alice to Bob. We shall shorten ${A'}^n$ to $A'$ and 
${B}^n$ to $B$.
\item  The maximally entangled state $\Phi^{T_A T_B}$, 
$\dim T_A = \dim T_B = 2^{n E }$,
shared between Alice and Bob.
 \end{itemize}
The resource to be simulated is the perfect
classical channel of size $D = 2^{n(R - \delta)}$
on any source, in particular on the
random variable $X$ corresponding to
the uniform distribution $\tau_{D}$.

\begin{figure}
\centerline{ {\scalebox{0.50}{\includegraphics{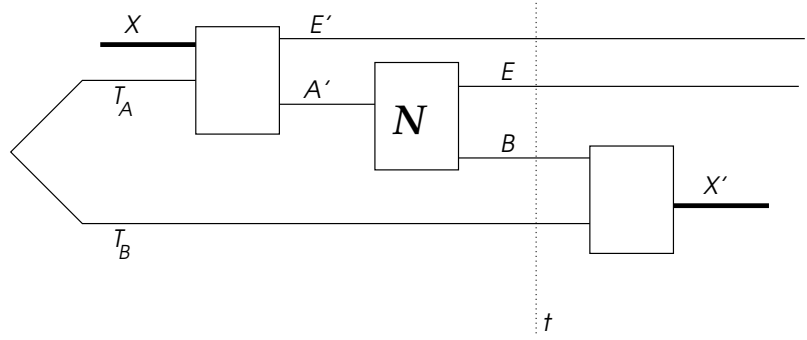}}}}
\caption{A general protocol for entanglement assisted
classical communication.}
\label{fig:eafig}
\end{figure}

In the protocol (see \fig{eafig}), 
Alice performs a $\{ c q \rightarrow q \}$ 
encoding $(\cE_x: T_A \rightarrow A')_x$, 
depending on the source random variable, 
and then sends the $T_A$ system through the noisy channel
$\cN:{A' \rightarrow BE}$.
After time $t$ Bob performs a POVM 
$\Lambda: T_B B \rightarrow X'$, 
on the system $T_B B$, yielding the random variable $X'$.
The protocol ends at time $t_f$.
Unless otherwise stated, the entropic quantities below
refer to the state of the system at time $t$. 

\par
Since at time $t_f$
the state of the system $XX'$ is supposed to be $\epsilon$-close to 
$\bar{\Phi}_{D}$,
\lem{fano} implies
$$
I(X;X')_{t_f} \geq n (R - \delta) - \eta(\epsilon) - K \epsilon n R.
$$ 
By the Holevo bound
$$
I(X;X')_{t_f} \leq I(X; T_BB).
$$
Using the chain rule twice, we find
\ben
I(X; T_BB) &=& I(X;B|T_B) + I(X;T_B)
\\ &=& I(XT_B;B) + I(X;T_B) - I(T_B;B)
\een
Since  $I(T_B;B) \geq 0$  
and in this protocol $I(X;T_B) = 0$, this becomes
$$
I(X; T_BB) \geq  I(XT_B;B).
$$
These all add up to
$$
R \leq \frac{1}{n}I(XT_B;B) + \delta + Kd \epsilon
+ \frac{\eta{\epsilon}}{n},
$$
while on the other hand,
$$
nE \geq H(T_B|X).
$$
As these are true for any $\epsilon, \delta > 0$ and 
sufficiently large $n$, we have thus shown a variation
on the converse with the state $\sigma$ from
(\ref{sigea}) replaced by $\tilde{\sigma}$,
$$
\tilde{\sigma}^{XABE'} = \sum_x p_x \proj{x}^X \otimes 
\cN\circ U_x^{A'' \rightarrow 
A' E'}(\phi^{AA''}),
$$
defining $A := T_B$  and letting $U_x: T_A \rightarrow 
A' E'$ be the isometric extension of $\cE_x$.

However, this is a weaker result than we would like; the converse we
have proved allows arbitrary noisy encodings and we would like to show
that isometric encodings are optimal, or equivalently that the $E'$
register is unnecessary.  We will accomplish this, following Shor
\cite{private-shor-04}, by using a standard trick
of measuring $E'$ and showing that the protocol can only improve.
If we apply the dephasing map $\bar{\id}: E' \rightarrow Y$ 
to $\tilde{\sigma}^{ABE'}$, we obtain a state of the form
$$
{\sigma}^{XYAB} = \sum_{xy} p_{xy} \proj{x}^X \otimes \proj{y}^{Y}
 \otimes \cN (\psi_{xy}^{AA'}).
$$
The converse now follows from
\ben
I(B;AX)_{\tilde{\sigma}} & \leq & I(B;AXY)_{\sigma} \\
H(A|X)_{\tilde{\sigma}} & \geq & H(A|XY)_{\sigma}.
\een
\end{proof}

\vfill\pagebreak

\section{Conclusion}
\label{sec:conclusion}
We have shown how to set up a systematic theory of quantum information
resources. We restricted attention to communication
scenarios with two active protagonists 
connected by unidirectional channels with passive feedback.
After mastering the formal foundations, this theory allows for
fairly flexible play with existing protocols, and derivation of
new ones. The main tools for the latter turned out to be derandomization
and coherification. Then we went on to prove trade-off converses for a family
of protocols. Again the general resource calculus came in handy
to save work, and to organize the converse proofs.

The primary limitation is that our approach is most successful when
considering one-way communication and when dealing with only one noisy
resource at a time.  These, and other limitations, suggest a number of ways in which we
might imagine revising the notion of an asymptotic resource 
given in Definition \ref{def:asy-resource}.  For example, if we were to explore
unitary and/or bidirectional resources more carefully, then we would
need to reexamine our treatments of depth and of relative resources.
Recall that % in %\defn{asy-ineq} 
we (1) always simulate the depth-1
version of the output resource, (2) are allowed to use a depth-$k$
version of the input resource where $k$ depends only on the target
inefficiency and not the target error.  These features were chosen
rather delicately in order to guarantee the convergence of the error
and inefficiency in the Composability Theorem
\ref{thm:composability}, which 
in turn gets most of its depth blow-up from the double-blocking of the
Sliding Lemma \ref{lemma:sliding}.  However, it is possible that a
different model of resources would allow protocols which deal with
depth differently.  This won't make a difference for one-way resources
due to the Flattening Lemma \ref{lemma:flattening}, but there is
evidence that depth is an important issue in bidirectional
communication \cite{Klauck01}; on the other hand, it is unknown how
quickly depth needs to scale with $n$.

Relative resources are another challenge for studying bidirectional
communication.  As we discussed in Section \ref{sec:finite-res}, if $\rho^{AB}$
cannot be bilocally prepared then $\<\cN : \rho^{AB}\>$ fails to
satisfy \eq{asy-monotone} and is  thus not a valid resource.  The
problem is that being able to simulate $n$ uses of a channel on $n$
copies of a correlated or entangled state is not necessarily stronger
than the ability to simulate $n-1$ uses of the channel on $n-1$ copies
of the state.  The fact that many bidirectional problems in classical
information theory \cite{Shannon61} remain unsolved is an indication
that the quantum versions of these problems will be difficult.  On the
other hand, it is possible that special cases, such as unitary gates
or Hamiltonians, will offer simplifications not possible in the
classical case \cite{BHLS02,CLL05}.

Another challenge to our definition of a resource comes from unconventional
``pseudo-resources'' that resemble resources in many ways but fail to
satisfy the quasi-i.i.d. requirement \eq{asy-quasi-iid}.  For
example, the ability to remotely prepare an arbitrary $n$ qubit state cannot be
simulated by the ability to remotely prepare $k$ states of
$n(1+\delta)/k$ qubits each.  There are many fascinating open
questions surrounding this ``single-shot'' version of remote state
preparation (RSP); for
example, is the RSP capacity of a channel ever greater than its quantum
capacity?\footnote{Thanks to Debbie Leung for suggesting this
question.} The case of a noiseless channel was treated
in~\cite{BHLSW03}. Another example comes from the 
``embezzling states'' of \cite{HV03}.  The $n$-qubit embezzling state
can be prepared from $n$ cbits and $n$ ebits (which are also
necessary \cite{HW02}) and can be used as a resource for entanglement
dilution and for simulating noisy quantum channels on
non-i.i.d. inputs \cite{BDHSW05}; however, it also cannot be prepared
from $k$ copies of the $n(1+\delta)/k$-qubit embezzling state.
These pseudo-resources are definitely useful and interesting, but it
is unclear how they should fit into our resource formalism.

Other extensions of the theory will probably require less
modification.  For example, it will not a priori be hard to extend the
theory to multi-user scenarios.  Resources and capacities can even be
defined in non-cooperative situations pervasive in cryptography (see
e.g.~\cite{WNI03}), which will mostly require a more careful
enumeration of different cases.  We can also consider privacy to be a
resource.  Our definitions of decoupled classical communication are a
step in this direction. Also there are expressions for the private 
capacity of quantum channels \cite{Devetak03} and states \cite{DW03b},
and there are cryptographic versions of our Composability
Theorem \cite{BM04, Unruh04}.

Our expressions for trade-off curves also should be seen more as first
steps rather than final answers.  For one thing, we would ultimately
like to have formulae for the capacity that can be efficiently
computed, which will probably require replacing our current
regularized expressions with single-letter ones.  This is related to
the additivity conjectures, which are equivalent for some channel
capacities \cite{Shor03}, but are false for others \cite{DSS97}.

A more reasonable first goal is to strengthen some of the
converse theorems, so that they do not require maximizing over as many
different quantum operations.  As inspiration, note that \cite{BKN98}
showed that isometric encodings suffice to achieve the optimal rate of
quantum communication through a quantum channel.
However, the analogous result for entanglement-assisted quantum
communication is not known.  Specifically, in \fig{dadfig}, we suspect
that the $E'$ register (used to discard some of the inputs) is only
necessary when Alice and Bob share more entanglement than the protocol
can use.  Similarly, it seems plausible to assume that the optimal
form of protocols for noisy teleportation (\fig{ntpfig}) is to perform
a general CPTP preprocessing operation on the shared entanglement,
followed by a unitary interaction between the quantum data  and Alice's
part of the entangled state.  These are only two of the more obvious
examples and there ought to be many possible ways of improving
our formulae.

\section*{Acknowledgements}
This work grew over the course of several years. The authors are
indebted to many people for conversations, encouragement and criticism.
We want to thank everybody who had the patience to listen to us.
In particular, we are grateful to Anura Abeyesinghe for discovering
several errors in an earlier version of the manuscript.
ID was partially supported by the NSF under grant no. CCF-0524811,
and conducted part of this research while at the IBM T.J. Watson Research Center.
AWH was partially supported by the NSA and ARDA under ARO contract
DAAD19-01-1-06.
AW was supported by the U.K. Engineering and Physical Sciences Research
Council's ``IRC QIP'', and by the EC project RESQ (contract IST-2001-37759).

\bibliography{ref}
\bibliographystyle{plain}

\end{document}